\documentclass[journal]{IEEEtran}

\usepackage{graphicx}
\usepackage{float} 
\usepackage{textcomp} 
\usepackage[nospace,noadjust]{cite} 
\usepackage{amsmath,amssymb,amsfonts} 
\usepackage{amsthm} 
\usepackage{mathrsfs} 
\usepackage[mathscr]{euscript} 
\usepackage{epstopdf} 
\usepackage[table,xcdraw]{xcolor}
\usepackage{pgfplots}
\pgfplotsset{compat=newest} 
\usepackage{tikz, tikz-3dplot}
\usepgfplotslibrary{groupplots}
\usetikzlibrary{plotmarks}
\usepackage[normalem]{ulem} 
\usepackage{gensymb} 
\usepackage{balance}
\usetikzlibrary{intersections,backgrounds, patterns}
\usepackage{array}
\usepackage[nolist]{acronym} 
\usepgflibrary{arrows}
\usepackage{multirow} 
\usepackage{makecell}
\renewcommand{\arraystretch}{1.2} 
\usepackage{url}
\usepackage{soul}

\usepackage{diagbox}  

\usepackage{authblk}

\usepackage{amssymb}
\usepackage{pifont}

\usepackage[cmintegrals]{newtxmath} 
\usepackage{silence}
\usepackage[font=footnotesize]{subcaption}
\usepackage[font=footnotesize]{caption}
\WarningsOn

\usepackage[c2]{optidef}
\usepackage{algorithm}
\usepackage{algorithmic}

\definecolor{mycolor1}{RGB}{19, 133, 189}
\definecolor{mycolor2}{RGB}{230, 112, 32}
\definecolor{mycolor3}{RGB}{130, 173, 98}
\definecolor{mycolor4}{rgb}{0.49412,0.18431,0.55686}%
\definecolor{myhist}{RGB}{73, 80, 87}
\definecolor{hgreen}{rgb}{0, 0.5, 0}

\definecolor{tab1}{RGB}{189, 215, 238}
\definecolor{tab11}{RGB}{222, 235, 247}
\definecolor{tab2}{RGB}{248, 203, 173}
\definecolor{tab22}{RGB}{251,229,214}
\definecolor{tab3}{RGB}{197, 224, 180}
\definecolor{tab33}{RGB}{226, 240, 217}
\definecolor{tab4}{RGB}{255, 230, 153}
\definecolor{tab44}{RGB}{255, 242, 204}
\definecolor{tab5}{RGB}{228, 207, 241}
\definecolor{tab55}{RGB}{242,232,248}
\definecolor{tab6}{RGB}{254, 206, 206}
\definecolor{tab66}{RGB}{255, 229, 229}
\definecolor{tab7}{RGB}{217, 217, 217}
\definecolor{tab77}{RGB}{242, 242, 242}

\usepackage{hhline}

\usepackage{enumitem}

\pgfplotsset{every axis/.append style={
		scaled x ticks = false,
		label style={font=\footnotesize},
		tick label style={font=\footnotesize},
		tick scale binop=\times
	}
}
\pgfkeys{/pgf/number format/.cd,
	1000 sep={},
}
\def\plos{\mathcal{P}_\text{LOS}}

\begin{document}

\title{On the Ground and in the Sky: A Tutorial on Radio Localization in Ground-Air-Space Networks}
\author{Hazem~Sallouha,~\IEEEmembership{Member,~IEEE}, Sharief~Saleh,~\IEEEmembership{Member,~IEEE}, Sibren~De~Bast, Zhuangzhuang~Cui,~\IEEEmembership{Member,~IEEE}, Sofie~Pollin,~\IEEEmembership{Senior~Member,~IEEE}, and Henk~Wymeersch,~\IEEEmembership{Senior~Member,~IEEE} 
    \thanks{Hazem Sallouha, Sibren De Bast, Zhuangzhuang Cui, and Sofie Pollin are with the Department of Electrical Engineering, KU Leuven, 3001 Leuven, Belgium. Emails: \{firstname.lastname\}@kuleuven.be.}
    \thanks{Sharief Saleh and Henk Wymeersch are with the Department of Electrical Engineering, Chalmers University of Technology, 412 58 Gothenburg, Sweden. Emails: \{sharief, henkw\}@chalmers.se.}
    \thanks{Sibren De Bast is also with Septentrio, 3001 Leuven, Belgium. Email: sibren.debast@septentrio.com.}
    \thanks{This work was funded by the Hexa-X-II project. Hexa-X-II project has received funding from the Smart Networks and Services Joint Undertaking (SNS JU) under the European Union’s Horizon Europe research and innovation programme under Grant Agreement No 101095759. This work was also partially funded by the Research Foundation – Flanders (FWO) under projects No G0C0623N and G098020N, and by the Swedish Research Council (VR grant 2022-03007).}
    \thanks{The work of Hazem Sallouha was funded by the Research Foundation – Flanders (FWO), Postdoctoral Fellowship No 12ZE222N.}
    }

\maketitle

\begin{abstract}
The inherent limitations in scaling up ground infrastructure for future wireless networks, combined with decreasing operational costs of aerial and space networks, are driving considerable research interest in multisegment ground-air-space (GAS) networks. In GAS networks, where ground and aerial users share network resources, ubiquitous and accurate user localization becomes indispensable, not only as an end-user service but also as an enabler for location-aware communications. This breaks the convention of having localization as a byproduct in networks primarily designed for communications. To address these imperative localization needs, the design and utilization of ground, aerial, and space anchors require thorough investigation. In this tutorial, we provide an in-depth systemic analysis of the radio localization problem in GAS networks, considering ground and aerial users as targets to be localized. Starting from a survey of the most relevant works, we then define the key characteristics of anchors and targets in GAS networks. Subsequently, we detail localization fundamentals in GAS networks, considering 3D positions, orientations, and velocities. Afterward, we thoroughly analyze radio localization systems in GAS networks, detailing the system model, design aspects, and considerations for each of the three GAS anchors. Preliminary results are presented to provide a quantifiable perspective on key design aspects in GAS-based localization scenarios. We then identify the vital roles 6G enablers are expected to play in radio localization in GAS networks.
\end{abstract}

\IEEEpeerreviewmaketitle
\begin{acronym}

\acro{1G}{first generation}
\acro{3GPP}{3rd generation partnership project}
\acro{5G}{fifth generation}
\acro{6G}{sixth generation}

\acro{LEO}{low Earth orbit}
\acro{MEO}{medium Earth orbit}
\acro{GEO}{geostationary orbit}

\acro{AP}{access point}
\acro{A2A}{air-to-air}
\acro{A2G}{air-to-ground}

\acro{ADC}{analog-to-digital converter}
\acro{ADS-B}{automatic dependent surveillance-broadcast}
\acro{AI}{artificial intelligence}
\acro{AOA}{angle-of-arrival}
\acro{AOD}{angle-of-departure}
\acro{ATM}{air traffic management}

\acro{BS}{base station}
\acro{BPSK}{binary phase-shift keying}
\acro{BOC}{binary offset carrier}

\acro{CDF}{cumulative distribution function}
\acro{CRLB}{Cram\'{e}r-Rao lower bound}
\acro{CSI}{channel state information}
\acro{CC}[C\&C]{command and control}
\acro{CF}{cell-free}
\acro{CF-mMIMO}{cell-free massive MIMO}
\acro{CFO}{carrier frequency offset}

\acro{DOF}{degree of freedom}
\acrodefplural{DOF}{degrees of freedom}


\acro{GDOP}{geometric dilution of precision}
\acro{G2A}{ground-to-air}
\acro{G2G}{ground-to-ground}
\acro{GPS}{global positioning system}
\acro{GNSS}{global navigation satellite system}
\acro{GAS}{ground-aerial-space}
\acro{GCS}{global coordinate system}
\acro{CDMA}{code-division multiple access}

\acro{HAP}{high altitude platform}

\acro{IoT}{Internet of things}
\acro{IOC}{intersection of circles}
\acro{IMU}{inertial measurement unit}

\acro{JCAS}{joint communication and sensing}

\acro{KF}{Kalman filter}
\acro{KPI}{key performance indicator}
\acro{KVI}{key value indicator}

\acro{LAP}{low-altitude platform}
\acro{LOS}{line-of-sight}
\acro{LCS}{local coordinate system}
\acro{LR-FHSS}{long-range frequency hopping spread spectrum}

\acro{MIMO}{multiple-input multiple-output}

\acro{MISO}{multiple-input single-output}

\acro{ML}{machine learning}
\acro{MLAT}{multilateration}
\acro{mmWave}{millimeter-wave}

\acro{NTN}{non-terrestrial network}
\acro{NLOS}{non-line-of-sight}

\acro{OFDM}{orthogonal frequency division multiplexing}
\acro{OCXO}{oven controlled crystal oscillator}
\acro{OISL}{optical inter-satellite link}

\acro{PRS}{positioning reference signal}
\acro{PRN}{pseudo-random noise}
\acro{PNT}{positioning, navigation, and timing}
\acro{PF}{particle filter}

\acro{RF}{radio frequency}
\acro{RIS}{reconfigurable intelligent surface}
\acro{RSS}{received signal strength}
\acro{RMSE}{root-mean-square error}
\acro{RTK}{real-time kinematic}
\acro{RTT}{round-trip time}

\acro{SNR}{signal-to-noise ratio}
\acro{SRS}{sounding reference signal}
\acro{SOP}{signals-of-opportunity}
\acro{S2A}{space-to-air}
\acro{S2G}{space-to-ground}
\acro{SISO}{single-input single-output}
\acro{SIMO}{single-input multiple-output}
\acro{SLAM}{simultaneous localization and mapping}
\acro{STEC}{slant total electron content}

\acro{TDOA}{time differences of arrival}
\acro{THz}{terahertz}
\acro{TN}{terrestrial network}
\acro{TOA}{time of arrival}
\acro{TOF}{time of flight}
\acro{TLE}{two-line element}

\acro{UAV}{unmanned aerial vehicle}
\acro{URA}{uniform rectangular array}
\acro{ULA}{uniform linear array}
\acro{UTM}{universal transverse Mercator}
\acro{UE}{user equipment}

\end{acronym}

\section{Introduction}
\IEEEPARstart{O}{ver} the past decade, acquiring location information in wireless systems has shifted from being an optional feature to becoming an imperative functionality that drives the innovation of new wireless applications. Location information serves as the essence of many technological and socially impactful applications, in the areas of mobile devices, \ac{IoT}, autonomous (aerial) vehicles, precision agriculture, as well as live augmented and virtual reality experiences. In applications such as autonomous vehicles \cite{bresson2017}, assisted living \ac{IoT} \cite{witrisal2016high}, \ac{UAV} traffic management \cite{vinogradov2020wireless}, and the vast majority of \ac{IoT} applications \cite{iot,sallouha2019localization,c4hazem}, the dependency on location information is pivotal. Any error in the location information could lead to serious or even life-threatening consequences. For instance, imagine losing the track position in a self-driving car or having the \ac{GPS} signal jammed during a commercial flight journey or a UAV mission.

\begin{table*}[!ht]
    \centering
    \caption{Topic-wise comparison with existing surveys and tutorials with \checkmark and $\partial$ denoting covered and partially covered, respectively.}
    \label{tab:SotA}
    \resizebox{\textwidth}{!}{
    \begin{tabular}{|c|c||c|c|c|c|c|c|c|c|c|c|c|c|}
    \hline
        \multirow{2}*{\textbf{Ref.}} & \multirow{2}*{\textbf{Year}} & \multicolumn{2}{c|}{\textbf{Sec. IV Ground Anchors}} & \multicolumn{2}{c|}{\textbf{Sec. V Aerial Anchors}} & \multicolumn{2}{c|}{\textbf{Sec. VI Space Anchors}} & \multicolumn{6}{c|}{\textbf{Sec. VII Towards 6G GAS Localization}} \\ 
        \cline{3-14}
      	&	& \textbf{Ground targets}  & \textbf{Aerial targets} &\textbf{Ground targets}  & \textbf{Aerial targets}  & \textbf{Ground targets}  & \textbf{Aerial targets}  & \textbf{RIS} & \textbf{JCAS} & \textbf{AI} & \textbf{CF} & \textbf{THz} & \textbf{KVIs} \\ \hline \hline
        \cite{azari2022evolution}		& 2022	&   			& $\partial$  		& \checkmark	&  &  		& 	&   & $\partial$	&	&	& &  \\ \hline
        \cite{vaezi2022cellular} 		& 2022	& $\partial$		&   	   		& \checkmark  	&  & 		& 	&   	   & 	&	&	&  & \\ \hline
        \cite{xiao2022overview}			& 2022	& \checkmark	& $\partial$  		& \checkmark  	&  &   		& 	&  	  & \checkmark	&	&	&  & \\ \hline
        \cite{liu2022survey} 			& 2022	& \checkmark	& 	   	   		& $\partial$   	&  &   		&   &   & \checkmark 	&	&	&  & \\ \hline
        \cite{chen2022tutorial}			& 2022	& \checkmark  	&  		    	& $\partial$ 	 	&  &  		&   & \checkmark  	 & 	&	&	&  \checkmark & \\ \hline
        \cite{khelifi2022loc}	& 2022	& 			  	&  		    	& \checkmark 	 			& $\partial$  &  		&   &  	 & 	& $\partial$	&	&   & \\ \hline
        \cite{shastri2022review}		& 2022	& \checkmark  	&  		    	& 	 	&  &  		&   &  	 & $\partial$ 	& $\partial$	& 	&   & \\ \hline
        \cite{placed2023survey}			& 2023	& \checkmark  		&   	&   		  	&  & 		&   &   	   & $\partial$	&	&	& &  \\ \hline
        \cite{janssen2023survey}			& 2023	&   $\partial$		&   	&   		  	&  & 	\checkmark	&   &   	   & $\partial$	&	&	& &  \\ \hline
        \cite{trevlakis2023localization}& 2023	& \checkmark	& $\partial$  		& 	 	&  & $\partial$	&		&   $\partial$	   & $\partial$	& \checkmark	&	& $\partial$ & \\ \hline
       	\multicolumn{2}{|c||}{\textbf{This work}}  &  \checkmark	& \checkmark & \checkmark & \checkmark & \checkmark & \checkmark & \checkmark & \checkmark & \checkmark	& \checkmark	& \checkmark & \checkmark  \\ \hline
    \end{tabular}}
\end{table*}

While localization in terrestrial networks is well incorporated in the different generations of cellular networks, from the \ac{1G} to the \ac{5G} \cite{del2017survey}, the integration of \acp{NTN} in current \ac{5G} and future \ac{6G} wireless networks, necessitate a paradigm shift in localization systems, promoting novel \ac{GAS} multisegment localization systems, in which terrestrial and non-terrestrial segments operate seamlessly, harnessing both horizontal and vertical dimensions. On the ground, current 5G wireless systems handle the increasing connectivity and spectral efficiency demands by employing base stations with a massive number of antennas \cite{marzetta2016fundamentals} and exploring new spectrum bands, such as \ac{mmWave} and Terahertz bands \cite{guo2021quasi}. This combination of a large number of antennas and large radio bandwidth offers exceptionally sharp angle and time measurements, enabling unprecedentedly accurate localization. Departing from the ground upward, the non-terrestrial aerial and space segments follow, which include \acp{LAP}, \acp{HAP}, and satellite networks. Wireless networks in the non-terrestrial segments, which are commonly known as \acp{NTN}, are included in recent study items of \ac{3GPP} \cite{3GPP38811} as well as in \ac{3GPP} Release 17 and Release 18 packages \cite{lin2022overview}. Recent developments of aerial/space technologies coupled with reduced manufacturing costs have enabled various \ac{NTN}-based advanced applications focusing on providing continuous, ubiquitous, and high-capacity connectivity across the globe \cite{azari2022evolution}. Employing \acp{NTN} for connectivity naturally unlocks the potential of also providing \ac{RF}-based localization as a joint service by design. In this tutorial, we provide a systematic analysis of the radio localization system design in the different segments of the \ac{GAS} networks, concentrating on the scope detailed in the following.

\subsection{Scope}
In this work, we focus on the localization system design within the following scope. 
\begin{itemize}
	\item \textit{Radio technologies}: The theory of localization is well mature from the geometry-based perspective. However, key challenges lie in the ways the range and direction estimates are realized and combined, which can be done via \ac{RF}, computer vision, and inertia. The main scope of this work is RF-based localization in wide-area networks, and hence, it excludes short-range technologies such as near-field communications, ultra-wideband, and Bluetooth.
	\item \textit{\ac{GAS} anchors}: Satellite-based localization, e.g., \ac{GNSS} is the most popular way to localize a device, where \ac{GNSS} constellations are generally operated in \ac{MEO} and \ac{GEO}. In this paper, we look beyond GNSS and investigate localization using ground, aerial, and \ac{LEO} satellites.  
	\item \textit{Ground and aerial targets/users}\footnote{Throughout this work, we use \textit{target} and \textit{user} interchangeably to refer to the target to be localized.}: Given that satellites are launched to a predefined orbit, in this work, we focus on the ground and aerial target that emits and/or receives RF signals, and as such, passive devices are out of the scope.
	\item \textit{Outdoor environments}: Over the past decade non-\ac{GNSS}-based localization systems were mainly focused on indoor environments. In this paper, we address the immense demand for GNSS alternative outdoor localization methods, breaking from the confines of indoor short-range localization systems.
\end{itemize}

\subsection{Related Work}
The vast majority of related tutorials and surveys in the literature addressed either \ac{GAS} networks mainly from wireless communications perspective \cite{azari2022evolution,vaezi2022cellular}, or localization systems mainly from the ground segment perspective \cite{xiao2022overview,liu2022survey,chen2022tutorial,khelifi2022loc,shastri2022review,placed2023survey,janssen2023survey,trevlakis2023localization,kuutti2018survey}. A detailed topic-wise comparison between this tutorial and recent relevant tutorials and surveys is presented in Table \ref{tab:SotA}. The table shows that the majority of recent surveys and tutorials in the literature mainly focus on terrestrial localization systems, highlighting the gap regarding tutorials and surveys on non-terrestrial aerial and space localization systems. Broadly speaking, the technical content of related surveys and tutorials can be categorized into terrestrial localization systems, 6G-enabled localization systems, and non-terrestrial localization systems.

Surveys and tutorials studying terrestrial localization systems, which address the localization requirements of different ground users, e.g., people, vehicles, and IoT nodes, can be found in \cite{kuutti2018survey,shastri2022review,placed2023survey}. The authors of \cite{kuutti2018survey} presented an outlook on autonomous vehicle localization methods, along with a comparison and assessment of respective localization methods. A survey on the state-of-the-art in device-based localization and device-free sensing using mmWave communication and radar devices is provided in \cite{shastri2022review}. Active \ac{SLAM} approaches were surveyed in \cite{placed2023survey}. Despite their high relevancy, the relatively short-range ground-based localization systems become rather limited when considering diverse demands not only from the ground but also from the aerial \acp{UE}.
Starting from 2022, there was a noticeable shift in the scope of localization surveys and tutorials towards 6G enablers. In \cite{xiao2022overview}, the advancement of localization ability in 6G networks was emphasized, and a unified study on integrated localization and communication was conducted in \cite{liu2022survey}. The authors of \cite{shen2023five} discussed five facets of 6G, one of which focused on radio localization and sensing. A tutorial on \ac{THz} and mmWave localization systems was presented in \cite{chen2022tutorial}. The added value of \acp{RIS} in 6G localization systems was discussed in \cite{umer2023role,ma2023reconfigurable,chen2022reconfigurable}, detailing use cases, challenges, and the road ahead. However, the localization service in the aforementioned works on 6G was mainly tailored for ground targets by using ground anchors, such as cellular \acp{BS}.  

Regarding non-terrestrial localization systems, in \cite{azari2022evolution} and \cite{vaezi2022cellular}, the evolution of \acp{NTN} in terms of infrastructure and network architecture, transitioning from 5G to 6G, are comprehensively discussed, emphasizing recent advances, scenarios, and envisioned role of \acp{NTN} in 5G and 6G networks. The localization accuracy obtained using aerial anchors in IoT networks was succinctly highlighted in \cite{azari2022evolution} without elaborating on fundamental aspects of non-terrestrial localization systems. Moreover, localization aspects were not fully covered for ground and space segments. In \cite{janssen2023survey}, the authors presented a survey on state-of-the-art localization techniques for IoT applications using ground and space anchors. In particular, they discussed the localization performance in \ac{IoT} networks when using solutions based on terrestrial anchors, \ac{GNSS} anchors, and \ac{LEO} satellite anchors.
Localization in the air segment requires more \acp{DOF} to improve the localization accuracy, especially in the vertical planes where aerial nodes fly. The authors of \cite{khelifi2022loc} provided a survey on \ac{UAV} and swarm \acp{UAV} localization. As an important evolution, space anchors enabled by LEOs were briefly discussed in \cite{trevlakis2023localization} for positioning ground targets without comprehensive emphasis on aerial anchors and targets. Because of the steadily increasing density of aerial vehicles in the sky, aerial target localization has become more prevailing than ever before \cite{trevlakis2023localization,liu2022survey,vaezi2022cellular}.

Finally, in addition to the fact that most existing surveys and tutorials focus on terrestrial localization systems, from a system design point of view, only selected works detail mathematical modeling for the discussed localization technologies, such as \cite{trevlakis2023localization,chen2022tutorial,liu2022survey}. However, these works do not address localization in \acp{NTN}, stressing the need for tutorials to comprehensively investigate the fundamentals and practical implementations of various anchors and targets in integrated \ac{GAS} networks. To our best knowledge, there is no such thorough study that comprehensively investigates the localization functionality using \ac{GAS} networks. Therefore, this work, which presents relevant mathematical modeling and formulation, will facilitate the research community to better understand the localization system design with different \ac{GAS} segments and thus deal with on-demand localization services realized by the most superior and suitable methods.

\subsection{Contribution}
Considering the tremendous potential to realize precise localization services using modern ground, aerial, and space anchors, there is an urgent need to provide comprehensive tutorials on how to fully utilize the advantages of such \ac{GAS} segments. Unlike existing works in the literature, this tutorial is distinguished by introducing localization systems in \ac{GAS} networks, detailing system modeling, key design aspects, main considerations for system optimization, as well as the corresponding localization algorithms used. In particular, this tutorial aims to introduce the design foundations and fundamental aspects of localization systems in \ac{GAS} network. We put a spotlight on \ac{RF} localization in all three segments in \ac{GAS} networks by elaborating on key challenges to be addressed and opportunities to be leveraged in future deployments. Furthermore, we detail design aspects and considerations of integrating the main 6G enabler, namely \acp{RIS}, \ac{JCAS}, \ac{AI}, \ac{CF}, and \ac{THz}, in \ac{GAS} localization systems. Finally, we discuss localization \acp{KVI} in the context of \ac{GAS} localization systems, extending the perspective of the conventional localization \acp{KPI} to value-driven indicators.

\begingroup
\renewcommand{\arraystretch}{1.4}
\begin{table}[t]
    \centering
    \caption{The overall structure of the tutorial.}
    \label{tab:toc}
    \resizebox{0.49\textwidth}{!}{
    \begin{tabular}{|ll|}
    \hline
    \multicolumn{2}{|c|}{\cellcolor{tab7}\textbf{I. Introduction}}                                                                      \\ \hline
    \rowcolor{tab77} 
    \multicolumn{1}{|l|}{\cellcolor{tab77}A. Scope}                                        & B. Related Work                   \\ \hline
    \rowcolor{tab77} 
    \multicolumn{1}{|l|}{\cellcolor{tab77}C. Contribution}                                 & D. Organization                   \\ \hline
    \multicolumn{2}{|c|}{\cellcolor{tab1}\textbf{II. Beyond GNSS: GAS Anchors and Targets}}                                            \\ \hline
    \rowcolor{tab11} 
    \multicolumn{1}{|l|}{\cellcolor{tab11}A. GNSS Limitations}                             & B. Characteristics of GAS Anchors \\ \hline
    \rowcolor{tab11} 
    \multicolumn{1}{|l|}{\cellcolor{tab11}C. Characteristics of Ground and Aerial Targets} & D. Localization Scenarios in GAS  \\ \hline
    \multicolumn{2}{|c|}{\cellcolor{tab2}\textbf{III. Localization Fundamentals}}                                                      \\ \hline
    \rowcolor{tab22} 
    \multicolumn{1}{|l|}{\cellcolor{tab22}A. System Geometry in 3D}                        & B. Localization Measurables       \\ \hline
    \rowcolor{tab22} 
    \multicolumn{1}{|l|}{\cellcolor{tab22}C. Snapshot Localization Methods}                & D. From Snapshot to Tracking      \\ \hline
    \rowcolor{tab22} 
    \multicolumn{2}{|l|}{\cellcolor{tab22}E. Key Performance Metrics}                                                          \\ \hline
    \multicolumn{2}{|c|}{\cellcolor{tab3}\textbf{IV. Localization with Ground Anchors (5G)}}                                           \\ \hline
    \rowcolor{tab33} 
    \multicolumn{1}{|l|}{\cellcolor{tab33}A. System Model}                                 & B. Ground Targets: Design Aspects \\ \hline
    \rowcolor{tab33} 
    \multicolumn{1}{|l|}{\cellcolor{tab33}C. Aerial Targets: Special Considerations}       & D. Key Takeaways                  \\ \hline
    \multicolumn{2}{|c|}{\cellcolor{tab4}\textbf{V. Localization with Aerial Anchors}}                                                 \\ \hline
    \rowcolor{tab44} 
    \multicolumn{1}{|l|}{\cellcolor{tab44}A. System Model}                                 & B. Ground Targets: Design Aspects \\ \hline
    \rowcolor{tab44} 
    \multicolumn{1}{|l|}{\cellcolor{tab44}C. Aerial Targets: Special Considerations}       & D. Key Takeaways                  \\ \hline
    \multicolumn{2}{|c|}{\cellcolor{tab5}\textbf{VI. Localization with Space Anchors}}                                                 \\ \hline
    \rowcolor{tab55} 
    \multicolumn{1}{|l|}{\cellcolor{tab55}A. System Model}                                 & B. Ground Targets: Design Aspects \\ \hline
    \rowcolor{tab55} 
    \multicolumn{1}{|l|}{\cellcolor{tab55}C. Aerial Targets: Special Considerations}       & D. Key Takeaways                  \\ \hline
    \multicolumn{2}{|c|}{\cellcolor{tab6}\textbf{VII. Research Directions Towards 6G GAS Localization}}                                \\ \hline
    \rowcolor{tab66} 
    \multicolumn{1}{|l|}{\cellcolor{tab66}A. RIS}                                          & B. JCAS                           \\ \hline
    \rowcolor{tab66} 
    \multicolumn{1}{|l|}{\cellcolor{tab66}C. AI-Empowered Localization}                    & D. Cell-Free Paradigm             \\ \hline
    \rowcolor{tab66} 
    \multicolumn{1}{|l|}{\cellcolor{tab66}E. THz Band}                                     & F. 6G KVIs                        \\ \hline
    \rowcolor{tab66} 
    \multicolumn{2}{|l|}{\cellcolor{tab66}G. Key Takeaways}                                                                    \\ \hline
    \multicolumn{2}{|c|}{\cellcolor{tab7}\textbf{VIII. Conclusion}}                                                                    \\ \hline
    \end{tabular}}
\end{table}
\endgroup

\subsection{Organization}

This tutorial paper introduces the fundamentals of localization systems in \ac{GAS} networks. Unlike existing tutorials, we detail the system modeling of each segment in \ac{GAS} networks and the key design aspects and considerations with respect to both ground and non-terrestrial targets. Moreover, we position GAS segments into the perspective of the future that 6G networks enable, elaborating on the role and advantages to be gained from each of the enablers. In particular, the rest of the paper is organized as follows. Section \ref{sec:BndGNSS} details the rationale that urges for \ac{GNSS} localization alternatives and introduces the characteristics of GAS anchors and targets. In Section \ref{sec:locFun}, we present the \ac{RF}-based localization fundamentals, and we put a special emphasis on the 3D space associated with GAS networks. Sections \ref{sec:G2X}, \ref{sec:A2X}, and \ref{sec:S2X}, respectively, present ground, aerial, and space anchors in GAS networks. In each one of these sections, we detail the corresponding system model, design aspects, and special considerations for ground and aerial target localization. Section \ref{sec:GAS6G} discusses the integration of 6G enablers in GAS networks from a localization point of view and presents localization \acp{KVI} in the context of GAS localization systems. Key takeaways are summarized at the end of Section \ref{sec:G2X}-\ref{sec:GAS6G}. Finally, we conclude this tutorial in Section \ref{sec:conc}. The overall structure of the tutorial is presented in Table \ref{tab:toc}.

The \textit{acronyms} list is presented in Table \ref{table:acronyms}. \textit{Notation}: Italic, simple bold, and capital bold letters represent scalars, vectors, and matrices, respectively. We use $(a_1, a_2, \dots)$ to represent a sequence and $\boldsymbol{a} = [a_1, a_2, \dots]^{\top}$ to represent a column vector, with $[.]^{\top}$ being the transpose operator.

\begin{table*}[t!]
	\caption{List of acronyms}
	\label{table:acronyms}
	\centering
	\begin{tabular}{| l | l || l | l |}
		\hline
		\textbf{Acronym} & \textbf{Description}	& \textbf{Acronym} & \textbf{Description} \\
		\hline
		\hline
		1G 		& 1st generation of wireless telecommunication	&	3GPP 	& 3rd generation partnership project \\
		5G 		& 5th generation of wireless telecommunication	&	6G 		& 6th generation of wireless telecommunication \\
		A2A		& Air-to-air									&	A2G		& Air-to-ground \\
		ADC		& Analog-to-digital converter					&	ADS-B 	& Automatic dependent surveillance-broadcast \\
		AI		& Artificial intelligence						&	AP 		& Access point \\
		AOA 	& Angle-of-arrival								&	AOD		& Angle-of-departure \\
		ATM 	& Air traffic management						&	BS 		& Base station \\
		C\&C 	& Command and control							&	CDF 	& Cumulative distribution function \\
		CDMA	& code-division multiple access					&	CF		& Cell-free \\
		CF-mMIMO& Cell-free massive MIMO						&	CFO		& Carrier frequency offset \\
		CRLB 	& Cram\'{e}r-Rao lower bound					&	CSI 	& Channel state information \\
		DOF		& Degree(s) of freedom							&	G2A 	& Ground-to-air \\
		G2G 	& Ground-to-ground								&	GAS 	& Ground-aerial-space \\
		GCS 	& Global coordinate system						&	GDOP 	& Geometric dilution of precision \\
		GEO 	& Geostationary orbit							&	GNSS 	& Global navigation satellite system \\
		GPS 	& Global positioning system						&	HAP 	& High altitude platform \\
		IMU		& Inertial measurement unit						&	IOC 	& Intersection of circles \\
		IoT 	& Internet of things							&	JCAS	& Joint communication and sensing \\
		KF 		& Kalman filter									&	LAP		& Low-altitude platform \\
		LCS 	& Local coordinate system						&	LEO 	& Low Earth orbit \\
		LOS 	& Line-of-sight									&	LR-FHSS	& Long-range frequency hopping spread spectrum \\
		MEO 	& Medium Earth orbit							&	MIMO 	& Multiple-input multiple-output \\
		MISO	& Multiple-input single-output					&	ML		& Machine learning \\
		MLAT 	& Multilateration								&	mmWave 	& Millimeter-wave \\
		NLOS 	& Non-line-of-sight								&	NTN 	& Non-terrestrial network \\
		OCXO 	& Oven controlled crystal oscillator			&	OISL	& Optical inter-satellite link \\
		OFDM 	& Orthogonal frequency division multiplexing	&	PNT		& Positioning, navigation, and timing \\
		PRN 	& Pseudo-random noise							&	PRS 	& Positioning reference signal \\
		RF 		& Radio frequency								&	RIS		& Reconfigurable intelligent surface \\
		RMSE	& Root-mean-square error						&	RSS 	& Received signal strength \\	
		RTT 	& Round-trip time								&	S2A 	& Space-to-air \\
		S2G 	& Space-to-ground								&	SIMO	& Single-input multiple-output \\
		SISO	& Single-input single-output					&	SLAM	& Simultaneous localization and mapping \\
		SNR 	& Signal-to-noise ratio							&	SOP 	& Signals-of-opportunity \\
		SRS 	& Sounding reference signal						&	STEC	& Slant total electron content \\
		TDOA 	& Time difference of arrival					&	THz		& Terahertz \\
		TLE 	& Two-line element								&	TOA 	& Time of arrival \\
		UAV 	& Unmanned aerial vehicle						&	UE 		& User equipment \\
		URA 	& Uniform rectangular array						&	UTM		& Universal transverse Mercator \\
		\hline
	\end{tabular}
\end{table*}

\section{Beyond GNSS: GAS Anchors and Targets}\label{sec:BndGNSS}
This section presents the main limitations of \ac{GNSS}-based localization systems. Subsequently, we introduce the main characteristics of \ac{GAS} anchors and targets, followed by the corresponding localization scenarios.

\subsection{GNSS Limitations}
Owing to the precise atomic clocks, \ac{GNSS} was developed and first commercialized in the 1990s as a time-based localization system. The past decades have witnessed a sharp rise in GNSS-based applications and particularly \ac{GPS}-based ones \cite{janssen2023survey}. Nowadays, a \ac{GPS} receiver is a common element in almost all mobile devices, offering navigation capability around the globe. Moreover, a standalone \ac{GPS}-based localization system can provide accuracy in the range of tens of meters under ideal channel conditions. Augmenting \ac{GPS} systems with reference signals, such as real-time kinematic, can bring the localization accuracy to centimeter-level \cite{kaplan2005GPS}. Despite the high accuracy, \ac{GPS} systems suffer from some critical shortcomings that can compromise the main objective of wireless applications. In outdoor rich-scattering environments and urban canyons, \ac{GPS} receivers experience degraded performance, failing to provide a position fix. In the following, the main \ac{GPS} drawbacks are highlighted. 

\begin{itemize}[leftmargin=*]
	\item \textit{Vulnerable to jamming}: Each \ac{GPS} satellite broadcasts a radio signal containing its current location and the transmission time information. Under normal conditions, a \ac{GPS} receiver can process the signals from multiple satellites to obtain a fix of its position. However, the reception of \ac{GPS} signals can be severely degraded by either unintentional interference or intentional jamming signals. The received power of \ac{GPS} signals is typically around $-130$\,dBm on the Earth's surface, which is rather low against potential high power interference signals \cite{alshrafi2014compact}. This will result in an intermittent reception at the \ac{GPS} receiver, failing to provide a position estimate stably.
	
	\item \textit{Prone to spoofing}: Spoofing is an intelligent form of malicious attacks in which the attacker broadcasts false \ac{GPS} positions or fakes the \ac{GPS} signal itself \cite{jansen2018crowd}. For instance, false \ac{GPS} positions sent from a fake \ac{ADS-B} transmitter can force an aircraft to deviate from its planned trajectory. Furthermore, an attacker can send fake \ac{GPS} signals, resulting in wrong position estimates for nearby \ac{GPS} receivers. In order to detect spoofing attacks and verify the real position of the target, complementary localization methods such as time-based \cite{sallouha2021aerial} and \ac{RSS}-based methods \cite{liu2019} are employed.
	
	\item \textit{Extra power and size}: The size of \ac{GPS} receivers, including the antenna, has been reduced over time. Nevertheless, for compact low-power devices such as \ac{IoT} nodes in low-power wide-area networks \cite{sallouha2017ulora}, adding a \ac{GPS} receiver will take a significant part of the overall size. This is due to the fact that \ac{GPS} works on a different \ac{RF} band (e.g., 1575.42~MHz) compared to the \ac{IoT} band (i.e., 868~MHz). This means that an extra antenna is always needed for the \ac{GPS} receiver. For example, the Telecom Design \ac{IoT} modem with \ac{GPS} is almost double the size of the one without \ac{GPS} \cite{TD}. In addition to the extra size, \ac{GPS} receivers are known to be power-hungry with respect to battery-powered \ac{IoT} nodes \cite{sallouha2017ulora}, considerably reducing their lifetime. To obtain a position fix, a \ac{GPS} receiver could take a few seconds of continuous processing, consuming a relatively large amount of power. Therefore, due to power and size limitations, many \ac{IoT} deployments avoid including a \ac{GPS} receiver in every node \cite{sallouha2019localization,aernouts}.	
\end{itemize}

\begin{figure*}[t]
	\centering
	\includegraphics[width=0.8\textwidth]{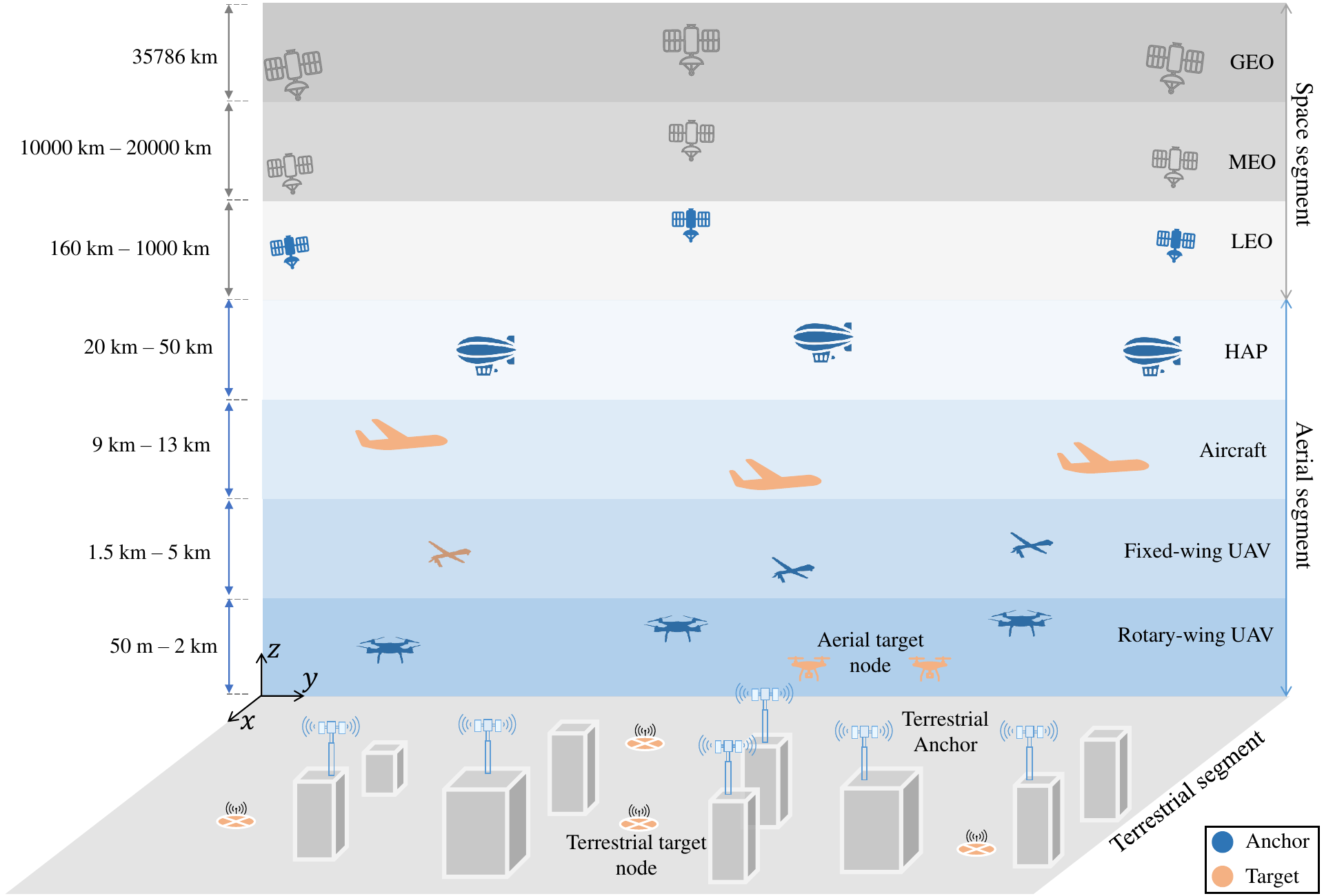}
	\caption{A representation of the different segments and elements of integrated GAS networks and the typical altitude of the various non-terrestrial elements. }
	\label{GASmodel}
\end{figure*}

\subsection{Characteristics of GAS Anchors}
The anchors in \ac{GAS} networks can be broadly classified into terrestrial anchors, which include cellular \acp{BS} and \acp{AP}, and non-terrestrial anchors, which include aerial anchors and space anchors, as visualized in Fig.~\ref{GASmodel}. Anchor nodes are the reference nodes in the localization system. Anchors can be either used to cooperatively localize a target (e.g., uplink localization), or a target can use them as references to localize itself (e.g., downlink localization). To this end, anchors need to have a precise known location. With the advances in the \ac{MIMO} technology and with aerial and space anchors entering the picture, some substantial novel characteristics can be spotted in the \ac{GAS} localization system design.

\begin{itemize} [leftmargin=*] 
	\item \textit{Ground Anchors}: The majority of ground anchors are fixated on the ground with a precisely known location as in the case of cellular network's \acp{BS} and \acp{AP} \cite{del2017survey}. There exist cases where ground mobile robots are used as anchors; however, these cases are mainly indoor \cite{hu2022ltrack}. Modern ground anchors in \ac{5G} networks are designed to have a multitude of antennas, boosting AOA and AOD accuracy. Massive \ac{MIMO} is one of the key technologies used in current 5G networks, in which terrestrial \acp{BS} are equipped with massive antenna arrays. The primary role of these massive antenna arrays is to increase spectral efficiency by providing high beamforming gain and spatial multiplexing for users. From the localization system perspective, such massive antenna arrays play a pivotal role in improving \ac{AOA} and \ac{AOD} measurements by significantly elevating angle estimation accuracy. Ground anchors can work as transmitters and receivers. When working as receivers, they shift the processing cost from the target to the anchors or network infrastructure.
	
	\item \textit{Aerial Anchors}: While terrestrial anchors tend to have fixed positions, aerial anchors have the flexibility needed to constantly adapt their positions, enabling dynamic anchor placement to achieve better localization performance. Aerial anchors can be deployed at altitudes spanning a few tens of meters up to 50~km as detailed in Fig.~\ref{GASmodel}. Aerial anchors are represented by \acp{LAP} and \acp{HAP}. \acp{LAP}, a.k.a \acp{UAV}, cover altitudes from a few meters up to 5 km. \acp{HAP}, on the other hand, are quasi-stationary aerial platforms located at heights of 20--50~km above the Earth’s surface in the stratospheric region of the atmosphere \cite{azari2022evolution}. \acp{UAV} as aerial anchors can exploit their altitude to obtain higher \ac{LOS} probability when compared to their ground counterparts. This results in significant improvements for the localization techniques, in which \ac{NLOS} links are the primary source of error. Moreover, \ac{UAV}'s flexibility can be further exploited to minimize the number of anchors needed to realize a given localization accuracy \cite{sallouha2017aerial}. Using a mobile \ac{UAV}, one can use the measurements collected from multiple \textit{waypoints}\footnote{Stopping points during the \ac{UAV} flying mission, where it hovers at each of them for some time before moving to the next one.} to localize quasi-stationary targets \cite{ebrahimi2020autonomous}. In principle, these waypoints represent virtual anchor points. \ac{LAP} anchors can work as transmitters and receivers, whereas \ac{HAP} anchors are most commonly used as transmitters.
		
	\item \textit{Space Anchors}: The deployment of space anchors takes place at \ac{LEO}, \ac{MEO}, or \ac{GEO} orbits. However, \ac{LEO} satellites, particularly, are attracting considerable research focus due to their reduced latency, broad coverage, and the possibility of deploying mega-constellations with multiple wireless technologies \cite{kodheli2020satellite}. As a key component of \acp{NTN}, \acp{LEO} are not only expected to provide communication service thanks to the suitable altitudes (e.g., 550~km for Starlink) but also to act as a space anchor to localize users. Anchors in \ac{LEO} have the advantage of covering a large part of the Earth. By deploying a constellation of a couple of hundred satellites in LEO, global coverage can be reached for a localization system \cite{janssen2023survey}. However, the received power of such systems is typically a lot lower in comparison with terrestrial networks, making satellite signals mainly suitable for \ac{LOS} communications. Space anchors are generally used only as transmitters; hence, corresponding users need to localize themselves.
\end{itemize}

\begin{table*}[t]
	\centering
	\caption{Elements of the localization scenarios in GAS networks.}
	\begin{tabular}{|c||l|l|l|}
		\hline
		\multicolumn{1}{|c||}{} & \multicolumn{1}{c|}{\textbf{Ground Anchor}}  & \multicolumn{1}{c|}{\textbf{Aerial Anchor}} & \multicolumn{1}{c|}{\textbf{Space Anchor}} \\ \hline \hline
		\multirow{3}{*}{\rotatebox[origin=c]{90}{\parbox[c]{1.2cm}{\centering \textbf{Ground\,\,Target}}}} & 
		$\begin{array}{l}
			\textrm{}\\
			\textrm{\textbf{G2G scenario}, Subsection \ref{subsec:G2G}}\\
			\textrm{}\\
			\textrm{\textbf{Target}:\hspace{0.10cm} Connected \ac{UE},}\\
			\textrm{\hspace{0.98cm} Mobile devices/vehicles}\\
			\textrm{\textbf{Anchor}: Base stations, gNB}
		\end{array}$ & 
		$\begin{array}{l}
			\textrm{}\\
			\textrm{\textbf{A2G scenario}, Subsection \ref{subsec:A2G}}\\
			\textrm{}\\
			\textrm{\textbf{Target}:\hspace{0.10cm} IoT nodes,}\\
			\textrm{\hspace{0.98cm} Mobile devices}\\
			\textrm{\textbf{Anchor}: LAPs, HAPs }
		\end{array}$ &  
		$\begin{array}{l}
			\textrm{}\\
			\textrm{\textbf{S2G scenario}, Subsection \ref{subsec:S2G}}\\
			\textrm{}\\
			\textrm{\textbf{Target}:\hspace{0.10cm} IoT nodes,}\\
			\textrm{\hspace{0.98cm} Mobile devices}\\
			\textrm{\textbf{Anchor}: LEO Satellites}
		\end{array}$ \\ \hline
		\multirow{3}{*}{\rotatebox[origin=c]{90}{\parbox[c]{1cm}{\centering \textbf{Aerial\,\,Target}}}} & 
		$\begin{array}{l}
			\textrm{}\\
			\textrm{\textbf{G2A scenario}, Subsection \ref{subsec:G2A}}\\
			\textrm{}\\
			\textrm{\textbf{Target}:\hspace{0.10cm}  LAPs, Amateur drones}\\
			\textrm{\textbf{Anchor}: Base stations, gNB}\\
		\end{array}$ & 
		$\begin{array}{l}
			\textrm{}\\
			\textrm{\textbf{A2A scenario}, Subsection \ref{subsec:A2A}}\\
			\textrm{}\\
			\textrm{\textbf{Target}:\hspace{0.10cm} LAPs, Amateur drones}\\
			\textrm{\textbf{Anchor}: LAPs, HAPs}
		\end{array}$ & 
		$\begin{array}{l}
			\textrm{}\\
			\textrm{\textbf{S2A scenario}, Subsection \ref{subsec:S2A}}\\
			\textrm{}\\
			\textrm{\textbf{Target}:\hspace{0.10cm} LAPs, Amateur drones}\\
			\textrm{\textbf{Anchor}: LEO Satellites}
		\end{array}$ \\ \hline
	\end{tabular}
	\label{tb:locoElements}
\end{table*}

\subsection{Characteristics of Ground and Aerial Targets}
Given that satellites are launched to a specific orbit, and they are constantly tracked \cite{riebeek2009catalog}, the targets considered in this paper, as visualized in Fig.~\ref{GASmodel}, are the ones at ground and aerial segments, namely, terrestrial targets and aerial targets, respectively. Target nodes are typically power and size-limited when compared to anchors, e.g., mobile \ac{UE}, armature drone, or \ac{IoT} nodes \cite{sallouha2019localization,abu2018error}. Target localization requirements are usually set as constraints on performance metrics (cf. Subsection \ref{Fun_PR}), which may include accuracy, coverage, latency, complexity, power efficiency, stability, scalability, and update frequency.

\begin{figure}[t]
	\centering
	\includegraphics[width=0.45\textwidth]{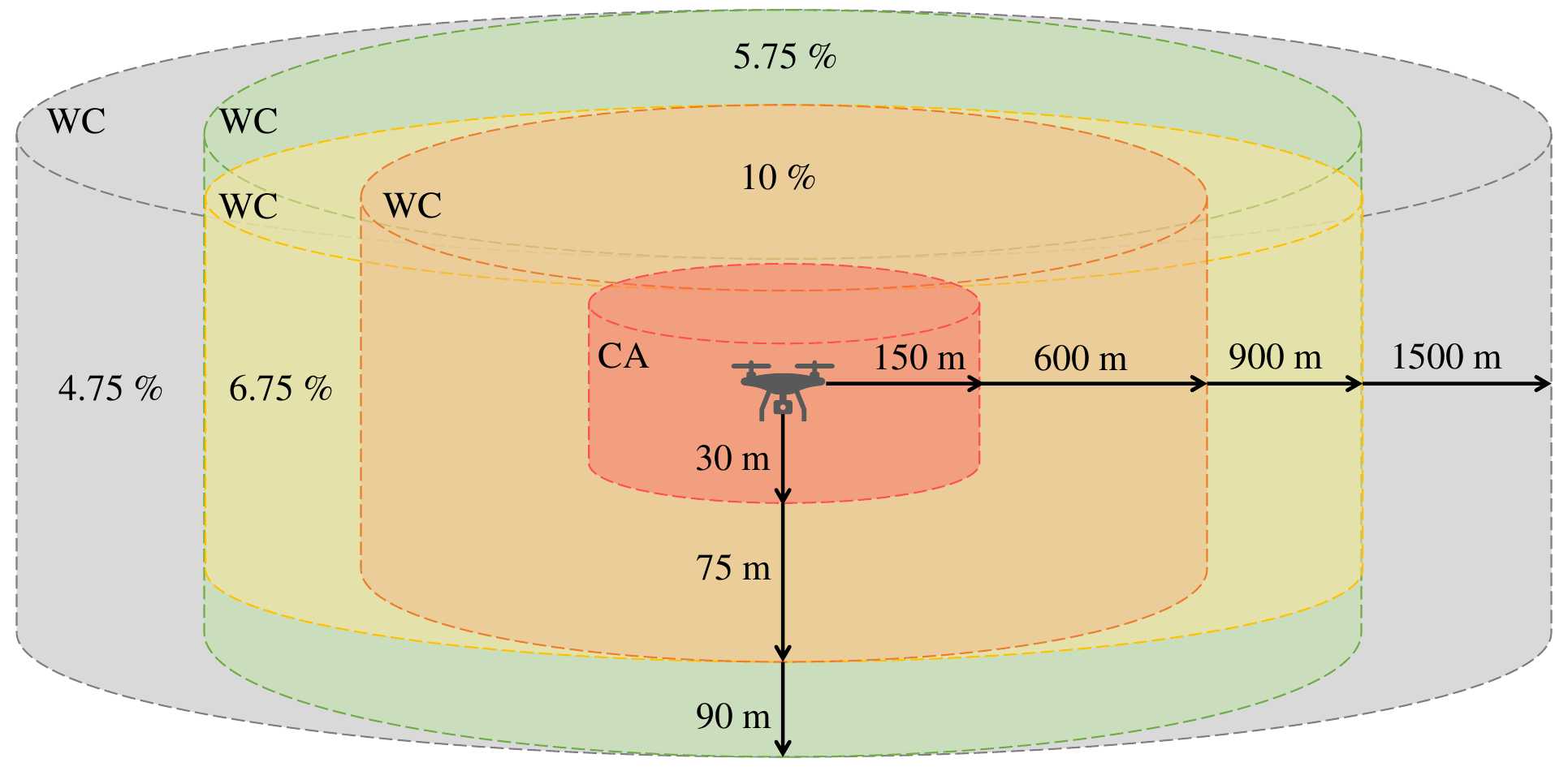}
	\caption{The levels of drones' well-clear (WC) zones each represented by a cylinder. The red cylinder represents the collision avoidance (CA) zone. For a given drone, the percentage values showed at its WC cylinders represent the probability of another drone present in a given cylinder violating/entering the next smaller CA cylinder \cite{vinogradov2020wireless}. E.g., the probability of a drone present in the yellow CV cylinder entering the orange CV cylinder is 6.75\%.}
	\label{droneLoc}
\end{figure}

\begin{itemize} [leftmargin=*]
\item\textit{Terrestrial Targets}: The main types of outdoor ground targets include \acp{UE} in cellular networks \cite{del2017survey}, connected autonomous vehicles \cite{saleh2021evaluation}, and \ac{IoT} nodes \cite{sallouha2019localization}. The localization requirements of ground targets vary depending on the application at hand. These applications may range from autonomous vehicles and search and rescue of connected \acp{UE}, which have stringent localization requirements, to surveying, automated agriculture, and automated train operation, which require relatively relaxed localization requirements. For instance, autonomous vehicle applications require cm-level of positioning accuracy and sub-$0.5$ degrees for $95\%$ of the time on average \cite{AVRequirements}, whereas in some \ac{IoT} applications such as package tracking, an accuracy of 100~meters is sufficient \cite{sallouha2019localization}. Moreover, depending on the application, ground targets might require positioning information only, like surveying and automated train operation applications, while others require both positioning and orientation information, like autonomous vehicles and \acp{UE} using directional beamforming in \ac{mmWave} \cite{abu2018error}.

\item\textit{Aerial Targets}: Main examples of aerial targets are rotary-wing or fix-wing drones, which are regarded as \acp{LAP}, which generally have high dynamics and relatively low altitudes (lower than 5~km). Localizing aircraft at higher altitudes such as \acp{HAP} can be neglected because of their quasi-stationary characteristics. In terms of \ac{RF} connectivity, \acp{LAP} can have links with ground \acp{BS} for data exchange, \ac{ATM}, and \ac{CC}, and hence, might act as an aerial connected \ac{UE}. In addition, they could also have side links among each other in the form of device-to-device or broadcast messages. Localizing aerial targets can ensure security within a \textit{no-fly} zone in cases involving \textit{non-cooperative} aerial targets or serve \ac{ATM} in cases where GPS is not available. Non-cooperative aerial vehicles include targets that do not share their \ac{GPS} locations, such as amateur drones, which may raise privacy and security concerns by penetrating a no-fly zone, e.g., close to airports \cite{azari2018key}. In non-cooperative target localization, the closer the target is to the no-fly zone, the stricter the localization requirements. In applications related to \ac{ATM}, localizing requirements depend on the distance-based guidelines for \ac{UAV} systems \cite{weinert2018well}. An overview of the different \ac{ATM} zones is presented in Fig.~\ref{droneLoc}, which can be categorized into well-clear and collision avoidance \cite{vinogradov2020wireless}. These zones can be applied to UAVs flying freely or those flying with a given route, i.e., \textit{corridors} \cite{karimi2023analysis}, aiding \ac{UAV} \ac{ATM}.
\end{itemize}

\subsection{Localization Scenarios in GAS}
Considering the three anchor types (ground, aerial, and space) and the two target types (ground and aerial), six positioning scenarios can be envisioned. These scenarios are \ac{G2G}, \ac{G2A}, \ac{A2G}, \ac{A2A}, \ac{S2G}, and \ac{S2A}. These scenarios, along with their corresponding elements, are summarized in Table \ref{tb:locoElements}. While terrestrial anchors and terrestrial targets are the typical elements considered in the localization systems literature, aerial and space, particularly at the LEO orbit, introduce fundamental novelties to the conventional localization problem. For instance, a \ac{UAV} can act as a supporting base station or as a user \cite{azari2022evolution}. Moreover, using thousands of LEO satellites at very low LEO, below 300 km, in what is known as a mega-constellation increases the geometrical diversity of the localization system \cite{janssen2023survey}.

\section{Localization Fundamentals}\label{sec:locFun}
In this section, we detail the key relevant localization fundamentals and position them in the context of \ac{GAS} networks.

\subsection{System Geometry in 3D}\label{sec:geometry}
In \ac{GAS} localization systems, it is crucial to characterize anchors and targets in a 3D space. In addition to their 3D positions, their orientation angles and velocity become highly relevant to capture their relative motion, e.g., in applications such as aerial and terrestrial autonomous vehicles \cite{kuutti2018survey,zheng2023jrcup}.
 
\subsubsection{Reference Coordinate System}\label{refFrame}
\begin{figure}[t]
	\centering
	\includegraphics[width=0.49\textwidth]{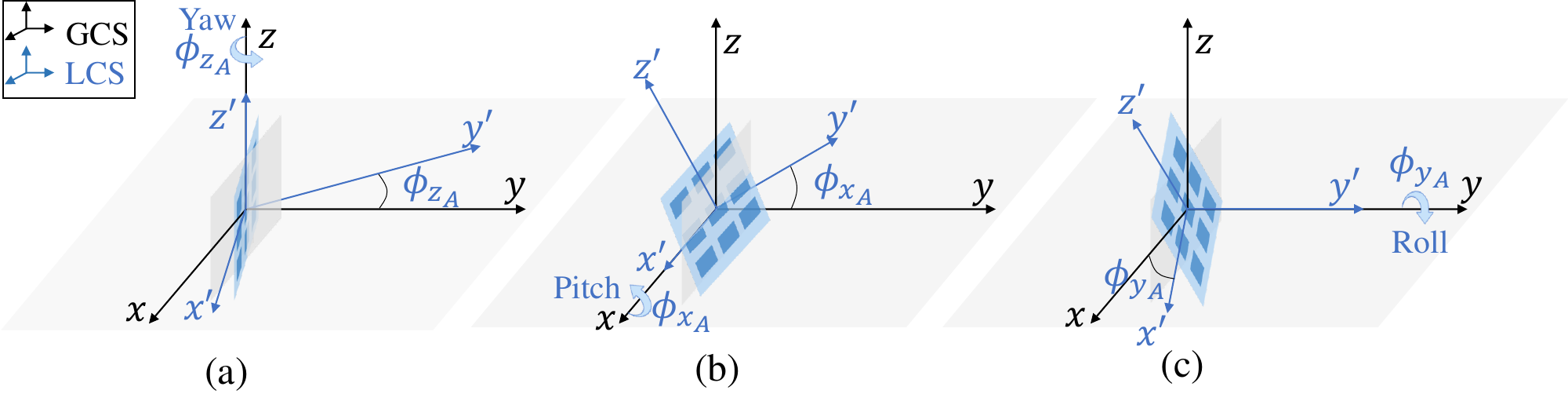}
	\caption{An example of (a) yaw, (b) pitch, and (c) roll rotation angles performed, with right-hand rule in mind, on a uniform rectangular array of antennas. The figure shows a yaw, pitch, and roll in their corresponding positive rotation direction.}
	\label{rotationAngle}
\end{figure}
The position of a given target/anchor can be defined using the \ac{GCS}, whereas its orientation can be described using its \ac{LCS} along with the \ac{GCS} \cite{chen2022tutorial,abu2018error}. A widely adopted \ac{GCS} is the \ac{UTM} coordinate system, which originates at the surface of the earth \cite{langley1998utm}. The $x$ and $y$ axes of the \ac{UTM} system point toward the east and north of the earth, respectively, whereas the $z$ axis is perpendicular to the earth's surface, pointing upwards, and originates at sea level. The \ac{LCS}, on the other hand, has an origin centered at the antenna (array) of the corresponding anchor/target, as shown in Fig.~\ref{rotationAngle}. The $x$ and $z$ axes of the \ac{LCS} point toward the anchor/target array's own horizontal and vertical directions, respectively. Subsequently, the $y$-axis of the \ac{LCS} is perpendicular to the $xz$-plane, following the right-hand rule, pointing towards the boresight of the array. Usually, we are interested in localizing the target in the \ac{GCS}. Yet, measurements are practically taken in the \ac{LCS} with respect to the anchor or the target. Hence, extra processing is usually needed to transform the \ac{LCS} measurements to the \ac{GCS}. The transformation from one reference frame to another is done through the translation and rotation processes \cite{noureldin2012fundamentals}. 

The translation between the two reference frames is computed by measuring the displacement between their origins. On the other hand, the rotation between the two frames can be fully captured by the attitude angles of the \ac{LCS} with respect to the \ac{GCS}, known as the \textit{pitch} $(\phi_x)$, \textit{roll} $(\phi_y)$, and \textit{yaw} $(\phi_z)$ angles \cite{noureldin2012fundamentals}. To demonstrate these angles, consider the \ac{URA} shown in Fig.~\ref{rotationAngle}, the pitch angle describes the vertical tilt between the $y$-axis of the \ac{LCS} and the $xy$-plane of the \ac{GCS}. This means that a \ac{URA} with a $y$-axis parallel to the \ac{GCS}'s $xy$-plane is said to have a zero pitch angle. Likewise, if the antenna array is either tilted upwards or downwards, then it is said to have a positive or negative pitch angle, respectively. The roll angle, on the other hand, describes the vertical tilt between the \ac{LCS}'s $x$-axis and the $xy$-plane. Hence, the counterclockwise rotation of the antenna array along its $y$-axis corresponds to a positive roll angle and vice versa. Finally, the yaw angle describes the horizontal tilt between the \ac{LCS}'s $y$-axis and the \ac{GCS}'s $y$-axis (i.e., the north), which means that directing the antenna array westward will result in a positive yaw angle, and vice versa. The rotation between the \ac{GCS} and the \ac{LCS} can be represented by a matrix multiplication of three rotation matrices that correspond to the individual attitude angles. These three orientation matrices represent the pitch around the $x$-axis, the roll around the $y$-axis, and the yaw around the $z$-axis, as illustrated in Fig.~\ref{LocMeasure}. It is worth noting that the order of applying these rotations matters and that the standard order that is conventionally followed is applying the roll first, then the pitch, and finally the yaw \cite{noureldin2012fundamentals}. The rotation matrices are expressed as \cite{vince2011rotation}

\begin{equation}
	{\textbf{R}_{x_j}(\phi_{x_j})} = 
	\begin{bmatrix}
			1 & 0 & 0 \\
			0 & \cos \phi_{x_j} & -\sin \phi_{x_j} \\
			0 & \sin \phi_{x_j} & \cos \phi_{x_j}
	\end{bmatrix}\,,
	\label{angleRotateX}
\end{equation}
\begin{equation}
	{\textbf{R}_{y_j}(\phi_{y_j})} = 
	\begin{bmatrix}
			\cos \phi_{y_j} & 0 & \sin \phi_{y_j} \\
			0 & 1 & 0 \\
			-\sin \phi_{y_j} & 0 & \cos \phi_{y_j}
	\end{bmatrix}\,,
	\label{angleRotateY}
\end{equation}
\begin{equation}
	{\textbf{R}_{z_j}(\phi_{z_j})} =
	\begin{bmatrix}
			\cos \phi_{z_j} & -\sin \phi_{z_j} & 0 \\
			\sin \phi_{z_j} & \cos \phi_{z_j} & 0 \\
			0 & 0 & 1
	\end{bmatrix}\,,
	\label{angleRotateZ}
\end{equation}
\begin{equation}
    \textbf{R}_j = \textbf{R}_{z_j}(\phi_{z_j})\textbf{R}_{x_j}(\phi_{x_j})\textbf{R}_{y_j}(\phi_{y_j})\,,
\end{equation}
where $j \in \{A, U\}$ indicates either anchor's or target's rotation, $\textbf{R}_{x_j}$, $\textbf{R}_{y_j}$, and $\textbf{R}_{z_j}$ denote the pitch, roll, and yaw rotation matrices, respectively, and $\textbf{R}_j$ denotes the collective rotation matrix.

\subsubsection{Anchor-Target Distance and Angle Geometry}
\begin{figure}[t]
	\centering
	\includegraphics[width=0.45\textwidth]{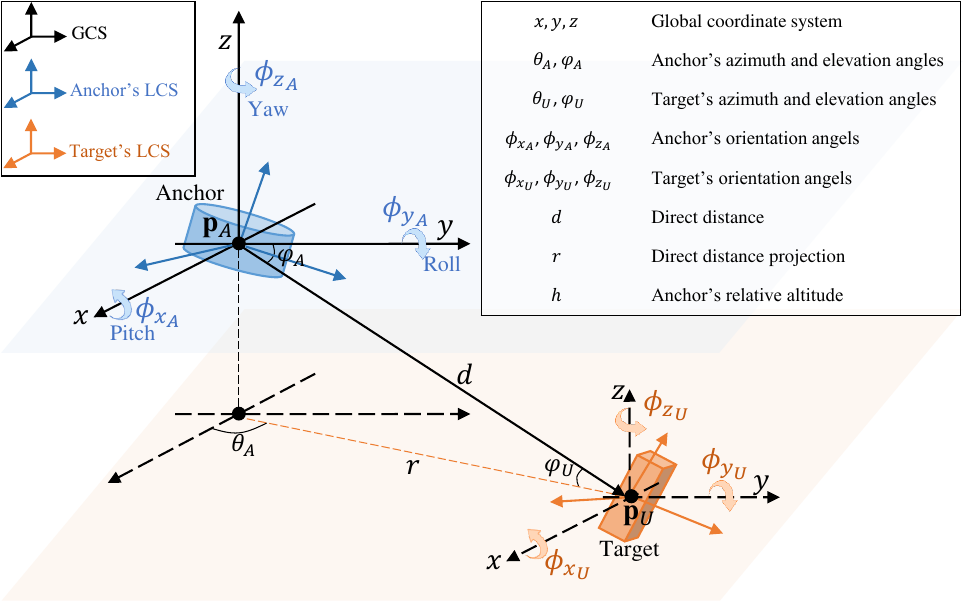}
	\caption{A representation of the localization measurables, which include distances and angles.}
	\label{LocMeasure}
\end{figure}

Throughout this paper, we use $\textbf{p}_A \in \mathbb{R}^3$, $\boldsymbol{v}_A \in \mathbb{R}^3$, and $\boldsymbol{\phi}_A \in \mathbb{R}^3$ to denote the anchor's true position, velocity, and orientation, respectively. For instance, considering a Cartesian \ac{GCS}, we have $\textbf{p}_A = (x_A, y_A, z_A)$, $\boldsymbol{v}_A = (v_{x_A}, v_{y_A}, v_{z_A})$, and $\boldsymbol{\phi}_A = (\phi_{x_A}, \phi_{y_A}, \phi_{z_A})$. Here, as depicted in Fig.~\ref{LocMeasure}, $\phi_{x_A}$, $\phi_{y_A}$, and $\phi_{z_A}$, respectively, correspond to the pitch, roll, and yaw attitude angles of the anchor with respect to the \ac{GCS} right-hand-based frame's $x$, $y$, and $z$ axes. Furthermore, we use $\textbf{p}_U \in \mathbb{R}^3$, $\boldsymbol{v}_U \in \mathbb{R}^3$, and $\boldsymbol{\phi}_U \in \mathbb{R}^3$ to denote the user's true position, velocity, and orientation, respectively, e.g., in a Cartesian coordinate system $\textbf{p}_U = (x_U, y_U, z_U)$, $\boldsymbol{v}_U = (v_{x_U}, v_{y_U}, v_{z_U})$, and $\boldsymbol{\phi}_U = (\phi_{x_U}, \phi_{y_U}, \phi_{z_U})$.
A target location can be defined in the \ac{GCS} with respect to an anchor position via the distance and angles between them, irrespective of their orientation, as shown in Fig.~\ref{LocMeasure}. The direct distance $d$, in 3D, between a target located at $\textbf{p}_U = (x_U, y_U, z_U)$ and an anchor at $\textbf{p}_A = (x_A, y_A, z_A)$ can be written as
\begin{equation}\label{Ddistance}
	\begin{split}
	  d & = \lVert \textbf{p}_A - \textbf{p}_U \rVert \\
		& = \sqrt{(x_{A}-x_{U})^2+(y_{A}-y_{U})^2+(z_{A}-z_{U})^2}.
	\end{split}
\end{equation}
The projection of $d$ on the $xy$-plane of the \ac{GCS} provides the 2D distance $r = (z_{A}-z_{U})/d$, as shown in Fig.~\ref{LocMeasure}.
The \ac{GCS} 3D angles, represented by the horizontal azimuth angle ${\theta}_A$ and the vertical elevation angle ${\varphi}_A$ with respect to the anchor in the \ac{GCS}, can be geometrically expressed as follows
\begin{equation}
	\begin{bmatrix}
		{\theta_A}  \\
		{\varphi_A} 
	\end{bmatrix}
	=
	\begin{bmatrix}
		\tan^{-1}\left( \frac{y_U - y_A}{x_U - x_A} \right) \\
		\sin^{-1} \left( \frac{z_U - z_A}{ d } \right)
	\end{bmatrix}\,,
	\label{angleMeasure}
\end{equation}
where $\theta_A \in (- \pi, \pi)$ is the horizontal counterclockwise angle from the $x$-axis, and $\varphi_A \in (-\pi/2, \pi/2)$ is the vertical counterclockwise angle from the $xy$-plane. From a target perspective, these geometric azimuth and elevation angles would thus be 
\begin{equation}
    \theta_U=\begin{cases}
    \theta_A-\pi, & \theta_A>0\\
    \theta_A+\pi, & \theta_A\leq0,
    \end{cases}
\end{equation}
\begin{equation}
    \varphi_U= -\varphi_A\,,
\end{equation}
in the case of LOS operation. While (\ref{angleMeasure}) presents the \ac{GCS} geometric angles, irrespective of the orientation, the geometric azimuth and elevation angle measurements, in practice, will be affected by the orientation of the measuring antenna array at the anchor/target since they are measured in the local coordinate system. Hence, we need to estimate or know the orientation of the measuring device to extract the \ac{GCS} azimuth and elevation angles from the \ac{LCS} AOA/AOD measurements, cf. Section \ref{subsec:AngleMeasurements}. Here, we will consider an example where we are interested in extracting the \ac{GCS} position of the UE from a range and an AOA measurement that is measured at the anchor. The same method can be used for AOD measurements at the anchor and for AOA/AOD measurements at the user, assuming that the orientation of the user is known. The anchor AOA measurement can be theoretically computed as
\begin{equation}
	\renewcommand{\arraystretch}{1.9}
	\begin{bmatrix}
		{\theta_A'}\\
		{\varphi_A'}
	\end{bmatrix}
	=
	\begin{bmatrix}
		\tan^{-1}\left( \frac{[\textbf{R}_{A}^{\top} (\textbf{p}_U - \textbf{p}_A)]_2}{[\textbf{R}_{A}^{\top} (\textbf{p}_U - \textbf{p}_A)]_1} \right) \\
		\sin^{-1} \left( \frac{[\textbf{R}_{A}^{\top} (\textbf{p}_U - \textbf{p}_A)]_3}{d} \right)
	\end{bmatrix}\,,
	\label{AOAMeasure}
\end{equation}
where $[\textbf{a}]_i$ indicates the $i$-th element of vector $\textbf{a}$. In (\ref{AOAMeasure}), $\theta_A'$ and $\varphi_A'$ correspond to the array's relative counterclockwise horizontal and vertical AOAs with respect to the array's $x$-axis and $xy$-plane, respectively. Hence, the user's position can be theoretically computed from anchor's \ac{LOS} range, $d$, and AOA measurements, $\theta_A' $ and $ \varphi_A'$, as

\begin{equation}
	\textbf{p}_U= \textbf{p}_A +d \,\textbf{R}_A
	\begin{bmatrix}
		\cos \theta_A' \cos \varphi_A'\\
		\sin \theta_A' \cos \varphi_A'\\
		\sin \varphi_A'
	\end{bmatrix}.
\end{equation}

\subsection{Localization Measurables}\label{subsec:LocPre}
Radio localization is fundamentally based on the spatial information implicitly conveyed by radio waves as they propagate through the wireless channel. This spatial information is impacted by the stochastic nature of the wireless channel as well as the transceiver noise and hardware imperfections. These effects can be observed via power, time, and angle measurements \cite{sayed2005network}. In the following, we first define measurement resolution, resolvability, and accuracy. Subsequently, we detail the localization measurables.

\subsubsection{Measurement Resolution, Resolvability, and Accuracy}
Before delving into the individual measurables, it is worth noting that in model-driven localization methods, a great emphasis is placed on the accuracy, resolution, and resolvability of the acquired measurables. In this paper, we use the following definitions.
\begin{itemize} [leftmargin=*]	
	\item \textit{Resolution} is the smallest distinguishable change in a quantity, e.g., time, power, or angle, that can be detected by a measuring instrument, such as a transmitter or receiver. Hence, it describes the degree of detail a measurement can have. Resolution is mainly affected by the receiver's hardware, such as the \ac{ADC}'s quantization error \cite{anttila2020full}.

	\item \textit{Resolvability} of components in a measurement is the ability to distinguish/resolve the various multipath components, i.e., \ac{LOS} and \ac{NLOS} components, that constitute the measured quantity. Resolvability is bounded by the available bandwidth for time-based measurements, the coherent integration time for frequency/Doppler measurements, and the size of the antenna array for angle-based measurements\footnote{Some works in the literature, particularly in the radar community, use the term measurement resolution to describe the resolvability/separability of radio paths in the time, Doppler, or angle domain \cite{keykhosravi2023leveraging}.}. If the measurements have low resolvability in the corresponding time, Doppler, or angle domain, then various path components, e.g., \ac{LOS} and \ac{NLOS}, will be amalgamated, limiting the localization performance, irrespective of the \ac{SNR}, unless super-resolution methods are applied \cite{pesavento2023three}. 
	
	\item \textit{Accuracy}, by definition, is a measure of how close the acquired measurement is to the true value of the measurable. The accuracy of measurements is affected by many factors, including both measurement resolution and resolvability, as well as the number of antennas used, the bandwidth, and the \ac{SNR}. To assess the quality of measurements, one needs to consider both accuracy (bias) and precision (variance), which are often combined in the form of \ac{RMSE} \cite{chen2022tutorial,trevlakis2023localization}.
\end{itemize}

\subsubsection{Power Measurements}
Localization using power measurements relies on \ac{RSS}, which is a measure of the received radio signal power. Measuring \ac{RSS} is a natively supported feature in modern wireless transceivers. Thanks to their intrinsic simplicity, \ac{RSS}-based methods have been widely adopted in the literature, offering cost and energy-efficient localization solutions \cite{zanella2016best}. \ac{RSS}-based localization can be roughly classified into range-free and range-based methods \cite{zanella2016best}. Range-free methods do not include distance measurements but rather rely on connectivity or signal propagation aspects between anchors and target nodes \cite{xiao2008distributed}. Range-based methods, on the other hand, use \ac{RSS} measurements to estimate the direct distance between an anchor and a target. The time-averaged received power, $\text{P}_\text{Rx}$, at an anchor located at a direct distance $d$ from a transmitting target node, can be generally expressed, in dBm, as \cite{zanella2016best, goldsmith}
\begin{equation}
\text{P}_\text{Rx} = \text{P}_\text{Tx} + \underbrace{\mathcal{C} - 10n_p\log_{10}\left(\frac{d}{d_0}\right)}_{\text{Path gain}, G_p} +\,\mathcal{X},\,\, \forall d \geq d_0,
	\label{Prx}
\end{equation}
where $\text{P}_\text{Tx}$ is the transmit power, in dBm, $n_p$ is the pathloss exponent, which reflects the way the received power decays with distance, and $d_0$ is the reference distance after which the far-field lies, and consequently, the received power model in (\ref{Prx}) is valid. In (\ref{Prx}), $\mathcal{X}$ is a zero-mean normal random variable with a standard deviation $\sigma_{\mathcal{X}}$ (in dB), representing the shadowing effect. The pathloss exponent, $n_p$, and shadowing standard deviation, $\sigma_{\mathcal{X}}$, depend on the considered scenario from Table \ref{tb:locoElements} (e.g., cf. Section \ref{subsec:AcMdl}). The term $\mathcal{C}$, in (\ref{Prx}) is a distance-independent term that collects constant factors between a given transmitter and a receiver such as the carrier frequency and the transmitter and receiver antenna gains, respectively denoted by $G_{\text{Tx}}$ and $G_{\text{Rx}}$.
The gain of the antenna differs with the type of antenna (antenna pattern), which is particularly formulated as\footnote{The dependency on the antennas' gains implies that $\mathcal{C}$ depends on the transmitter's antenna orientation relative to the receiver's antenna orientation \cite{dil2010rss}. The effect of the relative antenna orientation needs to be measured and calibrated.}
\begin{equation}
	G_j(\boldsymbol{\psi}'_j)=\eta \cdot \frac{4\pi A_e(\boldsymbol{\psi}'_j)}{\lambda^2},\, j \in \{\text{Tx}, \text{Rx}\}\,,
\end{equation}
where $\boldsymbol{\psi}'_j=[\theta'_j,\varphi'_j]$ comprises the array's relative horizontal (azimuth) and vertical (elevation) angles, $\lambda$ is the wavelength of the carrier frequency, $\eta$ is the antenna efficiency, and $A_e$ represents the effective aperture of the antenna.

In practice, receivers express the \ac{RSS} as a scaled version of $\text{P}_\text{Rx}$, i.e., $\text{RSS} = \mathcal{R}(\text{P}_\text{Rx})$, where $\mathcal{R}(.)$ is the received power to \ac{RSS} transduction function, which ideally provides a rescaled version of the received power \cite{zanella2016best}. In order to overcome any hardware impairments, a calibration process is needed to accurately rectify the power-to-RSS transduction function $\mathcal{R}(.)$ \cite{minucci2022measuring}. Assuming a calibrated receiver, the \ac{RSS}-based estimated direct distance between an anchor and a target is
\begin{equation}
	\hat{d} = d_0 10^{\frac{\mathcal{\text{P}_\text{Tx} + C - \text{P}_\text{Rx}}}{10n_p}} + e_{\text{RSS}}\,,
	\label{estdRSS}
\end{equation}
where $e_{\text{RSS}}$ is the RSS measurement error typically caused by the shadowing effect $\mathcal{X}$ and the uncertainty of antenna gains in the direction of departure and arrival. It is worth noting that (\ref{estdRSS}) provides a distance estimate for a given observation of $\text{P}_\text{Rx}$. However, for a given distance $d$, this observation is, in fact, a random variable statistically influenced by the shadowing effect $\mathcal{X}$.

\subsubsection{Time Measurements}
The core idea of a radio time measurement lies in the transmitter's and receiver's ability to define the exact timestamp of a radio signal's transmission and reception, respectively. Defining transmission timestamp at a transmitter requires a stable local clock with bounded drift, i.e., an oscillator that can provide an accurate and stable clock signal \cite{verbruggen2023enabling}, whereas defining the reception timestamp at the receiver, commonly known as \ac{TOA}, requires not only a stable clock at the receiver but also an accurate \ac{TOA} estimation technique. Examples of \ac{TOA} estimation techniques are peak detection techniques and statistical goodness-of-fit-based techniques \cite{wang2020performance}. The received \ac{SNR} and signal's bandwidth are key design parameters for better TOA estimation. Assume $t_0$ is the transmission time at the transmitter, and $t_r$ is the true \ac{TOA} at the receiver. The estimated direct distance between the transmitter and the receiver can be expressed as
\begin{equation}
	\hat{d} = c\, \Delta t + e_{\text{TOA}}\,,
	\label{ToAEstm}
\end{equation} 
where $c$ is the speed of light, $\Delta t = t_r - t_0$ is the true propagation time, and $e_{\text{TOA}}$ is the time measurement error resulting from multipath, e.g., positive bias due to the \ac{NLOS} propagation, as well as the clock offset between transmitter's and receiver's clocks. Synchronizing a transmitter and a receiver, e.g., an anchor and a target node, to realize accurate TOA measurements is not an easy task due to clock imperfections, which result in a drift that depends on environmental factors such as temperature and pressure \cite{verbruggen2023enabling}. One way to avoid the need to synchronize target nodes, e.g., non-cooperative targets, with anchors is \ac{TDOA}, which relies on synchronized anchors to measure the difference in distance using \ac{TOA} at several anchors. The corresponding difference in distance, $d_{12}$, from a target to two anchors can be written as \cite{sayed2005network}
\begin{align}
	\hat{d}_{12} = (t_{A_1} - t_{A_2})c + e_{\text{TDOA}} \,,
	\label{disDiff}
\end{align}
where $t_{A_1}$ and $t_{A_2}$ denote the \ac{TOA} at anchor $A_1$ and anchor $A_2$, respectively, and $e_{\text{TDOA}}$ is the TDOA measurement error, which similar to $e_{\text{TOA}}$, is influenced by the multipath propagation, and the clock offset between the anchor's clocks. Another widely adopted time-based ranging method is \ac{RTT} \cite{del2017survey}. Unlike \ac{TDOA} and \ac{TOA}, \ac{RTT} counteracted the synchronization requirements by performing back-and-forth time measurements per individual anchor. The main challenge with round-trip time is determining a good estimate of the processing duration at the receiver, which is hardware-dependent, making it only suitable for cooperative targets.

\subsubsection{Angle Measurements}\label{subsec:AngleMeasurements}
Antenna arrays can be exploited to measure the \ac{AOD} and the \ac{AOA} between an anchor and a target node. In order to localize a target in 3D space using angle measurements, one needs to estimate both azimuth and elevation angles, as discussed in Section \ref{sec:geometry} and as shown in Fig.~\ref{LocMeasure}. To estimate both angles at a single anchor, it should be equipped with a 2D array, e.g., \ac{URA}. This means that in order to measure both the azimuth and elevation angles of the uplink-AOA (UL-AOA) or the downlink-AOD (DL-AOD), the anchor node should be equipped with a 2D array. Likewise, if the DL-AOA or the UL-AOD are to be measured, then the user should be equipped with a 2D array. It is worth remembering that for these angles to be useful for positioning purposes, the attitude angles (i.e., pitch, roll, and yaw) of the measuring node should be known. In cases where the attitude angles of the measuring device are not known, then co-estimation of both the orientation and the position of the target should be conducted in what is commonly known as \textit{6D localization} \cite{nazari2023mmwave}. In these cases, extra information offered by other anchors can be utilized to solve the estimation problem.

The AOA is estimated by measuring the phase differentials across the array, as signals arrive at each antenna element with a slight delay compared to the array's phase center. These minor delays result in phase rotations, which can be measured when an RF chain is available at each antenna element \cite{van2018keeping}.
While there can be more than one propagation path, only the \ac{LOS} path is usually of interest for angle-based localization. This \ac{LOS} path can be extracted by, e.g., the  MUSIC algorithm \cite{schmidt1986multiple} or ESPRIT algorithm \cite{roy1989esprit}. 
The \ac{AOD} is often estimated by sending different directional beams in different directions/\acp{AOD} (e.g., with different azimuth and elevation angles when the transmitter is equipped with a 2D array). At the receiver's side, \ac{RSS} measurements are conducted for the various beams. The AOD is then estimated corresponding to the beam with maximum strength \cite{dwivedi2021positioning}. In case the receiver has access to the complex beam responses and beam weights, more sophisticated methods can be applied,  such as beamspace ESPRIT \cite{xu1994beamspace} or orthogonal matching pursuit \cite{lee2016channel}.

\subsubsection{Doppler Measurements}
In the case of a mobile localization system, the estimation of the 3D velocity of \acp{UE} is paramount for many applications, e.g., autonomous vehicles. To estimate the velocity of a mobile \ac{UE}, we usually rely on Doppler measurements, which measure the frequency shifts in the carrier frequency, and sub-carrier frequencies in OFDM-based systems, from the nominal frequency of operation (i.e., the original carrier frequency) due to the relative radial velocity between the \ac{UE} and the anchor node \cite{shi2023revisiting}. The relative radial velocity\footnote{The relative radial velocity is also commonly referred to as the \ac{LOS} velocity \cite{shi2023revisiting}.} between a given anchor and a \ac{UE} can be written as
\begin{equation}
    \Delta v = \left(\boldsymbol{v}_A - \boldsymbol{v}_U \right)^{\top} \frac{\textbf{p}_A - \textbf{p}_U}{\|\textbf{p}_A - \textbf{p}_U\|}\,.
    \label{radVelocity}
\end{equation}
In a \ac{LOS} scenario, the estimated frequency of the received signal, $\hat{f}_r$ can be formulated as
\begin{equation}
    \hat{f}_r=f_c \left(1+\frac{\Delta v}{c}\right) + e_{Dr}\,,
    \label{eqnDoppler}
\end{equation}
where $f_c$ is the carrier frequency and $e_{Dr}$ is the Doppler measurement error which is mainly caused by the \ac{CFO} between the transmitter and the receiver. Hence, in order to measure the contribution of the radial velocity to the frequency shift, we need to estimate and compensate the \ac{CFO}. In \ac{NLOS} scenarios, where the signal bounces off multiple objects before reaching the receiver, (\ref{eqnDoppler}) must be altered to include the relative radial velocities between the various objects in the environment.
Note that the Doppler measurements do not only possess information about the relative velocity of the \ac{UE}, but also its position. Hence, the availability of Doppler measurements can lead to enhanced positioning solutions \cite{shi2023revisiting,farhangian2020multi}.

\subsubsection{Joint measurables}
Methods exist that perform joint estimation in several domains, e.g., TOA and AOA \cite{li2022joint,rogel2021time,pan2021joint}, which provides improved accuracy at the cost of increased processing complexity. Compared to using a single measurement domain, multi-domain joint time and angle estimation methods can fully exploit both temporal and spatial diversity in resolving multipath components \cite{wen2018joint}. In order to jointly estimate time and angle information in multipath channels, the selection of bandwidth and array size can be jointly optimized to ensure time and angle resolvability. The joint estimation of angle and time can be done using sub-space-based methods or maximum-likelihood-based methods \cite{bellili2018maximum}. The former methods tend to be sub-optimal and require relatively high \ac{SNR}. In contrast, the latter methods usually impose major computational complexity as they need to estimate time and angle iteratively.

\begin{figure*}
	\centering
	\begin{subfigure}[b]{0.245\textwidth}
		\centering
		\includegraphics[width=\textwidth]{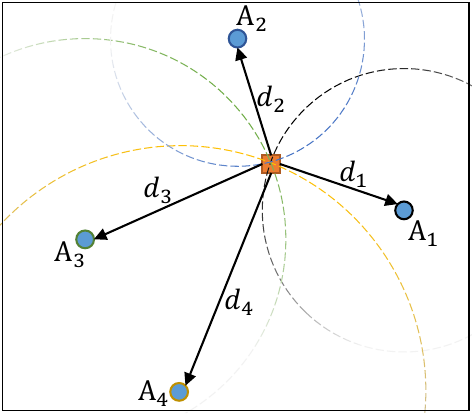}
		\caption{Distance MLAT}
		\label{fig:D_MLAT}
	\end{subfigure}
	\hfill
	\begin{subfigure}[b]{0.245\textwidth}
		\centering
		\includegraphics[width=\textwidth]{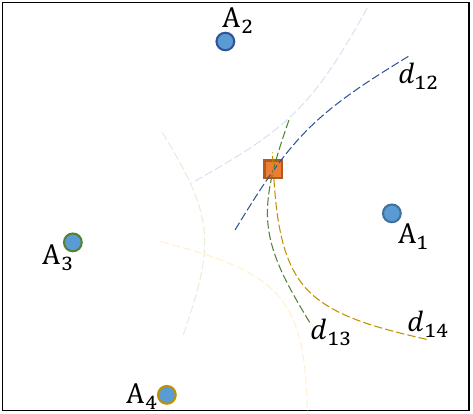}
		\caption{Distance difference MLAT}
		\label{fig:DD_MLAT}
	\end{subfigure}
	\hfill
	\begin{subfigure}[b]{0.245\textwidth}
		\centering
		\includegraphics[width=\textwidth]{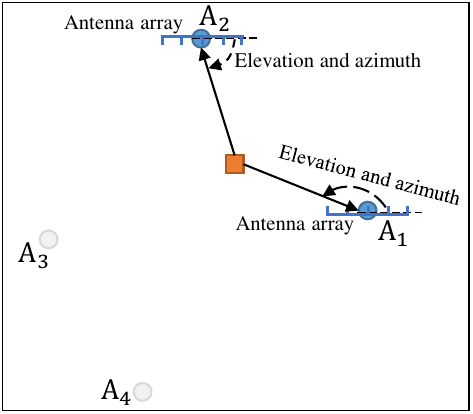}
		\caption{Triangulation}
		\label{fig:Triangulation}
	\end{subfigure}
	\hfill
	\begin{subfigure}[b]{0.245\textwidth}
		\centering
		\includegraphics[width=\textwidth]{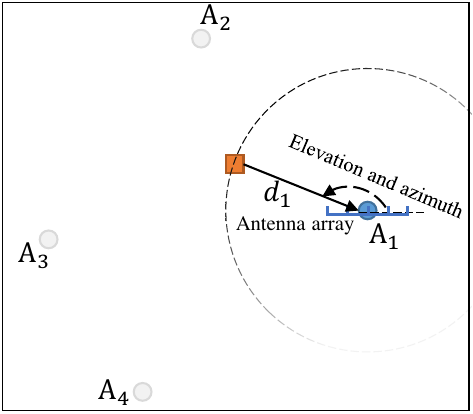}
		\caption{Distance and angle}
		\label{fig:D_A}
	\end{subfigure}
	\caption{Primary snapshot localization methods.}
	\label{fig:PrimLoc}
\end{figure*}

\subsection{Snapshot Localization Methods}
Once ranges, range differences, Doppler, and/or angles are measured as discussed in Section \ref{subsec:LocPre}, a data fusion step follows, in which measurements from/at different anchors are combined to obtain an estimate of the target node location. Broadly speaking, localization methods can exploit localization measurables in a data-driven or model-driven manner \cite{keykhosravi2023leveraging}. In data-driven methods, such as \acp{RF} fingerprinting, machine learning is employed to take advantage of the wireless channel's rich features to map received signal measurements spatially. In this context, no closed-form equation or easy-to-model mathematical framework is used. The model-driven methods, on the other hand, harness stochastic models describing the relation between the received radio signal and the signal propagation geometry from/to the target location. In general, the accuracy of model-driven methods depends on the ability to extract the \ac{LOS} path between an anchor and a target, making the multipath nature of the wireless channel one of the major localization error sources for \ac{LOS}-based methods \cite{sayed2005network,keykhosravi2023leveraging}.

The sensor fusion architectures in model-driven methods can be generally categorized into centralized (also known as tightly-coupled) or decentralized (also known as loosely-coupled) integration schemes. Centralized integration methods usually rely on a single central filter that fuses the positioning measurables from various anchors. In contrast, decentralized fusion algorithms fuse the individual/independent positioning estimates of the participating anchors into a single positioning estimate. Hence, by design, decentralized integration can only work for technologies that can estimate the position using a single anchor node, e.g., hybrid range and angle. A common practice in all measurement fusion methods is first to form a system of equations, e.g., in matrix form, linking the unknown target position to the measurements and anchors positions. Subsequently, the resulting set of system of equations can be solved to estimate the unknown target position. Broadly speaking, the system of equations' solution can be obtained either using a statistical approach based on the maximum likelihood estimator or using an algebraic approach based on the least-squares estimator \cite{sayed2005network,mantilla2015localization}. In the following, we detail the different model-driven data fusion methods used to construct the system of equations.
\begin{itemize} [leftmargin=*]
\item \textit{Ranging-based multilateration}: Ranging-based \ac{MLAT} uses the estimated distances between a target node and at least four different anchors in order to perform 3D localization, as illustrated in Fig.~\ref{fig:D_MLAT}. Once a sufficient number of distance estimates between anchors and the target are collected, one can formulate a system of equations representing the MLAT estimation problem \cite{sayed2005network}. Subsequently, the target's position is determined by solving the MLAT estimation problem using (non)linear least squares estimators or maximum likelihood estimators. \ac{MLAT} is usually utilized when only range measurements are available, i.e., no angle-based measurements. Also, \ac{MLAT} is preferred even if angle measurements are available in scenarios where the distance between the anchor node and the \ac{UE} is relatively large, as in \ac{GNSS}, where small angle measurement errors cause high positioning errors. It is worth noting that the relationship between range measurements and position states is non-linear, and hence, linearization errors are incurred when utilizing linear filters. These linearization errors are magnified when the \ac{UE} is close to the anchor node and can be neglected for long \ac{UE}-anchor distance \cite{saleh20225g}.

\item \textit{Range-difference-based multilateration}: In range-difference-based \ac{MLAT}, range-difference between at least four anchors is needed to estimate a 3D target's position, as shown in Fig.~\ref{fig:DD_MLAT}. Similar to ranging-based MLAT, a system of equations can be formulated using range difference measurements, e.g., using TDOA \cite{mantilla2015localization}. Range-difference-based \ac{MLAT} can be used in the same scenarios where \ac{MLAT} can be used, but requires tight synchronization between the anchor nodes. This method is more popular in cellular positioning, where \acp{BS} are expected to be communicating and synchronized with each other. Like ranging-based \ac{MLAT}, this method incurs linearization errors when linear filters are utilized. The linearization error increases when the \ac{UE} is close to either of the anchor nodes. Hence, the lowest linearization error occurs when the \ac{UE} is right in the middle between the anchor nodes \cite{saleh2021vehicular}.

\item \textit{Triangulation}: The basic principle of triangulation, illustrated in Fig.~\ref{fig:Triangulation}, is to measure the angle between the target and at least two anchors to perform 3D localization. In triangulation algorithms, we construct a set of linear equations as a function of the target position, $(x_U,y_U,z_U)$, where each anchor may contribute to two equations of the equations set \cite{steendam20173}. Triangulation is usually utilized when only angle measurements are available, e.g., no range-based measurements. Similar to its \ac{MLAT} counterparts, this method incurs high linearization errors when used with linear filters \cite{saleh_5g_2023}.

\item \textit{Hybrid distance-angle}: The main advantage of hybrid fusion methods is their ability to provide a target location estimate using a single anchor \cite{sayed2005network}, as shown in Fig.~\ref{fig:D_A}. For instance, an anchor with a \ac{URA} can perform AOA estimates and combine them with ranging-based measurements, e.g., TOA, to obtain a target's position. Additionally, hybrid positioning is preferred when using decentralized linear integration schemes, as it minimizes the linearization errors caused by their \ac{MLAT} and triangulation counterparts \cite{saleh_decentralized}.
\end{itemize}

\subsection{From Snapshot to Tracking}
Snapshot localization is realized when the state of interest $\boldsymbol{x}$, i.e., the position of the target, is assumed to be stationary and not in motion.
Tackling an estimation problem with dynamic states is known as a tracking problem. Discrete-time tracking implementations usually use a discrete-time index denoted by $k$. State estimation in tracking problems comprises a recursive iteration of state prediction and correction in what is commonly known as a \textit{filter} \cite{maybeck1982stochastic}. In the state prediction stage, the filter predicts the current epoch's state $\boldsymbol{x}^-_k$ through a transition model $f(\boldsymbol{x}^+_{k-1},\boldsymbol{u}_k)$ that utilizes the last epoch's estimated state $\boldsymbol{x}^+_{k-1}$ and the current epoch's control inputs, denoted by $\boldsymbol{u}_k$. Here, $\boldsymbol{u}_k$ usually comprises the control signals that control the motion of the UE, like the steering wheel and the gas pedal signals of land vehicles or the throttle and the wing control signals of aerial vehicles. In case these signals are not available, measurements from motion sensors that are attached to the UE, like accelerometers, gyroscopes, odometers, or magnetometers, can act as the control inputs $\boldsymbol{u}_k$. Note that the predicted and corrected states are usually distinguished by negative and positive superscripts, respectively. The correction stage is conducted through the utilization of the measurements vector $\boldsymbol{z}$, which might comprise range, angle, position, orientation, and/or velocity measurements. There are many estimators proposed in the tracking literature that are utilized for positioning purposes. The most famous among them is the family of \acp{KF} and particle filters \cite{chen2003bayesian}. One of the main differences between these filters is that the KF propagates a single belief/hypothesis of the state of the system, which comprises the mean, $\boldsymbol{x}$, and the variance, $\boldsymbol{P}$, of the state. On the other hand, the particle filter entertains multiple beliefs/hypotheses of the states of the system, i.e., the particles. Kalman filtering is a Bayesian estimator that is considered to be BLUE (best linear unbiased estimator) when the following criteria are met:
\begin{itemize}[leftmargin=*]
    \item Transition $f(\cdot)$ and measurement $h(\cdot)$ models are linear.
    \item The noise of the process and the measurements are independent zero-mean (unbiased) white Gaussian noises with perfectly known covariance.
\end{itemize}
If the criteria above are met, the \ac{KF} is optimal in the sense that it will result in the minimum state estimation covariance compared to other linear estimators. If any of the optimality criteria are breached, then the \ac{KF} is considered sub-optimal \cite{maybeck1982stochastic}.
The \ac{KF} has evolved over the years to tackle such scenarios of sub-optimality better. For instance, the extended \ac{KF} and the unscented \ac{KF} have been proposed to address the linearity restrictions. 
The particle filter, on the other hand, does not assume the linearity of the models nor a specific distribution of the noises \cite{elfring2021particle}; hence, there is no guarantee of optimality. Instead, it relies on the propagation, weighting, and redistribution of a large number of \textit{particles}, which hold different hypotheses of the state of the system \cite{elfring2021particle}. It is worth noting that the propagation of a high number of beliefs will lead to higher computational complexity compared to the various \ac{KF} implementations \cite{elfring2021particle}.

\subsection{Key Performance Metrics}\label{Fun_PR}
The main objective of any localization system, whether localizing an aerial or a terrestrial target, is to estimate the target's position with the minimum error possible. This objective intuitively makes localization accuracy a primary performance metric. Nonetheless, other localization \acp{KPI} play an essential role in the localization system design. In the following, we detail these localization \acp{KPI}.

\begin{itemize} [leftmargin=*]     
\item \textit{Localization Accuracy}:
The accuracy as a localization \ac{KPI} is often presented as a metric combining both the accuracy, i.e., bias, as well as the precision, i.e., variance. Throughout this tutorial, we use the term localization accuracy to refer to both bias and variance unless otherwise mentioned. The localization error in 3D is defined as the Euclidean distance between the true position, $\textbf{p}_U \in \mathbb{R}^3$ and its estimate, $\hat{\textbf{p}}_U$, which can be expressed as 
\begin{equation}
	\mathcal{E}_{\textbf{p}_U} =  \lVert\textbf{p}_U - \hat{\textbf{p}}_U \rVert \,,
	\label{locoErrGenral}
\end{equation}
where $\lVert\cdot\lVert$ is the L$^2$-norm. The equation in (\ref{locoErrGenral}) calculates a snapshot localization error from a single observation. In order to evaluate the localization system accuracy (bias and variance) from multiple spatial or temporal observations, the \ac{RMSE} is typically used, which is given by
\begin{equation}
	\mathcal{E}_{\textbf{p}} = \sqrt{ \mathbb{E}\left[\lVert\textbf{p}_U - \hat{\textbf{p}}_U \rVert^2\right]} \,,
	\label{locoErrMSE}
\end{equation} 
where $\mathbb{E}[\cdot]$ is the expected value. Alternatively, error percentiles can be used to represent the performance over multiple time/space location observations, which can be obtained using the \ac{CDF} of the localization error. For example, some applications care about the percentiles of the error, i.e., what error can be attained for 90\% or 99\% of the time/space. One of the most widely adopted benchmarks for localization accuracy is the \ac{CRLB}, which not only provides a lower bound for the location estimator but can also be used as an indicator for the design and optimization of localization systems. In particular, the \ac{CRLB} defines a lower bound for the variance of any (asymptotically) unbiased estimator \cite{kay1993fundamentals}.

\item \textit{Localization Coverage}:
Since most of the localization methods involve measurements from several anchors collectively working to estimate a target's position, the localization coverage could be defined as the region where localization (obtaining a position fix) is possible or the region where the \ac{GDOP} is sufficiently good, or the region where the localization error is below a defined threshold. A more strict definition of the localization coverage can be obtained by setting an upper bound on the range/angle estimator error per anchor, which could be a factor of the \ac{CRLB} \cite{sallouha2018energy}.

\item \textit{Localization Latency}:
The localization latency metric is the time duration between the trigger of the localization process and the time at which the target position is estimated. This time is mainly affected by three factors, namely, the propagation time between the transmitter and the receiver, the processing time, and the measurement aggregation time. Latency caused by propagation time is usually negligible for ground and aerial operation, as it is in the order of (sub)micro-seconds. On the other hand, \ac{S2G} and \ac{S2A} propagation time is usually in the order of milliseconds and thus cannot be neglected. Processing time is the time needed to process a set of measurables, which usually scales with the system's complexity and is a function of the signaling protocols deployed. Lastly, as the name suggests, aggregation time accounts for the time taken to accumulate enough poisoning measurables to estimate the position. For instance, in cases where only range measurements are available from a single \ac{UAV} or \ac{LEO} satellite, measurements should be accumulated over an extended period of time to compute a single anchor \ac{MLAT} solution. It is worth noting that the latency \ac{KPI} is mainly important for mobile target applications, in which an error caused by small localization latency scales up with the target's speed.

\item \textit{Localization Power-Efficiency}:
Tracking power consumption is crucial to ensure an energy-efficient localization process, especially when a mobile anchor is used or in case several iterations of signaling are needed between the anchor and the corresponding target. This metric is particularly important in \ac{IoT} applications, e.g., assets tracking, where prolonging the IoT node's battery life is a priority, even if it means compromising the localization accuracy by several meters \cite{sallouha2019localization}. The localization power efficiency as a metric is directly related to the communication computational complexity of the localization algorithm at hand, i.e., complex localization algorithms with long convergence time are power-hungry. 

\item \textit{Localization Stability}:
As we explore high-frequency bands such as mmWave and \ac{THz} that function in narrow beams, considerable attention must be paid to beam misalignment. Beam pointing errors can cause deafness, which results in localization outage or loss of tracking. The localization stability represents the rate at which beam misalignment occurs \cite{chen2022tutorial}—this misalignment rate increases in the case of a mobile anchor/target.

\item \textit{Localization Scalability}:
The scalability metric measures the localization system's ability to scale with the number of targets, considering anchors' placement, processing time, and power constraints \cite{taranto14,chen2022tutorial}. For instance, a localization system that uses \ac{TDOA} scales better with the number of targets compared to a localization system based on \ac{RTT}. This is due to the fact that \ac{RTT} requires dedicated two-way communications between anchors and the targets, as opposed to \ac{TDOA}, which can be done based on broadcast signals.
\end{itemize}
\section{Localization with Ground Anchors (5G)} \label{sec:G2X}
In the past, localization with ground anchors has been synonymous with indoor localization due to the short-range operation of contemporary technologies such as Wi-Fi, ultra-wideband, Bluetooth, and ZigBee. Such a short range of operation made outdoor deployment prohibitive in nature \cite{zafari2019survey}. Outdoor localization through cellular ground anchor nodes has been investigated using 3G and 4G technologies, yet it was limited to low-accuracy positioning. This is mainly due to the fact that 3G and 4G positioning measurables were of low quality due to the low bandwidth and number of antennas utilized \cite{del2017survey}. On the other hand, the \ac{5G} of cellular networks holds great promise to realize high-precision positioning thanks to the use of mmWaves and massive antenna arrays \cite{del2017survey}. Hence, this section will discuss the general 5G system model and the various system design aspects, research topics, opportunities, and challenges when localizing ground and aerial UEs using terrestrial 5G anchors.

\begin{table}[t]
	\centering
	\caption{5G radio technology.}
	\label{tab:5Gradio}
        \resizebox{0.49\textwidth}{!}{
	\begin{tabular}{|l|l|l|l|}
		\hline
		\textbf{Technology} & \textbf{Frequency band} & \hspace{0.2cm}\textbf{Bandwidth} & \hspace{0.4cm}\textbf{Range} \\ \hline \hline
		5G-NR FR1           & sub-6~GHz               & 20 - 100~MHz       & $\sim$ 1~km \cite{nokiaBellRep}           \\ \hline
		5G-NR FR2           & 27-71~GHz               & 400~MHz            & $\sim$ 190~m \cite{nokiaBellRep}          \\ \hline
	\end{tabular}}
\end{table}

\subsection{System Model}
\subsubsection{Radio Technology}
5G-NR (new radio) have access to two frequency ranges, FR1 (sub-6~GHz) and FR2 (mmWave) as presented in Table \ref{tab:5Gradio}. FR1 is designed for medium to long-range communications while FR2 ($27-71$~GHz) focuses on short-range, deep-urban environment communications \cite{chen2022standardization}. Note that distance ranges in Table \ref{tab:5Gradio} assume an \ac{SNR} of 0~dB \cite{nokiaBellRep}. FR1 radios have access to a bandwidth ranging from $20$ to $100$~MHz. On the other hand, thanks to mmWaves, FR2 5G \acp{TN} have a large bandwidth that may reach up to $400$~MHz, which allows for accurate time-based measurements (\ac{TOA}, \ac{TDOA}, and \ac{RTT}) and the ability to resolve multipath signals in the time domain \cite{wymeersch20175g}. Moreover, 5G FR2 \acp{BS} are equipped with a large number of antennas, enabling accurate UL-AOA and DL-AOD measurements and resolving multipath signals in the angular domain \cite{wymeersch20175g}.
It is worth noting that UE antenna arrays are expected to house significantly fewer antenna elements than BS arrays \cite{huo_5g_2017}. Hence, they might be able to perform DL-AOA and UL-AOD measurements but at a lower accuracy. In the case of having access to both UE-based and BS-based angle measurements, the orientation of the UE can be estimated.
5G-NR is set to achieve a maximum latency of $1$ ms, which is crucial for delay-sensitive applications like autonomous vehicle navigation \cite{wymeersch20175g}. Low latency communication is even more important when the UE is moving at high speed. For instance, if the UE is in a vehicle that is driving at $100$ km/hr, then a latency of $10$ ms would result in a constant positioning error of $27.7$ cm, not counting other sources of error. Finally, 5G FR2 small cells are designed to be densely deployed with an inter-cell distance of $200-500$~m \cite{3GPP38855}, which gives more chances to have a \ac{LOS} communication with multiple BSs simultaneously, reducing positioning errors.

In order to estimate the various 5G localization measurements, two pilot signals were specifically designed, namely the downlink \ac{PRS} and the uplink \ac{SRS}. The \ac{PRS} and SRS were originally introduced by the \ac{3GPP} for LTE localization \cite{zafari2019survey}.
\ac{3GPP} Release 16 enhanced the functionality of \ac{PRS} and \ac{SRS} by allocating higher bandwidth to them (i.e., $400$~MHz) \cite{etsi2022138}. PRS and SRS can have four different sub-carrier comb sizes/patterns, i.e., $(2, 4, 6,\text{ and } 12)$. The comb pattern indicates the spacing between the resource elements (sub-carriers) utilized by the reference signals within a single time slot.
A denser comb pattern, e.g., comb size of 2, means that the UE will have access to multiple measurements from multiple sub-carriers per \ac{OFDM} symbol in a given resource block, enhancing the accuracy of the positioning measurements \cite{dwivedi2021positioning}.

\subsubsection{Channel Model}
In order to introduce a general model for a 5G FR2 localization system, we consider a MIMO system with $N_{\text{Tx}}$ transmitting antennas, $N_{\text{Rx}}$ receiving antennas, $N_{\text{RF}}$ RF chains, and $G$ paths between the transmitter and the receiver. Moreover, we consider the transmission of 5G \ac{OFDM} frames comprising $K_s$ sub-carriers, and $L_s$ symbols. The individual paths, sub-carriers, and \ac{OFDM} symbols are denoted by $g=0,\dots,G-1$, $k=0,\dots,K_s-1$ and $l=0,\dots,L_s-1$, respectively. The duration of each \ac{OFDM} symbol is $T_{\text{Sym}}=T_{\text{CP}}\,+\,T_p$, where $T_{\text{CP}}$ and $T_p$ are the cyclic prefix period and the actual symbol period, respectively. The OFDM sub-carrier spacing is thus $\Delta f=1/T_p$. The received signal at the receiver, denoted by $\boldsymbol{y}_{k,l}\in \mathbb{C}^{N_{\text{RF}}}$ constitutes the transmitted complex pilot symbol, denoted by $s_{k,l} \in \mathbb{C}$, modulated by the precoding vector $\boldsymbol{f}_l \in \mathbb{C}^{N_{\text{Tx}}}$, the channel impulse response $\boldsymbol{H}_{k,l}\in \mathbb{C}^{N_{\text{Rx}}\times N_{\text{Tx}}}$, and the combining matrix $\boldsymbol{W}_l\in \mathbb{C}^{N_{\text{Rx}}\times N_{\text{RF}}}$, and corrupted by the thermal noise $\boldsymbol{n}_{k,l} \sim \mathcal{CN}(\boldsymbol{0},\boldsymbol{W}_l^{\mathsf{H}}\boldsymbol{W}_l\sigma_{\text{Rx}}^2)$ at the receiver \cite{wymeersch2022radio1}: 
\begin{equation}\label{ch-model}
\boldsymbol{y}_{k,l}=\boldsymbol{W}_l^{\mathsf{H}}\boldsymbol{H}_{k,l}\boldsymbol{f}_ls_{k,l}+\boldsymbol{n}_{k,l}\,.
\end{equation}
The precoding vector $\boldsymbol{f}_l$ and the combining matrix $\boldsymbol{W}_l$ are designed and optimized by the operator to either serve communications or localization purposes. On the other hand, the channel between the transmitter and the receiver is mainly determined by the surrounding environment. It is worth noting that such a notion of complete uncontrollability over the channel is currently being challenged by the emergence of \ac{RIS}, which will be discussed in Section \ref{sec:GAS6G}. The general channel model between a ground UE and a \ac{BS} for the $k$-th sub-carrier and the $l$-th OFDM symbol while aggregating $G$ paths in a MIMO system is
\begin{multline}\label{Channel}
\mathbf{H}_{k,l}=\\
\sum_{g=0}^{G-1} \alpha_g 
\underbrace{e^{-\jmath 2 \pi (f_c+k\Delta f) \tau_g}}_{\text{path delay}}
\underbrace{e^{\jmath 2 \pi f_c \nu_g l T_{\text{Sym}}}}_{\text{frequency shift}}
\underbrace{\mathbf{a}_{\mathrm{\text{Rx}}}(\boldsymbol{\psi}'_{\text{Rx}_g}) \mathbf{a}_{\mathrm{\text{Tx}}}^{\top}(\boldsymbol{\psi}'_{\text{Tx}_g})}_{\text{steering vectors}},
\end{multline}
 where $\alpha_g \in \mathbb{R}$ is the channel attenuation factor, $\tau_g$ is the delay of the $g$-th path, $f_c$ is the carrier frequency, $\nu_g$ is the frequency shift factor caused by the $g$-th path, $\mathbf{a}_{\mathrm{\text{Rx}}}$ and $\mathbf{a}_{\mathrm{\text{Tx}}}$ are the receiver and transmitter steering vectors, respectively, $\boldsymbol{\psi}'_{\text{Rx}_g}=[\theta'_{\text{Rx}_g},\varphi'_{\text{Rx}_g}]$ comprises the horizontal (azimuth) and vertical (elevation) angles of arrival relative to the receiver's orientation, and $\boldsymbol{\psi}'_{\text{Tx}_g}=[\theta'_{\text{Tx}_g},\varphi'_{\text{Tx}_g}]$ comprises the horizontal and vertical angles of departure relative to the transmitter's orientation. In the following, we use $g=0$ to indicate the LOS path, if any, and $g\neq0$ to indicate the NLOS components. The channel gain, $\alpha_g$, is modeled as follows
\begin{equation}
    \alpha_g=10^{\left(\frac{G_p}{20}\right)}, 
\end{equation} 
where $G_p$ is the path gain in dB from (\ref{Prx}).
The path delay is computed as $\tau_g=d_g/c+\tau_{b}$, where $d_g$ is the length of the $g$-th path and $\tau_b$ is the time synchronization bias between the transmitter and the receiver. In the case of LOS communications, the LOS path length is the direct distance between the anchor and the target and is computed as given in (\ref{Ddistance}).
In case of an NLOS path $g$ with a number of bounces $M_g\geq2$, the distance will be
\begin{equation}\label{dist_g}
    d_g=\Vert\textbf{p}_A-\boldsymbol{s}_{g,1}\Vert + \Vert\boldsymbol{s}_{g,M_g}-\textbf{p}_U\Vert+\sum_{m=1}^{M_g-1} \Vert\boldsymbol{s}_{g,m}-\boldsymbol{s}_{g,m+1}\Vert,
\end{equation}
where, $\boldsymbol{s}_{g,m}$ is the position of the $m$-th scatterer of the $g$-th path, $\textbf{p}_A$ is the position of the BS, and $\textbf{p}_U$ is the position of the UE as stated in Section \ref{sec:geometry}. The frequency shift component in (\ref{Channel}) is caused by the net difference of the radial velocity between the receiver and the last incident point of the given path, i.e., the Doppler shift, and the \ac{CFO} factor between the oscillators of the transmitter and the receiver \cite{wymeersch2022radio1}. The radial velocity is a measure of how the transmitter and the receiver are moving away/closer from/to each other. Hence, relative tangential motion between the two does not contribute to the Doppler shift. In the case of localization with ground anchor nodes, the BS can be assumed to be stationary. Hence, only the velocity of the UE, denoted by $\boldsymbol{v}_U=[v_{x_U}, v_{y_U}, v_{z_U}]^{\top}$, is taken into consideration. The \ac{CFO} factor is, in principle, unknown and is denoted by $\delta_f$. Based on (\ref{eqnDoppler}), the frequency shift factor of the $g$-th path, assuming that the reflectors are stationary, can be computed as
\begin{equation}
    \nu_g= \frac{\boldsymbol{v}_U^{\top} \boldsymbol{u}_g}{c} + \delta_f,
\end{equation}
where
\begin{equation}
\begin{split}
    \boldsymbol{u}_g&=\boldsymbol{u}(\theta_g,\varphi_g)\\
    &=[\cos(\theta_g)\cos(\varphi_g), \sin(\theta_g)\cos(\varphi_g), \sin(\varphi_g)]^{\top}\\
    &=\frac{\boldsymbol{s}_{g,M_g}-\textbf{p}_U}{\Vert\boldsymbol{s}_{g,M_g}-\textbf{p}_U\Vert}\,.
\end{split}
\end{equation}
Here, $\theta_g$ and $\varphi_g$ are the relative horizontal and elevation angles between the receiver and $\boldsymbol{s}_{g,M_g}$, the last incident point of the $g$-th path, in the \ac{GCS}.

The steering vectors $\mathbf{a}_{\mathrm{\text{Rx}}}(\boldsymbol{\psi}'_{\text{Rx}_g})$ and $\mathbf{a}_{\mathrm{\text{Tx}}}(\boldsymbol{\psi}'_{\text{Tx}_g})$ can be formulated by assuming a local coordinate system, centered around the center of the given array, as discussed in Section \ref{sec:geometry}. Hence, the various array elements will transmit/receive the signal with a positive or a negative time delay with respect to the reference transmission/reception time at the center of the array. The array elements will span the $x$ and $z$ axes of the local system, where the $i$-th element will have the following local coordinates $\boldsymbol{p}_i=[p_{i_x},0, p_{i_z}]^{\top}$. Accordingly, the generalized form of the $i$-th element of the receiver/transmitter steering vector given the relative local coordinate system and the $g$-th path is formulated as follows \cite{abu2020single}:
\begin{equation}\label{steering}
    \text{a}_{j}(\boldsymbol{\psi}'_{j_g},i) = e^{\frac{\jmath 2 \pi}{\lambda} \boldsymbol{p}^\top_i \boldsymbol{u}'_{j_g}},
\end{equation}
where $j \in \{\text{Tx, Rx}\}$, $\lambda=c/f_c$ is the wavelength of the carrier frequency, and $\boldsymbol{u}'_{j_g}=\boldsymbol{u}(\theta'_{j_g},\varphi'_{j_g})$. Note that (\ref{steering}) can be reduced to a 2D coordinate system by only accounting for $\theta'_g$ and by reducing $\boldsymbol{p}_i=[p_{i_x},0]^{\top}$ and $\boldsymbol{u}'_{j_g}=[\cos(\theta'_{j_g}), \sin(\theta'_{j_g})]^{\top}$. The multiplication of the steering vectors $\mathbf{a}_{\mathrm{\text{Rx}}}(\boldsymbol{\psi}'_{\text{Rx}_g})$ and $\mathbf{a}_{\mathrm{\text{Tx}}}^{\top}(\boldsymbol{\psi}'_{\text{Tx}_g})$ results in a matrix of phase shifts of the size $N_{\text{Rx}}\times N_{\text{Tx}}$. If either the transmitter or the receiver does not have access to multiple antennas (i.e., a \ac{MISO}, \ac{SIMO}, or \ac{SISO} system), the respective $\mathbf{a}_{\mathrm{\text{Rx}}}(\boldsymbol{\psi}'_{\text{Rx}_g})$ and/or $\mathbf{a}_{\mathrm{\text{Tx}}}(\boldsymbol{\psi}'_{\text{Tx}_g})$ should be eliminated from the equation. Hence, the matrix will reduce to a vector in the case of a \ac{MISO} or \ac{SIMO} system or be totally removed in the case of a \ac{SISO} system.

As can be seen in (\ref{ch-model})-(\ref{steering}), the FR2 channel model comprises elements that are directly related to the UE's position, velocity, and orientation, such as $\tau_0$, $d_0$, $\nu_0$, $\boldsymbol{\psi}'_{\text{Tx}_0}$, and $\boldsymbol{\psi}'_{\text{Rx}_0}$. Additionally, it contains elements that are related to the environment between the UE and the BS, which are captured in the multipath components of $\tau_g$, $d_g$, $\nu_g$, $\boldsymbol{\psi}'_{\text{Tx}_g}$, and $\boldsymbol{\psi}'_{\text{Rx}_g}$, where $g\neq0$. Hence, the accuracy of localizing the UE and mapping its surroundings is heavily dependent on the accuracy of the channel estimation algorithms used as well as the accuracy of the algorithms utilized to extract the positioning and mapping elements out of the channel response. Such algorithms will be discussed in the next subsection. In cases where FR1 is utilized, the bandwidth and number of antennas are limited compared to FR2, making the various paths' resolvability considerably challenging in complex environments \cite{wymeersch2022radio1}. In that case, stochastic channel models like Rician, Rayleigh, and Nakagami distributions can be used to model small-scale fading caused by multipath. To overcome the limited localization accuracy obtained in such scenarios, these channel models can be coupled with \acp{RF} fingerprinting and AI-based positioning techniques to coarsely estimate the position of the user \cite{witrisal2021localization}.

\subsection{Ground Targets: Design Aspects}\label{subsec:G2G}

Over the past years, research endeavors in 5G \ac{TN} positioning have exhibited substantial growth, encompassing a diverse spectrum of topics and methodological approaches. At a more abstract level, the domain of 5G positioning investigation can be partitioned into three principal domains: the estimation of 5G positioning measurables, the optimization of communication infrastructures to augment the quality of 5G positioning measurables, and the formulation of 5G positioning algorithms. Within each of these domains, researchers are engaged in furthering an array of sub-disciplines with the overarching aim of augmenting the precision and reliability of 5G positioning systems \cite{saleh_5g_2023}. Before delving into the individual fields, it is worth noting that researchers in all of the aforementioned fields tend to derive error bounds for the various stages of the 5G positioning system. For instance, researchers involved with estimating the 5G positioning measurables will derive \ac{CRLB} for these measurables to measure the optimality of their methods. On the other hand, researchers concerned with optimizing 5G systems for positioning purposes utilize these bounds as a theoretical framework to formulate their optimization problems \cite{keskin2022optimal}. Finally, researchers who developed 5G positioning algorithms utilize intermediate measurable bounds as well as the position error bound and the orientation error bound to provide an uncertainty quantification for their algorithms. Additionally, they utilize position and orientation error bounds as means to benchmark the optimality of their methods \cite{abu2018error}.

\subsubsection{Estimation of 5G Positioning Measurables}\label{sec:Est. of 5G meas.}
The first research area focuses on estimating various measurables used in 5G positioning, including time-based and angle-based measurables and \ac{CSI} \cite{fokin2022direction, pan2021joint,chen2021carrier}. The accuracy of these estimation algorithms has a direct impact on the accuracy of the positioning estimate. Additionally, within this domain, some researchers are interested in estimating the time synchronization bias between the UE and the BS \cite{cui2016vehicle, koivisto2016joint, ceniklioglu2022error} and the detection of NLOS signals \cite{bader2022nlos, wang2017multipath, wang2018joint, zhang2022online}. Estimating synchronization bias would enable the use of TOA, reducing the reliance on TDOA and RTT measurements. NLOS signal detection is vital for improving the accuracy of positioning algorithms that solely depend on LOS signals. A specific aspect of NLOS detection algorithms is identifying the number of signal bounces encountered by the NLOS signal \cite{li2023iterative}. This classification is valuable for algorithms relying on NLOS signals for positioning, as they primarily utilize single-bounce reflections \cite{wen20205g}.

\subsubsection{Optimization of Communication Systems to Enhance 5G Positioning Measurables}
The second domain of research is centered on enhancing the design and performance of the communication system to improve the quality of measurements used in 5G positioning. Subcategories within this domain encompass strategies like optimizing BS placement \cite{guo2022performance, saleh2021evaluation}, enhancing power allocation and beamforming \cite{kakkavas2021power, gao2022wireless,keskin2022optimal}, and refining signal design \cite{zhao2020beamspace, khan2021doppler, wei20225g}. Optimizing BS placement entails identifying the most advantageous BS locations within the network to enhance measurement quality, often by minimizing the \ac{GDOP}. Power allocation and beamforming optimization involve adjusting the transmission power of BSs and beam direction to enhance measurement quality \cite{keskin2022optimal}. These approaches consider the utilization of both LOS and NLOS signals, optimizing power allocation per beam or direction \cite{keskin2022optimal}. Signal design optimization aims to create new signals tailored for positioning, like PRS and SRS, which can yield more precise measurements through improved autocorrelation properties \cite{zhao2020beamspace, khan2021doppler, wei20225g}. Additionally, some designs seek to achieve higher signal-to-noise and interference ratios, further enhancing measurement performance.

\subsubsection{5G Positioning Algorithms}
The third research area comprises the development and enhancement of positioning and localization algorithms that rely on 5G measurements. This field is primarily divided into LOS positioning and other minor sub-fields. In each sub-field, both snapshot and tracking algorithms are present, albeit more attention has been given to snapshot techniques. As the name suggests, LOS 5G positioning refers to the use of LOS 5G measurables and their uncertainties\footnote{Uncertainties help with optimal weighting of the measurables. Lower uncertainty measurables should have higher weight compared to others, and vice versa.} to determine the position of the UE that is either standing still or in motion. These algorithms can be categorized based on the measurements they use. Trilateration algorithms utilize TOA, TDOA, and RTT measurements from multiple BSs \cite{talvitie2018positioning_1,talvitie2019positioning, ko2022high}, triangulation algorithms utilize AOA and AOD measurements from multiple BSs \cite{rastorgueva2018beam}, and hybrid algorithms utilize a mix of both range and angle measurements \cite{talvitie2018positioning, talvitie2019radio, talvitie2019positioning_1, trivedi2021localization, garcia2017direct,  gertzell20205g, xhafa2021evaluation, koivisto2016joint}. All LOS algorithms assume that LOS signals are available, detectable, and resolvable (i.e., from NLOS/multipath components). Such an assumption is valid, particularly when considering the wide bandwidth available in 5G NR, as there are other works that tackle such a problem, as mentioned in \ref{sec:Est. of 5G meas.}. Nevertheless, there are certain assumptions in the literature that require careful attention from a practical point of view, as detailed in the following.
\begin{itemize}[leftmargin=*]
	\item \textit{Tight BS-UE synchronization} is practically challenging, cost-wise, as time synchronization bias is expected to exist between the \acp{BS} and the \ac{UE} due to the use of low-cost local oscillators \cite{abu2020single}. Hence, direct utilization of \ac{TOA} measurements without estimating such bias will be erroneous. One solution to address this issue is to jointly estimate the time synchronization bias and the position of the UE, as seen in \cite{werner2015joint,bai2021toa,koivisto2017joint}.
	
	\item \textit{Constant connection to multiple BSs} is challenging to realize in the communication-focused network deployment, where a single \ac{BS} suffices. A more practical approach would be to follow the standards stipulated by \ac{3GPP} with regard to micro BS deployment in urban scenarios \cite{3GPP38855}. Additionally, utilizing quasi-real simulations where 3D maps of real-world cities are utilized to dictate LOS and NLOS operations helps with realizing a more accurate testing scenario \cite{sallouha2023rem,saleh_5g_2023}. The distributed ultra-dense deployment of \acp{BS}, as in the case of \textit{cell-free} networks, offers a promising solution to tackle this challenge (cf. Section \ref{sec:GAS6G}). 
\end{itemize}

Minor 5G positioning sub-fields include NLOS positioning, \acp{RF} fingerprinting techniques, cooperative (co-op) positioning, and integrating 5G with other sensors. 

\begin{itemize}[leftmargin=*]
    \item \textit{NLOS positioning} leverages NLOS and, if available, LOS measurables that may include reflections, diffractions, and scattering to estimate the position of the UE \cite{kim20185g,mendrzik2018harnessing,ge20205g,wen20205mapping,barneto2022millimeter}. To do that, these methods tend to also co-estimate the position of the scatterers surrounding the UE/BS. Such methods are known as \ac{SLAM} techniques and are rigorously investigated in the radar-, LiDAR-, and vision-based positioning literature \cite{placed2023survey}.
    
    \item \textit{RF Fingerprinting} techniques are mainly utilized in indoor scenarios, where the environment is less dynamic \cite{li2021indoor, al2022deep, ji2021multivariable}. 5G indoor \acp{RF} fingerprinting methods utilize both \ac{RSS} and \ac{CSI} measurements/prints to estimate the UE location through various end-to-end \ac{ML} techniques \cite{sallouha2019localization}. Such techniques perform poorly in outdoor settings, as the environment is more dynamic compared to indoor settings.
    
    \item \textit{Cooperative (co-op) positioning} techniques are concerned with exploiting vehicle-to-vehicle communications to share data and positioning measurables to enhance the accuracy and the robustness of the positioning solutions \cite{zhao2020fundamental, kim20205g, chu2021vehicle, chukhno2021d2d}.
    
    \item  \textit{Sensor fusion} of 5G and external sensors like GNSS, \acp{IMU}, and perception-based systems have been established in \cite{saleh_5g_2023,mostafavi2020vehicular,luo2021research,wang2022simulation}. Such integration is essential to realize a seamless positioning solution during inevitable 5G outages. It is worth noting that this promising research area is expected to garner more attention as real-world 5G-based experimental setups/datasets start to become readily available for researchers around the world.
\end{itemize}

\subsection{Aerial Targets: Special Considerations} \label{subsec:G2A}
This subsection focuses on the special considerations associated with positioning aerial targets within 5G terrestrial networks in an \ac{G2A} localization system. Aerial UE poses unique challenges due to geometrical factors such as limited \ac{SNR} caused by downward-tilted BS panels, poor \ac{GDOP} depending on BS deployment, and the requirement for 2D arrays for altitude and attitude estimation  \cite{geraci2022will, strohmeier2018k}. Moreover, special considerations should be given to radio aspects like the reduced multipath effects and the potential applicability of FR1 for positioning purposes \cite{khawaja2019survey, azari2017ultra}. Finally, aerial targets will face challenges arising from high mobility, such as the need for frequent positioning updates and low-latency communications. Accordingly, this subsection is further divided into three aspects, namely, geometrical, radio, and mobility aspects.

\begin{figure}[t]
	\centering
	\input{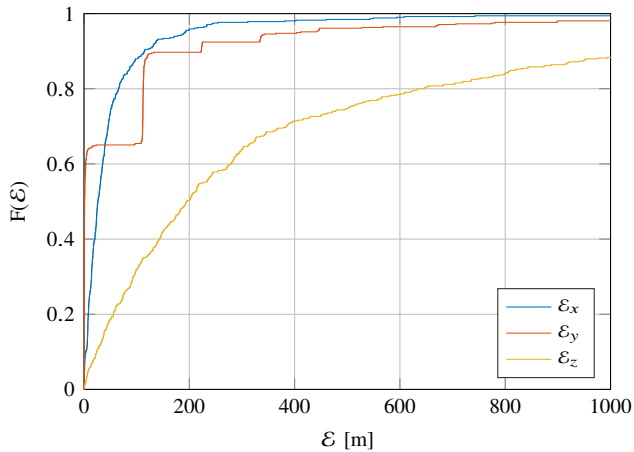}
	\caption{An example CDF of TDOA-based aircraft localization error (in meters) in 3D along the horizontal ($x$ and $y$) and vertical ($z$) axes using terrestrial anchors. The figure demonstrates the poor localization performance on the vertical axis compared to horizontal axes due to the high vertical GDOP. Target aircraft are at altitudes ranging from 8~km to 12~km.}
	\label{Vertical_GDOP}
\end{figure}

\subsubsection{Geometrical Aspects}
In 5G terrestrial networks, \ac{BS} panels are often tilted downward to optimize ground coverage, leading to limited SNR for aerial targets \cite{geraci2022will}. This reduced SNR can adversely affect aerial user equipment's positioning accuracy and reliability. Additionally, the positioning accuracy of aerial targets can be influenced by the deployment of \acp{BS}, resulting in poor GDOP \cite{strohmeier2018k}. This is especially the case for altitude GDOP. Since aerial \acp{UE} are expected to navigate in 3D, information about horizontal and vertical relative angles is paramount. Hence, 2D antenna arrays at the \acp{BS} will be necessary to compute the 3D position and orientation of aerial \acp{UE}. In order to accommodate the relatively high speed and long distances aerial target travel, compared to ground targets, large-scale deployments, e.g., tens or thousands of kilometers, of ground anchors are needed to attain sufficient localization coverage for the aerial segment \cite{2014bringing,strohmeier2018k}. One type of network that can achieve such large-scale coverage is crowdsourced wireless networks, which rely on off-the-shelf massively distributed software-defined radios to provide large-scale localization coverage. OpenSky-network \cite{2014bringing} is one example of an established crowdsourced wireless network initiative, covering the majority of western Europe, focusing on aircraft monitoring by capturing \ac{ADS-B} broadcasts. By capturing a given \ac{ADS-B} message at four anchors, one can use \ac{TDOA}-based MLAT to localize the corresponding aerial target \cite{sallouha2021aerial}. While a crowdsourced wireless network offers large-scale localization coverage, such large-scale deployment makes the anchors synchronization needed for \ac{TDOA} a challenging task \cite{verbruggen2023enabling}. Machine-learning-based clock models, such as LSTM (Long Short-Term Memory), can be used to predict and compensate for clock offset in crowdsourced wireless networks, boosting the accuracy of \ac{TDOA}-based \ac{MLAT} \cite{sallouha2021aerial}. Nonetheless, due to the fact that ground anchors are majorly deployed into a 2D plane, the poor vertical \ac{GDOP} remains an open challenge for aerial target localization with ground anchors \cite{sallouha2021aerial,strohmeier2018k}. Fig.~\ref{Vertical_GDOP} presents an example \ac{CDF} of TDOA-based aircraft localization error along the horizontal $x$ and $y$ axes and the vertical $z$ axis, respectively corresponding to $\mathcal{E}_x$, $\mathcal{E}_y$, and $\mathcal{E}_z$. In this figure, we consider four synchronized ground anchors aiming to localize a set of aircraft via \ac{TDOA} measurements. Each anchor measures the \ac{TOA} of the aircraft's \ac{ADS-B} messages in nanoseconds. Subsequently, these measurements are fused using \ac{MLAT} based on \ac{TDOA} to provide a snapshot position estimate followed by a \ac{KF} to enable aircraft trajectory tracking as detailed in \cite{sallouha2021aerial}. As the figure demonstrates, the error along the vertical direction is noticeably larger, which is mainly attributed to the poor \ac{GDOP} in the vertical dimension.

\subsubsection{Radio Aspects}
Unlike ground-based UEs, aerial targets experience fewer multipath effects, which imposes reduced requirements in terms of measurement resolvability \cite{khawaja2019survey}. This advantage is attributed to the limited bandwidth allocated for FR1 as well as the geometry of the environment. Therefore, the channel for aerial UEs using FR1 is expected to be narrowband with multipath following a Rician distribution \cite{azari2017ultra}. Furthermore, the usage of FR1 will provide wider coverage compared to FR2, increasing the availability of the positioning solution. Hence, positioning using FR1 may hold greater promise compared to FR2, which is more suitable for ground UEs. 

\subsubsection{Mobility Aspects}
Due to the higher dynamics and more stringent requirements of aerial targets, frequent positioning updates and low-latency communications are essential. The rapid movement of aerial \acp{UE} necessitates real-time positioning information. High-speed movement of aerial targets induces a high Doppler spread \cite{wang2022high}, resulting in challenges for positioning accuracy when multiple \acp{BS} are involved. Additional processing and advanced algorithms are required to compensate for the Doppler effect and maintain reliable positioning in multi-BS scenarios \cite{khuwaja2018survey}.

\subsection{Key Takeaways}
This segment delved into localization with ground anchors. While ground anchors have traditionally been utilized for indoor localization, recent advancements in cellular technology, particularly 5G, are extending this capability to outdoor spaces. Within terrestrial network positioning, research spans three primary domains: estimation of positioning measurables, optimization of communication systems, and development of positioning algorithms. Efforts in each domain focus on minimizing error bounds, optimizing system components, and devising algorithms for trilateration, triangulation, and hybrid positioning solutions. Open terrestrial localization challenges include mitigating synchronization bias (time and frequency), detecting LOS and NLOS operation, using the appropriate localization technique for each situation, and signal design to name a few. Furthermore, advances in 5G communications promoted niche research areas within the 5G terrestrial localization field, including NLOS positioning, RF fingerprinting, cooperative positioning, and sensor fusion with onboard motion sensors. In order to realize practical localization solutions, some pitfalls to be avoided. These pitfalls include assumptions about tight BS-UE synchronization and constant simultaneous connection to multiple BSs. Localizing aerial targets using terrestrial networks brings unique challenges caused by the limited SNR, poor GDOP, and high mobility, highlighting the need for frequent positioning updates and low-latency communications. Geometrical and radio aspects also play crucial roles, requiring 2D antenna arrays, large-scale ground anchor deployments, and optimization for frequency ranges to achieve wider coverage and reduced multipath effects.
\section{Localization with Aerial Anchors} \label{sec:A2X}
The integration of \ac{LAP} and \ac{HAP} in future wireless networks is becoming a reality, offering swift and on-demand connectivity for mobile users \cite{geraci2022will}, as well as data collection \cite{li2019joint} and wireless power transfer \cite{jeong2020simultaneous} for \ac{IoT} devices. This integration at the \ac{RF} communication level paves the way to exploit \acp{UAV} as aerial anchors for localization purposes. While Section \ref{sec:G2X} elaborated on the techniques used by ground anchors to handle the multipath wireless channel, exploiting LOS and NLOS links individually, in this section, we focus on aerial anchors, which enjoy the luxury of influencing their wireless channel by leveraging their flexible placement and maneuvering both on the horizontal and the vertical dimensions. This flexible placement of aerial anchors unlocks the possibility of designing a dynamic radio localization system with adaptive anchor positioning. This section covers the main aspects of aerial-anchor-based localization systems designed to localize ground and aerial targets.

\subsection{System Model}
Designing a localization system using aerial anchors shares several similarities with its ground counterpart in terms of the signal strength, time, or angle measurements needed, the primary methods used to estimate the target position, and the filtering employed for target tracking. However, aerial-based localization systems introduce crucial new design aspects, such as adaptive 3D placement and trajectory design. In the following, we detail key system models and design aspects of aerial-based localization systems.

\subsubsection{Radio Technology}
The \ac{RF} radios used in modern \acp{UAV} can be broadly categorized into \ac{ATM} and \ac{CC} radios and data communication radios. A summary of the main radio technologies used in modern UAVs for both categories is presented in Table \ref{UAVtech}. In the category of \ac{ATM} and \ac{CC}, the radios used mainly depend on the \ac{UAV} type and, by extension, on the availability of visual \ac{LOS}, cost, range, and reliability requirements. For instance, in rotary-wing \acp{UAV}, which typically fly at low altitudes ($<$ 500 m), a combination of remote control and Wi-Fi are typically employed, whereas in \acp{HAP} and fixed-wing \acp{UAV}, which tend to fly at high altitudes ($>$ 500 m), long-range protocols, such as \ac{ADS-B} or Flarm, are used \cite{vinogradov2020wireless}. In the category of data connectivity, the types of radios carried onboard depend on the UAV mission's application, which could be a data collection mission form \ac{IoT} nodes, e.g., using LoRa, a video streaming mission using LTE or 5G NR, or a connectivity mission in which the \ac{UAV} acts as an aerial base station \cite{azari2017ultra, khawaja2017uav, ghazali2021systematic}. The radio technology mounted onboard a \ac{UAV} is selected considering the tradeoff between data rate and range as well as constraints set by the application on reliability, cost, and security. Similar to the case of terrestrial anchors, the radio technologies carried onboard the \ac{UAV} have a direct influence on the localization system design in terms of available bandwidth, power, latency, and the frequency used. The same holds when the UAV is the target to be localized. In that case, whether the UAV target is cooperative or non-cooperative, its message/data rate directly relates to the \ac{RF}-based position update rate. It is worth noting that existing \ac{ATM} and \ac{CC} radios in UAVs are mainly focused on communication, location updates, and synchronization with ground stations. However, in order to realize the foreseen dense deployments of UAVs, control messages for time synchronization, location updates, and interference coordination between UAVs need to be standardized. A \ac{UAV} \ac{ATM} ecosystem is under development aiming for standardized autonomously controlled operations of \acp{UAV} beyond visual LOS, which is led by FAA and NASA in the USA \cite{kopardekar2016unmanned}, and in parallel by SESAR (Single European Sky ATM Research) in Europe \cite{undertaking2017u}.

\begin{table}[t]
	\centering
	\caption{Range and transmission rate of the main RF technologies used in UAVs for both ATM and C\&C as well as data.}
	\label{UAVtech}
	\begin{tabular}{l|l|c|cll|}
		\cline{2-6}
		& \multicolumn{1}{c|}{\textbf{Technology}} & \textbf{Max. Range} & \multicolumn{3}{c|}{\textbf{Message/Data Rate}} \\ \hhline{-:=====}
		\multicolumn{1}{|l|}{\cellcolor[HTML]{EFEFEF}}                                    & \textbf{ADS-B}                           & 100 km         & \multicolumn{3}{c|}{2 msg/s}                    \\ \cline{2-6} 
		\multicolumn{1}{|l|}{\cellcolor[HTML]{EFEFEF}}                                    & \textbf{Flarm}                           & 10 km          & \multicolumn{3}{c|}{0.1 msg/s}                  \\ \cline{2-6} 
		\multicolumn{1}{|l|}{\cellcolor[HTML]{EFEFEF}}                                    & \textbf{APRS}                            & 10 km          & \multicolumn{3}{c|}{0.2 msg/s}                  \\ \cline{2-6} 
		\multicolumn{1}{|l|}{\cellcolor[HTML]{EFEFEF}}                                    & \textbf{Wi-Fi SSID}                      & 800 m          & \multicolumn{3}{c|}{20 msg/s}                   \\ \cline{2-6}
		\multicolumn{1}{|l|}{\cellcolor[HTML]{EFEFEF}}                                    & \textbf{Cellular}                      &  2 km          & \multicolumn{3}{c|}{20 msg/s}                   \\ \cline{2-6} 
		\multicolumn{1}{|l|}{\multirow{-5}{*}{\cellcolor[HTML]{EFEFEF}{\rotatebox[origin=c]{90}{\parbox[c]{1.5cm}{\centering \textbf{ATM and C\&C} \cite{vinogradov2020wireless, besada2022modelling}}}}}}      & \textbf{RC}                              & 1 km           & \multicolumn{3}{c|}{25 msg/s}                   \\ \hline \hline
		\multicolumn{1}{|l|}{\cellcolor[HTML]{EFEFEF}}                                    & \textbf{Wi-Fi  2.4~GHz}                   & 500 m          & \multicolumn{3}{c|}{450 Mbps}                   \\ \cline{2-6} 
		\multicolumn{1}{|l|}{\cellcolor[HTML]{EFEFEF}}                                    & \textbf{Wi-Fi 5~GHz}                      & 50 m           & \multicolumn{3}{c|}{1.2 Gbps}                   \\ \cline{2-6} 
		\multicolumn{1}{|l|}{\cellcolor[HTML]{EFEFEF}}                                    & \textbf{LTE}                             & 1 km           & \multicolumn{3}{c|}{100 Mbps}                   \\ \cline{2-6}
		\multicolumn{1}{|l|}{\cellcolor[HTML]{EFEFEF}}                                    & \textbf{5G FR1}                             & 1 km           & \multicolumn{3}{c|}{100 500 Mbps}                   \\ \cline{2-6}
		\multicolumn{1}{|l|}{\cellcolor[HTML]{EFEFEF}}                                    & \textbf{5G FR2}                             & 500 m           & \multicolumn{3}{c|}{2 Gbps}                   \\ \cline{2-6}
		\multicolumn{1}{|l|}{\multirow{-6}{*}{\cellcolor[HTML]{EFEFEF}{\rotatebox[origin=c]{90}{\parbox[c]{1.5cm}{\centering \textbf{Data} \cite{khawaja2017uav, vinogradov2020wireless, azari2022evolution}}}}}} & \textbf{LoRa}                            & 10 km          & \multicolumn{3}{c|}{50 kbps}                    \\ \hline
	\end{tabular}
\end{table}

\subsubsection{Channel Model}\label{subsec:AcMdl}
As discussed in Section \ref{subsec:LocPre}, a major source of error that hampers the quality of localization measurables, in general, is the randomness resulting from the shadowing effect \cite{zanella2016best}. Aerial anchors can swiftly adapt their 3D position, which has a direct impact on the LOS probability and, hence, on the shadowing effect. In order to model the large-scale fading channel between an aerial anchor, e.g., a UAV, at an altitude $h$ above ground level, and a ground target, an altitude-dependent pathloss model is adopted in \cite{sallouha2017aerial,liu2019uav}. This pathloss model represents the received power as a function of the direct distance $d$ and the elevation angle $\varphi = \tan^{-1}{(h/r)}$. This model follows the formulation introduced in (\ref{Prx}) to superimpose pathloss and shadowing; however, it is revised to incorporate the dependencies of shadowing and the pathloss exponent on the elevation angle, as presented in  \cite{hourani,hourani2,amorim2017radio,al2017modeling}. Accordingly, the pathloss exponent $n_p$ and shadowing effect $\mathcal{X}$ in (\ref{Prx}) can be respectively replaced by $n_p(\varphi)$ and $\mathcal{X}(\varphi)$.
Here, $n_p(\varphi)$ is the elevation-angle-dependent pathloss exponent and $\mathcal{X}(\varphi) \sim \mathcal{N}(0,\,\sigma_{\mathcal{X}}^{2}(\varphi))$ represents the elevation-angle-dependent shadowing effect which is a normally distributed random variable with zero mean and variance $\sigma_{\mathcal{X}}^{2}(\varphi)$, in dB.
The elevation-angle-dependent standard deviation $\sigma_{\mathcal{X}}(\varphi)$ with an independently distrusted \ac{LOS} and \ac{NLOS}, is given as \cite{hourani}
\begin{equation}
	\sigma_{\mathcal{X},\,j}(\varphi) = a_j \exp{(-b_j\varphi)},~~~j \in \{\mathrm{LOS}\,,\mathrm{NLOS}\}\,,
	\label{sigj1}
\end{equation}
with $a_j$ and $b_j$ being frequency and environment-dependent parameters. Now, the overall average shadowing effect in the links can be represented by the variance written as
\begin{equation}
	\sigma_{\mathcal{X}}^2(\varphi) = \plos^2(\varphi)\cdot\sigma^2_{\mathcal{X},\,\text{LOS}}(\varphi) + [1-\plos(\varphi)]^2\cdot\sigma^2_{\mathcal{X},\,\text{NLOS}}(\varphi)\,,
	\label{shdow}
\end{equation}
where $\plos$ is the probability of \ac{LOS}. Equation (\ref{shdow}) implies that the shadowing effect gradually diminishes with increasing aerial anchor's altitude due to the decreased probability of encountering obstacles between a transmitter and a receiver \cite{azari2017ultra,amorim2017radio}. The \ac{LOS} probability can be calculated empirically based on measurements \cite{akdeniz2014millimeter}, deterministically based on ray tracing \cite{sallouha2023rem}, or stochastically based on geometry \cite{cui2019frequency}.

In terms of small-scale fading in \ac{UAV} communication channels, the Rician model is widely adopted since it accommodates \ac{LOS}, which has a relatively high probability, and multipath fading \cite{khuwaja2018survey}. In Rician fading, the Rician $K$ factor is a quantitative parameter to represent the severity of the \ac{NLOS} multipath fading. UAV altitude-dependent and elevation-angle-dependent Rician $K$ factors are considered in the literature for \ac{A2A} channels \cite{goddemeier2015investigation} and \ac{A2G} \cite{azari2017ultra}, respectively. Moreover, UAV channels tend to possess a higher Doppler effect compared to their terrestrial counterparts because the relative velocity and mobility of UAVs are higher \cite{khuwaja2018survey}. Large Doppler spread occurs when the transmitting UAV is relatively close to the ground or aerial user. However, suppose the transmitting UAV is further away, e.g., at a sufficiently high altitude. In that case, multipath components will experience a similar Doppler frequency since reflective objects in close proximity to the receiving user will all be seen under similar angles from the transmitting UAV. Such Doppler effect can be mitigated via frequency calibration and synchronization \cite{khawaja2019survey}.

\subsubsection{Anchor Placement}\label{subsec:UAV_trajectory}
\acp{UAV}' versatility and maneuverability enable flexible 3D placement, which is arguably the main distinguishing feature \ac{UAV} aerial anchors have over their terrestrial and celestial counterparts. Since aerial anchors are mobile in nature, their placement refers to \textit{trajectory design}. To fully exploit the benefits offered by this distinctive characteristic, extensive research efforts focused on trajectory design and optimization of UAVs integrated into wireless networks \cite{won2022survey}. In particular, several studies addressed the trajectory optimization of UAV anchors, aiming at leveraging their flexible positioning in designing aerial-based localization systems \cite{sallouha2017aerial,ebrahimi2020autonomous,esrafilian2020three,sallouha2018energy}. To obtain a tractable form of the UAV trajectory with a finite number of optimization variables, path discretization is widely adopted in the literature \cite{zeng2019energy}, in which the UAV path is discretized into $L$ line segments represented by $L+1$ waypoints. The length of a line segment is chosen, taking into account mission characteristics such as UAV velocity, total mission duration, and area of interest. Subsequently, a \ac{UAV} trajectory can be characterized by a set of waypoints with a known size and the duration the UAV spends flying sequentially between waypoints. The set of waypoints is denoted by $\{\textbf{w}_l\}_{l = 0}^{L}$, where $\textbf{w}_l \in \mathbb{R}^{3}$ is the 3D coordinates of the $l$-th waypoint. The duration the UAV spends within each line segment is denoted by  $\{{T}_l\}_{l = 0}^{L-1}$, together indicating a trajectory $\mathcal{T}(\{\textbf{w}_l\}, \{T_l\})$. It is worth noting that the path discretization presented here can also describe cases where a rotary-wing UAV hovers at a given location by having a line segment with a length equal to zero, i.e., $\textbf{w}_l = \textbf{w}_{l + 1}$. A generic formulation for \ac{UAV} trajectory design in an aerial-based radio localization system can be expressed as
\begin{mini!}
	{\mathcal{T}(\{\textbf{w}_l\}, \{T_l\})}{U(\mathcal{E}_{\textbf{p}})\label{eq:TrajO}} {\label{eq:Traj}}{}  
	\addConstraint{f_i(\{\textbf{w}_l\}, \{T_l\})}{\leq \gamma_{f_i}, \,\, i = 1, 2, \dots \label{eq:TrajC1}}
	\addConstraint{p_i(\{\textbf{w}_l\}, \{T_l\})}{\leq \gamma_{p_i}, \,\, i = 1, 2, \dots \label{eq:TrajC2}}
	\addConstraint{g_i(\{v_l\})}{\leq \gamma_{g_i}, \,\, i = 1, 2, \dots \label{eq:TrajC3}}
	\addConstraint{E}{\leq \gamma_E,\label{eq:TrajC4}}
\end{mini!}
where $U(\mathcal{E}_{\textbf{p}})$ is the objective function, which could describe the error of a single target or multiple targets averaged in the time or the spatial domain. The constraint functions (\ref{eq:TrajC1})-(\ref{eq:TrajC4}) are, respectively,
\begin{itemize}
	\item $f_i$ models communication-related constraints such as spectral efficiency, resource allocation, latency, as well as coverage, each represented by a threshold $\gamma_{f_i}$.
	\item $p_i$ represents localization-related constraints, which include the targeted \ac{GDOP}, type of measurements (e.g., \ac{RTT} needs more time to conduct than TOA), LOS/NLOS situations, and the number of waypoints per UAV (e.g., four waypoints if a single UAV is used).
	\item $g_i$ models mobility-related conditions as a function of the segment speed $v_l = (\textbf{w}_{l+1} - \textbf{w}_{l})/T_l\,, \forall\,l$, such as the maximum speed constraint of UAVs and the turning radius constraint upper bounded by the corresponding threshold $\gamma_{g_i}$.
	\item $E$ defines the total energy consumption of the \ac{UAV} anchors bounded by a limit $\gamma_E$.
\end{itemize} 
\begin{figure*}[t]
	\centering
	\includegraphics[width=\textwidth]{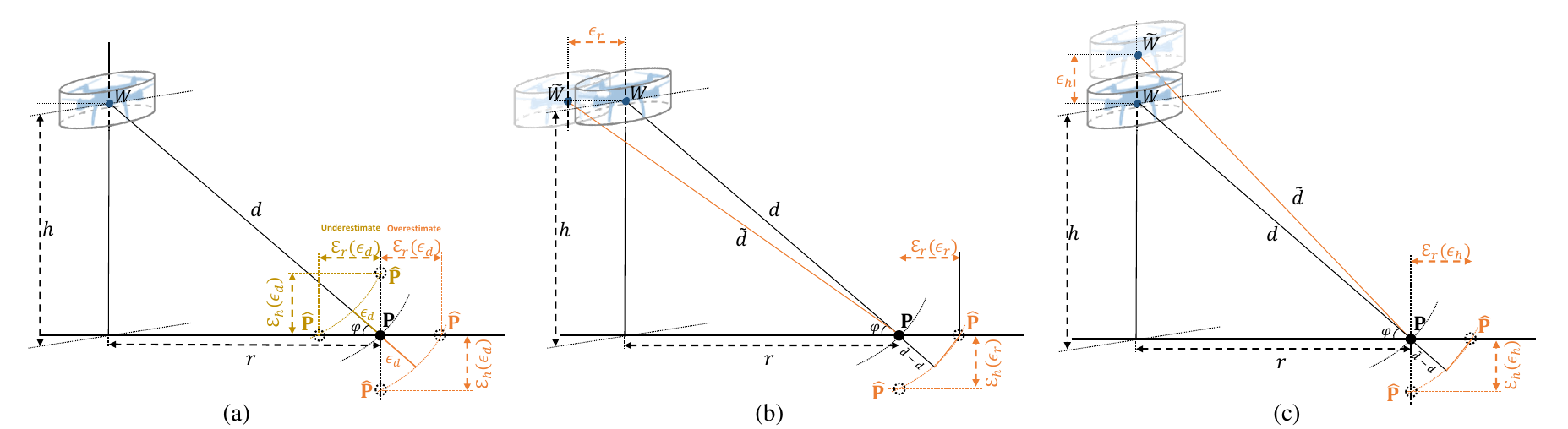}
	\caption{Ranging error bounds with aerial anchors caused by (a) range underestimation or overestimation errors, (b) anchor displacement error on the horizontal plane, resulting from GPS error, (c) anchor altitude error, resulting from barometer error. The point $W$ denotes the scheduled waypoint (the position where the UAV is supposed to be), and the point $\tilde{W}$ denotes the actual position where the UAV is present.}
	\label{A2Gerror}
\end{figure*}

The total energy consumption of a UAV is generally a sum of two components: the propulsion energy, which is required to ensure that the UAV remains aloft as well as to support its mobility, and the communication-related energy used for radiation/reception and signal processing. The propulsion energy depends on whether the UAV is fixed-wing \cite{Zeng} or rotary-wing \cite{sallouha2018energy,zeng2019energy} and is typically expressed as a function of the \ac{UAV}'s instantaneous velocity. The communication-related energy mainly depends on the wireless technology onboard. For instance, in large-size \acp{UAV} that are deployed in remote areas to collect information from IoT nodes, the communication-related can be neglected, whereas in small-size \acp{UAV} that carries a broadband communication technology onboard, the communication-related energy must be considered in the optimization problem.

Optimizing \acp{UAV}' trajectory using (\ref{eq:Traj}) minimizes the localization error and improves the localization accuracy. However, the tradeoff between this better accuracy, on the one hand, and the localization coverage and latency, on the other hand, must be carefully analyzed \cite{sallouha2018energy}. The localization coverage aspect is relevant whether (\ref{eq:Traj}) optimizes the trajectory of single or multiple aerial anchors. Since the communication coverage changes as the \ac{UAV} travels from one waypoint to another \cite{azari2017ultra}, it is important to bound the coverage to the area where reliable localization services are guaranteed. The latency aspect is more relevant to the case when a single UAV is used. In this case, a position fix is only possible after collected measurements at several waypoints, e.g., $\geq 4$ for 3D positioning, enabling a single UAV to act as multiple virtual anchors. The time needed to collect these measurements is bounded by the UAV mobility constraints such as maximum forward and turning speed. Accordingly, when optimizing an aerial anchor's trajectory, coverage and latency requirements can be incorporated into (\ref{eq:TrajC1}) and (\ref{eq:TrajC3}), respectively.

\subsubsection{Anchor Placement Inaccuracy}\label{subsec:UAV_inAccu}
Considering inaccuracy in anchor locations is particularly crucial in aerial-based localization systems. Unlike terrestrial anchors, which are typically stable on the ground, aerial anchors are exposed to weather conditions such as wind and rain, directly influencing their hovering and cruising stability. This weather influence causes a rapidly varying true anchor position, which cannot be accurately captured by the onboard barometer and the GNSS receiver, introducing an error in the local position estimate of the aerial anchor. 
Consequently, the ranging error between the aerial anchor and the target is a function of three independent components, namely, the direct distance estimation error $\epsilon_d$, the anchor's horizontal displacement error $\epsilon_r$, and the anchor's altitude error $\epsilon_h$ \cite{sorbelli2022measurement}. In Fig.~\ref{A2Gerror}a, \ref{A2Gerror}b, and \ref{A2Gerror}c, we visualize the effect on horizontal and vertical ranging error bounds caused by the total ranging error, horizontal UAV displacement error, and error in UAV's altitude, respectively. By applying geometric rules, one can derive the ranging error bounds for a target in 3D both on the horizontal $xy$-plane as well as the vertical plane. The maximum ranging error projected on the $xy$-plane for a given $\epsilon_d$, $\epsilon_r$, and $\epsilon_h$ can be approximated as \cite{sorbelli2022measurement}      
\begin{equation}
	\mathcal{E}_r(\epsilon_d, \epsilon_r, \epsilon_h) \approx \epsilon_r + \frac{h}{r}\epsilon_h + \epsilon_d \sqrt{1 + \frac{h^2}{r^2}}\,,
	\label{UAV_Er}
\end{equation}
where $h$ is the anchor's true altitude, and $r$ is the true distance projected on the horizontal plane between the anchors and the target. Following the derivation of ranging error bounds projected on the horizontal plane \cite{sorbelli2022measurement}, the maximum ranging error projected on the vertical plane can be expressed as
\begin{equation}
		\mathcal{E}_h(\epsilon_d, \epsilon_r, \epsilon_h) \approx \epsilon_h + \frac{r}{h}\epsilon_r + \epsilon_d \sqrt{1 + \frac{r^2}{h^2}}\,.
		\label{UAV_hr}
\end{equation}  
Note that (\ref{UAV_Er}) and (\ref{UAV_hr}) are sufficient to account for anchor placement error in case ranging-based localization is adopted, e.g., \ac{RSS}-based or time-based. However, in the case of angle-based localization such as AOA or AOD, the orientation error of the UAV should also be considered. In the following subsection, a simulated ranging example is presented to study the effect of anchor position inaccuracy, considering different values of $\epsilon_h$ and $\epsilon_r$.

\subsection{Ground Targets: Design Aspects} \label{subsec:A2G}
In this section, we focus on aerial anchors for ground target localization, namely \ac{A2G} localization. The research fields on ground target localization using aerial anchors mainly focus on exploiting adaptive aerial anchor placement. These fields can be broadly categorized into two fields, namely, enhanced localization measurables and localization algorithm design. 
\subsubsection{Enhanced Localization Measurements}
The randomness of the wireless channel resulting from shadowing, a.k.a. large-scale fading, and multipath, a.k.a. small-scale fading, is generally considered as a critical source of error in radio localization systems \cite{zanella2016best}. In \ac{A2G} localization systems, where aerial anchors are employed, the vertical dimension can be exploited to minimize the shadowing effect and improve the \ac{LOS} experience (cf. Section \ref{subsec:AcMdl}), affecting the localization performance positively \cite{perazzo,sallouha2017aerial}.

\begin{figure}[t]
	\centering
	\input{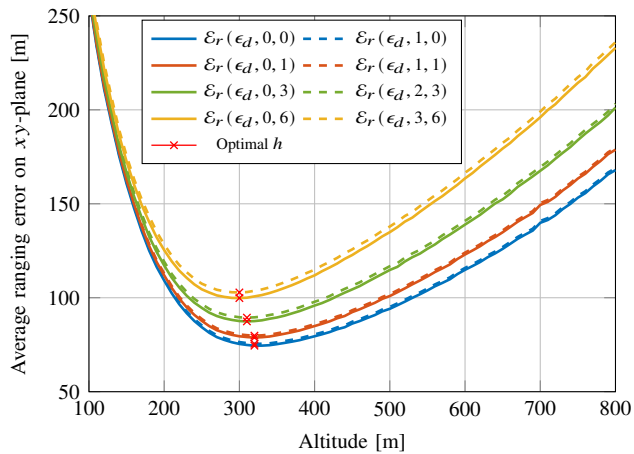}
	\caption{Average ranging error projected on the $xy$-plane versus aerial anchor's altitude $h$, considering distance estimation error as well as aerial anchor's position inaccuracy.}
	\label{AvLocErrUAV2IoT}
\end{figure}

\acp{UAV} favorable channel conditions due to their higher altitude triggered several recent works to investigate \ac{RSS}-based localization as a low-cost and simple solution for wide-range outdoor wireless networks \cite{sallouha2017aerial,sallouha2018energy,esrafilian2020three,ebrahimi2020autonomous}. In \cite{sallouha2017aerial}, the positioning of multiple UAVs flying at the same altitude is optimized with the objective of minimizing the average localization error with \ac{IoT} nodes being the targets to localize. The results showed that \acp{UAV}' altitude plays a significant role in the localization performance, and this positive effect is more pronounced in urban areas compared to suburban areas. Fig.~\ref{AvLocErrUAV2IoT} presents the impact of aerial anchors' altitude on \ac{RSS}-based ranging accuracy. In this figure, we used a Monte Carlo simulation to study the effect of a UAV's altitude, $h$, on the average ranging error of ground IoT nodes uniformly distributed in an urban area with a radius of 1000~m. \ac{RSS}-based ranging is done by incorporating $n_p(\varphi)$ and $\mathcal{X}(\varphi)$ into (\ref{Prx}) and solving the resulting equation for $d$, as detailed in \cite{sallouha2017aerial}. 
The figure shows that an optimal altitude exists at which aerial anchors considerably outperform their terrestrial counterparts. The decreasing trend of the average ranging error at low altitudes is attributed to the decreasing shadowing effect's variance. The increasing trend of the average ranging error is mainly due to the diminishing slope of the RSS-distance logarithmic curve at high altitudes, making distance estimation significantly more sensitive to any relatively small variations in the RSS measurements. The error resulting from the inaccuracy in the aerial anchor position when localizing ground nodes is studied in Fig.~\ref{AvLocErrUAV2IoT} by using the projected ranging error bound defined in (\ref{UAV_Er}). As shown in the figure, the ranging error projected on the ground is more sensitive to the vertical error in the anchor's position $\epsilon_h$ compared to the anchor's horizontal displacement error $\epsilon_r$. This sensitivity to the anchor's vertical error $\epsilon_h$ does not only introduce a ranging error, but it also influences the optimal altitude. In addition to raining accuracy gains, results in \cite{sallouha2017aerial} presented the benefits in terms of the required number of anchors, demonstrating that a lower number of aerial anchors are required to meet a given localization accuracy compared to localization solutions with terrestrial anchors.

The use of time-based localization methods with aerial anchors is investigated in \cite{sorbelli2022measurement,yuan2022uav,perazzo}. Unlike \ac{RSS}-based localization methods, which are mainly influenced by the propagation environment, time-based methods are also affected by the anchor's hardware, such as the local clock stability and the supported bandwidth. In the presence of a stable local clock and sufficient bandwidth, time-based localization methods are able to outperform \ac{RSS}-based methods by orders of magnitude \cite{sorbelli2022measurement,yuan2022uav}. In \cite{sorbelli2022measurement}, the authors demonstrated that by using \ac{TOA}, a localization accuracy in the scale of submeters can be achieved when choosing an elevation angle, $\varphi$, that maximizes the \ac{LOS} probability. However, the direct distance, $d$, was limited to a maximum of 60~m, which is relatively low for aerial anchors. To address the \ac{A2G} localization problem in \ac{NLOS} environments, the authors in \cite{yuan2022uav} introduced an NLOS bias into the \ac{TOA} measurement model and proposed an algorithm that estimates the average bias together with target location to improve localization accuracy. Reported results showed a noticeable localization performance gain, with the tradeoff being the computational complexity. The use of hybrid \ac{TOA}/\ac{AOA} measurements in aerial anchors is investigated in \cite{le2020hybrid}, where \ac{LOS} link of aerial anchors were exploited to localize targets in 3D space.  

\subsubsection{Trajectory Optimization}
As discussed in Section \ref{subsec:UAV_trajectory}, in addition to the waypoint coordinates $\{\textbf{w}_l\}_{l = 0}^{L}$, \ac{UAV}'s trajectory in 3D, $\mathcal{T}(\{\textbf{w}_l\}, \{T_l\})$, is characterized also by the number of waypoints, $L+1$, and the time spent flying between and at them, $\{{T}_l\}_{l = 0}^{L-1}$, \cite{sallouha2018energy,ebrahimi2020autonomous}. In \cite{sallouha2018energy}, an energy-constrained trajectory design of rotary wings UAVs is explored, aiming to minimize the localization error of ground IoT nodes. The reported results showed that a mobile UAV anchor could outperform ground anchors in terms of localization performance. While the results demonstrated the major effect of optimizing the UAV anchor's altitude, it was also shown that when the hovering time at waypoints is relatively low, trajectory optimization on the horizontal dimension is also crucial. In \cite{ebrahimi2020autonomous}, the authors introduced a reinforcement-learning-based framework to minimize the RSS localization error of ground nodes by optimizing the UAV trajectory, considering operational time, number of waypoints, trajectory shape, velocity, altitude, and hovering time. The results showed that there is a tradeoff between the localization error and path length. In particular, increasing the UAV's hovering time, number of waypoints, and velocity remarkably improves the localization performance at the cost of a longer path or longer operation time and hence, higher energy consumption. In \cite{esrafilian2020three}, the authors capitalized on the environment's 3D map to optimize the \ac{UAV} trajectory to localize terrestrial users using RSS measurements under a given mission duration.

Apart from ranging-based \ac{A2G} localization, the mobility of aerial anchors can be exploited in designing non-\ac{MLAT} range-free localization algorithms \cite{sorbelli2022measurement}. The essence of such range-free localization algorithms is the \textit{heard and not-heard} (HnH) technique, in which a stationary ground target exploits a periodically broadcast signal from an aerial anchor moving in a given trajectory to define an area where it may reside and places itself at the center. The broadcast signals from the aerial anchor include its current position (i.e., waypoint). Examples of such range-free algorithms are drone range-free (DRF) \cite{sorbelli2019range} and the intersection of circles (IOC) \cite{xiao2008distributed}. In general, range-free algorithms suffer from poor localization accuracy due to their dependency on the antenna radiation pattern quality \cite{sorbelli2019ground}. In \cite{sorbelli2022measurement}, the authors extended DRF and IOC algorithms to include \ac{TOA}-based distance measurements, resulting in localization error is 3--4 times lower than the original range-free versions.

\subsection{Aerial Targets: Special Considerations} \label{subsec:A2A}
The \ac{A2A} localization scenario, where aerial targets are localized using aerial anchors, which may include \acp{LAP} anchors and \acp{HAP} anchors, introduces several advantages and challenges when compared to localizing aerial targets using terrestrial anchors \cite{Abbas11}. On the one hand, the 3D placement of aerial anchors offers a promising solution to the vertical \ac{GDOP}, and links between aerial anchors and aerial targets are dominantly LOS. However, on the other hand, the high mobility of aerial anchors makes the Doppler effect more challenging, in addition to extra challenges related to the radio technology used, such as the 3D localization coverage and the need for relatively long-range communication. Aerial targets generally have high dynamics at relatively low altitudes (lower than 5~km). Such target include rotary-wing or fix-wing drones and other aircraft that can be regarded as \acp{LAP} \cite{azari2022evolution}. While the need to localize HAPs can be neglected because of their quasi-stationary characteristics \cite{aragon2008high}, they are highly expected to serve as anchors. In the following, we detail the key aspects that differentiate aerial targets from ground targets when using aerial anchors for target localization. 

\subsubsection{Geometrical Aspects}
\begin{figure}[t]
	\centering
	\includegraphics[width=0.45\textwidth]{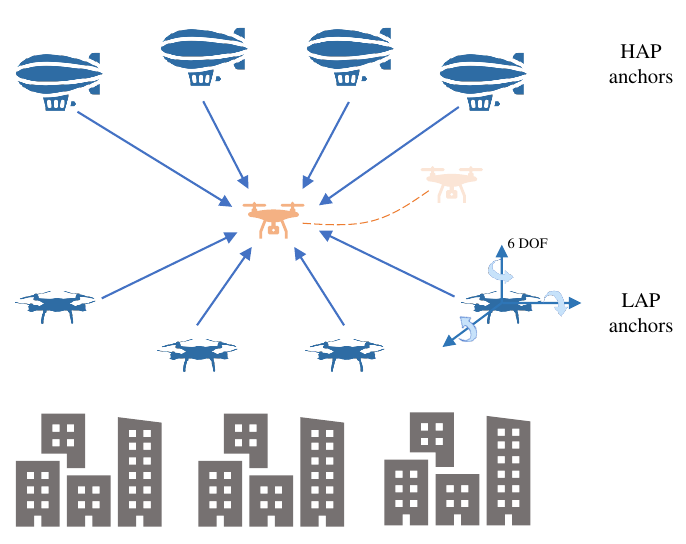}
	\caption{A2A localization: Different altitudes of aerial anchors offer a promising \ac{GDOP} experienced both from the horizontal and vertical perspective.}
	\label{aa_localization}
\end{figure}

Unlike ground and space anchors, aerial anchors can be present at higher and lower altitudes with respect to the aerial target, offering a better vertical \ac{GDOP} experience, as visualized in Fig.~\ref{aa_localization}. While using joint LAPs and HAPs to localize aerial targets improves GDOP, it brings deployment and coordination challenges in terms of both horizontal distances and height differences among aerial anchors. The joint deployment of \ac{LAP} and \ac{HAP} networks is investigated in \cite{jia2022hierarchical,goehar2023investigation}. A general model of collaborative UAV networks and corresponding position error bounds were proposed in \cite{anna2018}. The authors investigated the effect of the number of UAVs and the use of direct or multi-hop links between them on localization accuracy. It has been shown that increasing the number of hops used to share coordinates and distance measurements degrades the localization accuracy. 
In order to minimize the deployment and operation costs of \ac{A2A} localization systems, the localization coverage area can be confined in a given zone, e.g., secure \textit{no-fly} zone. The use of surveillance \acp{UAV} for detecting and localizing aerial targets in a defined no-fly zone is explored in \cite{azari2018key}.

\subsubsection{Radio Aspects}
\Ac{A2A} links are relatively more \textit{clean} than G2A links which experience serious multipath effects \cite{Cui2305}. Thus, the almost pure LOS channel characteristics facilitate LOS-based localization using multiple aerial anchors, thanks to the decreasing trend in the cost of deploying massive UAV swarms. On the one hand, LOS links are beneficial for \ac{A2A} radio localization, aiding both accuracy and coverage. However, on the other hand, they make the interference problem significantly more challenging \cite{reynders2020skysense}. This tradeoff between localization coverage and interference will be particularly explicit in setups adopting relatively low-frequency long-range communications such as FR1. The use of the mmWave band can address the interference problem by relying on narrow directive beams, which also serve angle-based localization. However, the communication range will be determined by mmWave propagation loss, which may result in a reduced \ac{A2A} localization coverage. Radio aspects in \ac{A2A} localization scenarios should also consider whether LAPs, HAPs, or both are used as anchors \cite{jia2022hierarchical,goehar2023investigation}. For instance, HAPs can be used as anchors as well as to regularly update the distance-angle information among LAP anchors. Hence, long-range communication can be used for HAP-LAP links, whereas shorter-range communication can be used among LAPs in an energy-efficient manner.

\subsubsection{Mobility Aspects}
In addition to the geometry and radio aspects in A2A localization, we herein emphasize the mobility aspects represented by the tracking of aerial targets with aerial anchors. Unlike the G2A localization, the A2A localization is more likely to deal with the dual mobility of anchor and target. The 6D mobility, which geometrically includes 3D translational mobility and 3D rotational mobility, of rotary-wing drones will induce many micro-Doppler components. 
Moreover, the placement and orientation of the antennas mounted on the aerial anchors play an important role in receiving signals from aerial targets due to the polarization and directivity. This antenna placement issue is more pronounced in \ac{A2A} localization scenarios since aerial targets may be above or below the aerial anchor. The inherent 6D mobility of UAVs has an influence on angular-based localization. In \cite{sheng16}, the authors utilized AOA for target localization, where the UAV six-\ac{DOF} dynamic model is formulated, and a diffusion extended \ac{KF} is used to implement distributed adaptive estimation. However, the trajectory optimization complexity of aerial anchors becomes very high when considering six-\ac{DOF} in the angular domain. To address this complexity, one potential solution is to employ HAPs, exploiting their quasi-static behavior to minimize fluctuations in angle estimation.

\subsection{Key Takeaways}
Aerial anchors use radio technology both for data as well as for \ac{ATM} and \ac{CC}. They enjoy adaptive on-demand placement, which can be optimized to improve the LOS experience and, hence, the localization performance. To optimize an aerial anchor's trajectory to minimize localization error, aspects to take into account include radio communication requirements, localization measurements type, anchor mobility, as well as anchor's energy budget. While several works in the literature addressed aerial anchors' trajectory optimization, solutions that jointly consider communication, localization, mobility, and energy aspects are still an open research direction. Despite its enormous advantages, this adaptive placement comes with a cost related to the fact that aerial anchor placement is prone to errors that translate into target localization errors. One interesting open research direction is to study and model the key environmental factors that cause this error, such as wind speed and rain, combined with a \ac{KF} to enable accurate placement calibration. Another potential solution may rely on harnessing a known reference point, such as a ground station, to act as a reference to calibrate aerial anchor placement. In addition to aerial anchors' advantages in ground target localization, their 3D placement enables addressing \ac{GDOP} both horizontally and vertically in aerial target localization. However, the tradeoff is the excessive Doppler effect that may result from target and anchor mobility, particularly when considering fixed-wing aerial anchors.
\section{Localization with Space Anchors} \label{sec:S2X}

In recent years, the space industry has changed dramatically. The new space industry is dominated by commercial players, which is a break from the past when space was almost only accessible by governments. Launching satellites has become more affordable, especially launching satellites to \ac{LEO} \cite{janssen2023survey}. This has led to a large number of LEO telecommunication satellite constellations being proposed and launched. The major advantage of using LEO satellites in comparison with satellites in a higher orbit is the lower communication delay and lower pathloss \cite{liu2021leo,lalbakhsh2022darkening,lin2021path}. On the downside, each satellite can only cover a smaller area of Earth's surface. As a result, a LEO constellation consists of a large number of satellites to reach global coverage. Therefore, LEO satellites are often small and built from commercial off-the-shelf components to reduce costs. This section explores localization using \acp{NTN} with space-based anchors. More specifically, we will focus on the use of LEO satellites for localization. \ac{LEO} satellites have enticing attributes that are useful for localization purposes \cite{prol2022position,janssen2023survey}. Current space-based localization is based on \ac{GNSS}, which resides in \ac{MEO} at around 20,000 km above the surface of the Earth. The shorter distance to Earth of LEO satellites leads to high signal power and a shorter delay. LEO satellites often use different frequency bands, leading to a higher frequency diversity, which hardens space-based localization against jamming and spoofing. In addition, the large constellation sizes and the wide variety of different orbits increase the geometrical diversity.

\begin{table*}[!t]
	\centering
	\caption{Radio technology used in commercial LEO NTNs.}
	\label{tab:ntn_leo_const_radio}
	\begin{tabular}{|l|l|l|l|l|l|l|}
		\hline
		\multicolumn{1}{|c|}{\textbf{Name}}   & \multicolumn{1}{c|}{\textbf{Use}} 	& \multicolumn{1}{c|}{\textbf{Height} [km]}  & \multicolumn{1}{c|}{\textbf{Modulation}} 	& \multicolumn{1}{c|}{\textbf{Frequency}}       	&  \multicolumn{1}{c|}{\textbf{Bandwidth}} & \multicolumn{1}{c|}{\textbf{Proposed/Current size}}  	\\ \hline \hline
		Iridium NEXT    & BB  			& 780					& PSK        			& 1618.25~MHz 				& 31.5~kHz			&   66 / 66 		\\ \hline
		StarLink      	& BB  			& 550					& OFDM       			& 10.7-12.7~GHz   			& 250~MHz 			&	11927 / 5636	\\ \hline
		Kuiper        	& BB			& 610					& -          			& 17.7-20.2~GHz   			& 100~MHz     		&	3236 / 0		\\ \hline
		OneWeb        	& BB			& 1200					& OFDM       			& 10.7-12.7~GHz   			& 250~MHz    		&	648 / 634		\\ \hline
		GlobalStar    	& Voice			& 1420					& CDMA       			& 1618~MHz        			& 1.23~Mhz    		&	42 / 25			\\ \hline
		Sateliot      	& IoT			& 505					& NB-IoT     			& 2000~MHz        			& 180~kHz      		&	250 / 0			\\ \hline
		OmniSpace     	& IoT			& 530					& NB-IoT     			& 1990~MHz        			& 180~kHz      		&	200 / 2			\\ \hline
		OQ Technology 	& IoT			& 480					& NB-IoT     			& 2005~MHz        			& 180~kHz      		&	10 / 10			\\ \hline
		Lacuna        	& IoT			& 500-550				& LR-FHSS    			& 915~MHz     				& 336~kHz      		&	240 / 6			\\ \hline
		Swarm         	& IoT			& 300-550				& LoRa       			& 149~MHZ         			& 20,8-62,5~kHz		&	150 / 164		\\ \hline
		Myriota       	& IoT			& 530					& GMSK       			& 400~MHz         			& 150~kHz     		&	50 / 3			\\ \hline
		OrbComm       	& IoT			& 710					& PSK        			& 149~MHz         			& 15~kHz      		&	50 / 50			\\ \hline
	\end{tabular}
\end{table*}

There are two types of LEO satellites that can be used for localization. The first is communication LEO satellites, which can be used as an opportunistic source of localization signals; this is called the \ac{SOP} method \cite{reid2016leveraging}. The main advantage is the reuse of already deployed infrastructure. However, using these proprietary systems for localization might not be trivial as the used technology might not be disclosed. The second type is the LEO-\ac{PNT} constellations, which are specifically designed to provide \ac{PNT} services from low Earth orbit. Several commercial companies have launched initial LEO-\ac{PNT} test satellites, for example, the PULSAR system from Xona Space Systems \cite{Xona2022}, the GeeSAT satellites from GeeSpace and the CentiSpace satellites from Future Navigation \cite{Yuan2021}. Governmental institutes are also interested in LEO-\ac{PNT}. For instance, the ESA (European Space Agency) is planning to launch several test satellites in the near future as part of their FutureNAV program \cite{FutureNav}. 

\subsection{System Model}
Localization using LEO satellites brings some challenges \cite{Reid2018}. In order to localize targets, the location of the anchors has to be known. In the case of satellites, the anchors are in an orbit around Earth. As a result, the anchors are constantly moving, albeit in a very predictable way \cite{janssen2023survey}. Still, the precise orbit needs to be known before the target can be localized, and errors in the orbit data result in localization errors. In addition, the stability of the onboard satellite clock is very important as this can lead to time and frequency errors \cite{prol2022position}. For time-based ranging methods, a small time offset results in a sizeable ranging error, and Doppler ranging methods rely on precise frequency offset measurements. The next challenge is the signal structure of the communication signal. Localization methods often rely on searching for known signals. However, often, the used signals of commercial satellite communication systems are proprietary, resulting in difficulties in tracking the signals. In addition, the signals might not be optimized for being used on localization systems, resulting in sub-optimal performance. All of the aforementioned system aspects are discussed in the following.

\subsubsection{Radio Technology}
The first parameter that influences radio technology and its signal design is the frequency of operation. The chosen band has to be available following the frequency plan. For different applications, different parts of the spectrum are reserved. Furthermore, the chosen frequency dictates the available bandwidth; more bandwidth is available at higher frequencies. In addition, higher bandwidth is generally beneficial for localization performance as the time resolvability and measurement accuracy can be increased when a larger bandwidth is available \cite{ferre2021comparison}. In Table \ref{tab:ntn_leo_const_radio}, we present an overview of the radio technology used in existing commercial LEO constellations. Note that the constellation sizes presented in this table are constantly changing as satellites are being launched at the highest rate ever in history. When designing the modulation of the signal, the ranging properties of the signal have to be taken into account. A ranging signal should have good auto-correlation properties. When correlating the signal with itself, the response shows a high peak when the signals completely overlap and no response in all other cases. The width of the peak depends on the symbol rate of the sequence, which in turn relates to the bandwidth of the signal. A narrow peak results in a higher ranging accuracy. \ac{GNSS} signals use \ac{PRN} codes, which have excellent auto- and cross-correlation properties. These \ac{PRN} codes are modulated on the carrier using binary phase-shift keying or binary offset carrier modulations. However, when considering \ac{LEO} communication satellites, the modulation also has to be optimized for communication purposes as well. Therefore, \ac{OFDM} waveforms have to be considered. With the recent inclusion of the \ac{PRS} in 5G networks, an excellent ranging signal is added to the \ac{OFDM}-based standard \cite{ferre2019positioning}. The \ac{PRS} is a downlink signal that can be scheduled on different resource blocks, resulting in a configurable frequency and time span. The used symbols are gold sequences with excellent auto-correlation properties. This signal is a good starting point for enabling pseudo-range measurement with space-based communication systems.

\subsubsection{Channel Model}
Several channel models have been developed for LEO communication satellites. However, there is not one complete channel model that covers all LEO constellations. These channel models depend on multiple factors linked to the LEO constellations considered, as well as the system's frequency and bandwidth. An overview of channel models is given in \cite{monzon2022overview}. Therefore, in the following, we will discuss some of the key aspects that have to be taken into account in modeling satellite-to-Earth channels.

Due to the high mobility of LEO satellites, the Doppler frequency and rate are high. The Doppler frequency (i.e., Doppler shift) is highest for satellites with a low elevation angle (e.g., 10\degree), while it approaches zero for satellites a high elevation angle (e.g., right overhead at 90\degree). When acquiring and tracking LEO signals, taking the Doppler frequency into account is essential. Assuming a perfectly known \ac{CFO} and received signal frequency $f_r$, from (\ref{eqnDoppler}), the Doppler frequency is defined as $f_D := f_r - f_c$. It can be written as a function of the relative radial velocity $\Delta v$ between the target and space anchor as follows
\begin{equation}
    f_D = \frac{\Delta v}{c} f_c\,,
\end{equation}
where $f_c$ is the carrier frequency used by the satellite. For broadband networks using K-band frequencies ($\sim12$GHz), the Doppler frequency can be up to 250kHz \cite{kozhaya2023multi}. 

When transmitting signals through the ionosphere, the signal experiences an ionospheric delay \cite{liu2021arctic}. Due to solar radiation, the ionosphere is a partially ionized medium. The delay of the signal depends on the total amount of free electrons along the signal's path, noted as the \ac{STEC} \cite{liu2021arctic}. As this is influenced by solar radiation, the ionospheric delay is the largest during the day. The phase and the modulation of the signal experience a different delay. The modulated code gets delayed, called the \textit{group delay} $\Delta_{group}^{iono}$, while the phase actually experiences an advance $\Delta_{ph}^{iono}$ \cite{spilker1996global}. The size of the delay and the advance can be calculated as
\begin{equation}
	\Delta_{ph}^{iono} = -\frac{40.3}{f_c^2} \cdot\, \textrm{STEC}, ~~ \Delta_{group}^{iono} = \frac{40.3}{f_c^2} \cdot \, \textrm{STEC}.
\end{equation}
The higher the frequency, the lower the ionospheric effect. Due to the frequency dependence, the ionospheric effect can be estimated when using two separate frequencies. For example, all GNSS constellations are transmitting ranging signals at multiple frequencies to be able to correct the ionospheric delay. 

Higher frequencies require smaller antennas, as the size of an omnidirectional antenna is coupled to the wavelength of the carrier frequency. This can be an advantage as this makes the system more compact. However, the received power will also decrease due to the decreased antenna surface. As the distance between the transmitter and receiver is relatively large, increasing the received signal power is critically important. One popular solution exploited in satellites is the use of antenna arrays to create directional beams to increase signal power.

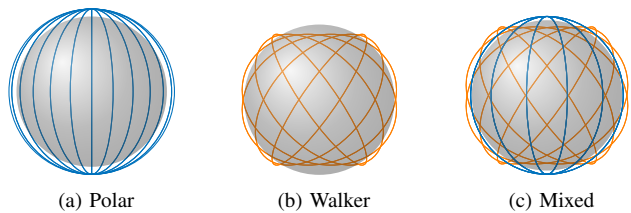
\begin{figure}[t]
	\centering
	\begin{subfigure}[b]{0.3\linewidth}
		\centering
		\definecolor{mycolor1}{rgb}{0.00000,0.44700,0.74100}%

\tdplotsetmaincoords{90}{15}
    \begin{tikzpicture}[tdplot_main_coords]%
        
        \foreach \plane in {0, 15, ..., 179}
            {\tdplotsetrotatedcoords{\plane}{90}{0}
             \begin{scope}[tdplot_rotated_coords]
                 \draw[color=mycolor1] (0,0,0) circle (1.1);
             \end{scope}}
             
        \shade[ball color = gray, opacity=0.4] (0,0,0) circle (1cm);
        
             

\end{tikzpicture}
		\caption{Polar}
		\label{fig:Polar}
	\end{subfigure}
	~
	\begin{subfigure}[b]{0.3\linewidth}
		\centering
		\tdplotsetmaincoords{90}{15}
\begin{tikzpicture}[tdplot_main_coords]%

    \foreach \plane in {30, 60, ..., 360}
        {\tdplotsetrotatedcoords{\plane}{56}{0}
         \begin{scope}[tdplot_rotated_coords]
             \draw[color=orange] (0,0,0) circle (1.05);
         \end{scope}}
         
    \shade[ball color = gray, opacity=0.4] (0,0,0) circle (1cm);

\end{tikzpicture}
		\caption{Walker}
		\label{fig:Walker}
	\end{subfigure}
	~
	\begin{subfigure}[b]{0.3\linewidth}
		\centering
		\definecolor{mycolor1}{rgb}{0.00000,0.44700,0.74100}%

\tdplotsetmaincoords{90}{15}
\begin{tikzpicture}[tdplot_main_coords]%

    \foreach \plane in {30, 60, ..., 360}
        {\tdplotsetrotatedcoords{\plane}{56}{0}
         \begin{scope}[tdplot_rotated_coords]
             \draw[color=orange] (0,0,0) circle (1.1);
         \end{scope}}

    \foreach \plane in {30, 60, ..., 360}
        {\tdplotsetrotatedcoords{\plane}{90}{0}
         \begin{scope}[tdplot_rotated_coords]
             \draw[color=mycolor1] (0,0,0) circle (1.05);
         \end{scope}}
         
    \shade[ball color = gray, opacity=0.4] (0,0,0) circle (1cm);

\end{tikzpicture}
		\caption{Mixed}
		\label{fig:Mixed}
	\end{subfigure}
	\caption{Polar orbits have a high inclination angle ($\theta_i \approx 90^\circ$) and have global coverage. However, they provide the best coverage in polar regions. Walker constellation has a lower inclination angle ($\theta_i \approx 55^\circ$). They optimize the coverage over the population hotspots at mid-latitudes but lack coverage at the poles. A mixed constellation can provide the best of both.}
	\label{fig:constellation}
\end{figure}

\subsubsection{Anchor Placement}

A satellite constellation is a group of satellites with the same function and with similar orbits. The way the orbits of the satellites are designed has a large influence on the coverage of the system. The most important parameters in constellation design are the orbit height, inclination angle, and constellation size \cite{prol2022position}. A higher orbit has the advantage of being able to cover a larger area but increases the free space pathloss, propagation delay, and launch cost. The inclination angle is the angle between the orbital plane and the equatorial plane. The inclination greatly influences where on Earth the coverage is optimized \cite{ma2020hybrid}. Polar orbits have an inclination angle close to 90\degree. They require the least amount of satellites to reach global coverage. However, the coverage is highest in the polar regions, while the population density of Earth is not concentrated around the poles. Using a Walker constellation with inclined planes of 50\degree\, to 60\degree\, optimizes the coverage at the major population centers at mid-latitudes but lacks coverage at the poles, as depicted in Fig.~\ref{fig:constellation}. The selection of the constellation type depends on the system's use case. A combination of mixed polar and inclined orbits can bring the best global coverage while optimizing the coverage at population density hotspots. The third parameter to optimize is the constellation size. The amount of satellites needed to reach global coverage depends on how many satellites must be visible at the same time for a user. For localization purposes, a minimum of four visible satellites is needed. The number of satellites needed to reach the targeted coverage can be calculated depending on the chosen orbit height and inclination. The authors of \cite{guan2020optimal} use a genetic algorithm to design a positioning constellation, optimizing for GDOP while minimizing constellation size.

Unlike aerial anchor placement, which can be controlled and adapted on-demand, space anchors follow predefined orbits that are prone to perturbations. For instance, effects such as atmospheric drag are particularly pronounced at LEO orbits. Therefore, even after designing the orbit height, inclination angle, and constellation size, orbit determination remains a key aspect of user localization. To localize users based on space anchors, the precise orbit of the satellites must be known. For LEO satellites, this can be achieved in two ways. The first method makes use of the already available GNSS constellations. A GNSS receiver is deployed in orbit on the LEO satellite. In this way, the precise orbit of the satellite can be determined. However, this makes the system reliant on GNSS, which means that if these constellations become unavailable, the orbit of the LEO anchors can not be determined. In addition, the accuracy of this method is limited unless the onboard GNSS receiver can be provided with correction signals needed to enable precise point positioning \cite{janssen2023survey}. To acquire more precise orbit information, a ground monitor network is needed. Such a network consists of static receivers that are continuously monitoring the signals coming from the satellites. Based on the measurements of the network, a more precise location of the satellites can be estimated.

\subsubsection{Clock Synchronization}

To exploit LEO satellite signals as a ranging reference, the clocks of the different satellites have to be mutually synchronized. GNSS satellites use large, heavy, and expensive atomic clocks to keep a very accurate timing reference on board the satellite. As a typical LEO constellation usually exists out of a large fleet of small and inexpensive satellites, such atomic clocks are not an option. Alternative options include:
\begin{itemize} [leftmargin=*]
	\item Use smaller, less accurate atomic clocks on board the LEO satellite. As the satellites orbit the planet several times a day, the clocks can be adjusted using correction signals transmitted from a fixed position on Earth. In this way, the period in which the clock has to be stable is limited to a couple of hours.
	\item Exploit higher orbits satellites disciplined clock, e.g., \ac{GNSS} disciplined clock. For this strategy, the satellite contains a \ac{GNSS} module and an \ac{OCXO} to keep the time. The \ac{GNSS} module is used to receive the signals of the \ac{GNSS} constellations and compute the time. This information is used to discipline the \ac{OCXO} and correct the drift of the clock. In \cite{Kunzinavi.531}, the authors present how they used an Ultra Stable Oscillator in combination with a triple frequency GNSS receiver to achieve sub-nanosecond time synchronization.
	\item Employ \acp{OISL} to enable communication and synchronization between the satellites in a constellation. \acp{OISL} make use of lasers deployed on the satellites to set up high-speed communication links with excellent ranging performance \cite{zech2019optical}. The authors of \cite{MICHALAK20214753} show that employing \acp{OISL} in a GNSS constellation can increase the accuracy of the precise orbit determination and decrease the signal-in-space ranging error.
\end{itemize}

\subsection{Ground Targets: Design Aspects} \label{subsec:S2G}
In this section, we focus on two aspects regarding the usage of LEO satellites for localization purposes. In particular, we discuss the various design aspects of LEO-based localization systems and the opportunistic usage of \ac{SOP} from communication-based LEO satellites. The design of space-based localization requires a careful trade-off between accuracy, cost, power, weight, and size, as discussed in the following.

\subsubsection{Constellation Selection}
When using \ac{NTN} LEO constellations for localization, constellation selection can be done based on use-case. Table \ref{tab:ntn_leo_const_radio} lists the major operational and proposed communication satellite constellations along with their corresponding localization use cases in the literature. We consider two key aspects, namely, high accuracy and low power. Applications considering high accuracy will benefit from choosing an NTN constellation using a large bandwidth signal such as StarLink, Kuiper, or OneWeb. These broadband NTN constellations use a very large bandwidth (100 to 250 MHz) and have a very high number of satellites in the constellation, leading to a very low GDOP \cite{Reid2018}. For low-power applications where accuracy is not critical, such as asset tracking, IoT-focused NTNs can be considered. These networks use a narrow bandwidth and use a simpler modulation, requiring less power to process the signals. As a consequence, the data rate and localization accuracy will decrease. A combination of multiple constellations can also be considered. In \cite{farhangian2020multi}, Iridium NEXT and Orbcomm satellites are used together as space anchors. The authors of \cite{kozhaya2023multi} combine even four different constellations.

\subsubsection{Signal Structure}

When using \acp{SOP} for localization, knowing the basic signal structure is imperative. The most straightforward approach is to use NTNs that use standardized modulations. Several IoT-focused satellite constellations use NB-IoT. This standard is regulated by 3GPP, and in the recent Release 17, space-based NB-IoT networks were presented \cite{Liberg_2020}. As a result, the signal structure is publicly known and could be exploited for localization. In addition, an altered version of LoRa is proposed for space-based networks, called \ac{LR-FHSS} \cite{Boquet_2021}, which will be used by the Lacuna constellation. Many other commercial communication satellite systems use proprietary signals which are not disclosed to the public. One strategy would be to reverse-engineer the signal structure. This can be achieved by recording the signal and figuring out the signal parameters, for example, sampling frequency, modulation type, and frame structure. In \cite{humphreys2023signal}, the authors recorded and analyzed StarLink signals. They described how they reverse-engineered the signals and documented the signal structure and synchronization sequences. They hope to exploit these sequences in the future as a means to measure the range to the receiver.

\subsubsection{Ephemeris}
In order to localize a target based on \acp{SOP}, the location of the anchors has to be known. The position of satellites is constantly changing. However, it is predictable as the satellite follows an orbit around Earth. Therefore, to calculate the position of the satellite, the orbit has to be known. The description of an orbit is called the \textit{ephemeris}. The ephemeris can be presented by a \ac{TLE}. These \acp{TLE} are publicly shared by the NORAD (North American Aerospace Defense Command) and are updated daily \cite{norad}. TLEs might be an easily available source of ephemeris data; however, they are not very precise, which results in reduced positioning accuracy. In order to obtain precise ephemeris data, a reference network can be set up. Such a network of receivers tracks the satellites from fixed and known positions. Based on the measurements, the orbits of the satellites can be precisely determined, improving the positioning performance. In \cite{khairallah2021ephemeris}, the ephemeris data of LEO satellites is estimated by tracking the satellites using Doppler and pseudo-range measurements to improve upon the TLE data. A datalink is needed in order to transfer the ephemeris of the satellites to the target. This datalink can be a terrestrial link or provided by the NTN itself. Providing the data through a satellite link has the advantage of only having to receive one signal and reaching global coverage. Following 3GPP Rel. 17, space-born NB-IoT satellites will provide the ephemeris data to the users of the network \cite{Liberg_2020}.

\begin{table}[t]
\centering
\caption{Localization methods used in the literature using commercial LEO NTNs.}
\label{tab:ntn_leo_const}
	\begin{tabular}{|l|l|l|}
		\hline
		\multicolumn{1}{|c|}{\multirow{2}*{\textbf{Name}}} 	 & \multicolumn{2}{c|}{\textbf{Localization user-cases}}  \\ \cline{2-3} 
		                          & \multicolumn{1}{c|}{\textbf{TDOA}} & \multicolumn{1}{c|}{\textbf{Doppler}} \\ \hline \hline
		Iridium NEXT              &  \cite{joerger2010analysis} &    \cite{kozhaya2023multi}\cite{Farhangian2021}\cite{janssen2023survey}                                   \\ \hline
		Starlink                  &  -    						&    \cite{kozhaya2023multi}\cite{Neinavaie2022}\cite{Psiaki2021}\cite{ardito2019performance}                                  \\ \hline
		OneWeb                    &  -    						&    \cite{kozhaya2023multi}\cite{Psiaki2021}                                   \\ \hline
		GlobalStar                &  -    						&    \cite{janssen2023survey}\cite{ardito2019performance}\cite{Neinavaie2021}                            \\ \hline
		Lacuna                    &  -   						&   \cite{janssen2023survey}                                     \\ \hline
		OrbComm                   &  -    						&  \cite{kozhaya2023multi}\cite{Farhangian2021}\cite{janssen2023survey}\cite{ardito2019performance}\cite{Neinavaie2022_unknown}\cite{Khalife2019}                                     \\ \hline
	\end{tabular}
\end{table}

\subsubsection{Localization Methods}

Localization using \acp{SOP} is a recent research domain that has attracted considerable attention. The recent addition of many new LEO communication satellites has led to a boom in research to use the new \acp{SOP} to localize terrestrial nodes. 

\begin{figure}[t]
	\centering
	\input{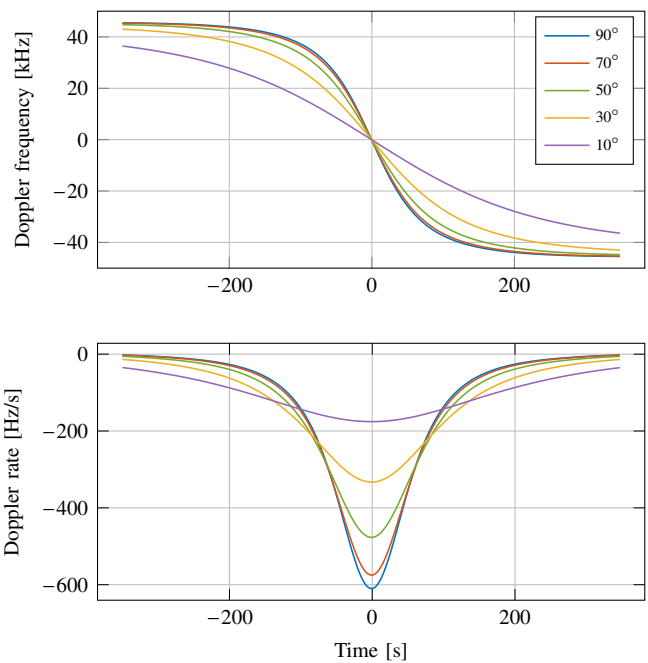}
	\caption{The Doppler frequency and rate for passes of a LEO satellite with various maximum elevation angles (10\degree-90\degree). The orbit height was set to 550 km and the carrier frequency is 2~GHz. By tracking the Doppler frequency of a LEO satellite over a short period, the pseudorange rate can be determined. The location of the receiver can be calculated by tracking multiple satellites. }
	\label{fig:doppler}
\end{figure}
\begin{itemize}[leftmargin=*]
\item \textit{Time-Difference-of-Arrival}: As this technique is very mature and well-researched, it is one of the prime candidates for being used with LEO \acp{SOP}.
In order for this technique to work, the space anchors have to be accurately synchronized with each other, and the signal structure has to be known in order to use ranging techniques. GlobalStar is a prime candidate to be used as a basis for a \ac{SOP} \ac{TDOA} navigation system as the GlobalStar signal structure is based on a \ac{CDMA} modulation similar to GNSS systems and has been studied before \cite{schiff2000signal} \cite{Neinavaie2021}. 

\item \textit{Doppler positioning}: LEO satellites move across the sky rapidly. Due to the significant relative motion between the satellites and the ground-based targets, the signal experiences a significant Doppler shift. The basic idea is to track the Doppler frequency, which has a different profile for different satellites and depends on the location of the satellite in the sky. Fig.~\ref{fig:doppler} shows the Doppler frequency and rate for LEO satellites with various maximum elevation angles. Depending on whether the satellite passes directly overhead (90\degree) or stays low at the horizon (0\degree), the maximum elevation angle of the satellite changes accordingly. The Doppler frequency relates to the relative speed of the satellite and the receiver. The values shown correspond to an orbit height of 550 km and a carrier frequency of 2 GHz, with a static receiver considered. When tracking the Doppler frequency of multiple anchors, a target can localize itself. This method was the basis of \textit{Transit}, the first satellite-based localization system \cite{noureldin2012fundamentals}. Moreover, the method is regaining popularity in research as a way to localize targets based on LEO \acp{SOP}. The main advantage compared to TDOA is that the exact signal structure is not needed to track the Doppler frequency of the signal. However, in order to reach a good positioning estimate, the signal has to be tracked for a longer time \cite{farhangian2020multi}.

The Doppler frequency measurement can be converted to a pseudorange rate measurement as the Doppler frequency relates to the relative motion between the satellite and the target receiver. The pseudorange rate $z_n[k]$ at timestep $k$ is calculated as
\begin{equation}
    z_n[k] = c \frac{f_{D_n}[k]}{f_{c_n}},
\end{equation}
where $f_{D_n}$ is the Doppler frequency of satellite $n$, $f_{c_n}$ is the carrier frequency of the $n$-th satellite \cite{farhangian2020multi}, and $c$ is the speed of light. When considering a static receiver, the pseudorange rate $\dot{\mathbf{p}}_{A,n}[k]$ relates to the location of the user $\mathbf{p}_U$ and the satellite/anchor $\mathbf{p}_{A,n}[k]$ as
\begin{equation}
    z_n[k] = \frac{\dot{\mathbf{p}}_{A,n}^{\top}[k]\left[\mathbf{p}_U - \mathbf{p}_{A,n}[k]\right]}{\|\mathbf{p}_U - \mathbf{p}_{A,n}[k]\|} + c\left(\dot{\delta t}_U-\dot{\delta t}_{A,n}\right) + \epsilon_{A,n}[k],
\end{equation}
 where $\dot{\delta t}_U$ and $\dot{\delta t}_{A,n}$ are the clock drift of the receiver and the the $n$-th satellite, respectively, and $\epsilon_{A,n}[k]$ includes errors induced by noise and atmospheric effects.

In \cite{farhangian2020multi}, Doppler positioning is deployed on OrbComm and Iridium signals. The Doppler frequency and carrier phase are tracked using Costas Loops. The position of the user is estimated using an Extended Kalman Filter. In \cite{kozhaya2023multi}, Starlink and OneWeb satellites are included in the Doppler receiver design to further improve the achievable accuracy; they reach an accuracy of 5.8~m. In \cite{Psiaki2021}, the authors develop a point-solution using a batch least-squares filter to estimate position, velocity, clock offset, and clock offset rate by tracking eight or more satellites simultaneously. Their simulations and \ac{GDOP} studies indicate that the possible accuracy is in the order of 1 to 5 meters.
\end{itemize}
Table \ref{tab:ntn_leo_const} reveals that until now, most of the literature on positioning with LEO satellites is based on Doppler positioning. The main reason for this is that the exact signal structure is often unknown, and Doppler positioning does not require this knowledge.

\subsection{Aerial Targets: Special Considerations} \label{subsec:S2A}

When using space anchors to localize aerial targets, some special considerations have to be taken into account, as detailed in the following. 

\subsubsection{Geometrical Aspects}
As aerial users move around in 3D space, estimating the altitude of the aerial targets is of utmost importance. When using LEO satellites as anchors, the vertical \ac{GDOP} tends to be double the Horizontal GDOP \cite{reid2016leveraging}. This leads to a halving of the accuracy when comparing the target's altitude estimation to its horizontal position estimation. The reason behind this poor vertical GDOP is mainly due to the fact that visible space anchors all have a positive elevation angle; therefore, the information of the vertical position only comes from above, while horizontal information comes from all four horizontal directions, i.e., north, south, west, and east. 

\subsubsection{Radio Aspects}

Aerial targets experience a high chance of open-sky visibility compared to their ground counterparts, increasing the localization availability using space anchors. Aerial targets can make use of all satellites above the horizon, even the ones at low elevation, which are often unavailable for terrestrial targets. Moreover, aerial users are not typically in a rich multipath environment, lowering the impact of possible multipath errors \cite{khuwaja2018survey}. Nevertheless, strong multipath effects can be present due to reflection from potential structural elements of the aerial user, e.g., the wings of an airplane. The high visibility of aerial targets also makes them very prone to interference and jamming events \cite{de2016gnss}\cite{sabatini2017global}. This increases the need for interference mitigation as well as anti-jamming and anti-spoofing technologies. 

\subsubsection{Mobility Aspects}

Aerial targets often exhibit a high mobility profile \cite{khawaja2019survey}, which includes not only 3D movement but also changes in the 3D orientation of aerial targets. These movements can lead to a changing orientation of the receiving antenna. For instance, typical antennas used in current GNSS applications are optimized for receiving signals arriving from a positive elevation angle. In this way, the antennas focus on the signals coming from the sky while suppressing ground reflection and interference. However, when the orientation of an aerial target antenna changes due to the high 6D mobility, the quality of the received signal will degrade for certain space anchors as the antenna is directed away from them \cite{sabatini2017global}. Furthermore, the received power of ground reflections and interference will increase.

\subsection{Key Takeaways}
Localization using LEO satellites represents a vibrant domain of research, propelled by the recent surge in LEO communication networks. A unique aspect of LEO satellites compared to ground and aerial anchors is their rapid movement in defined orbits. On the one hand, this movement unlocks Doppler-based localization and diversifies the set of satellites used in target localization, but on the other hand, it brings challenges related to synchronization, orbital errors, and dynamic resource allocation. Present research primarily focuses on harnessing LEO communication signals for localization purposes. However, these networks are not inherently tailored for positioning, necessitating receivers to opportunistically self-localize. Alternatively, dedicated LEO-PNT constellations are in developmental stages, exclusively engineered to provide precise and resilient positioning services. Future research should explore the synergy between these two approaches, combining communication capabilities with accurate positioning services within a unified \ac{NTN}. Moreover, as LEO constellations continue to expand, commercial positioning services will become available, opening the door to more comprehensive studies on localization systems coexistence and integration within the space segment and across the multiple \ac{GAS} segments.
\section{Research Directions Towards 6G GAS Localization} \label{sec:GAS6G}
With the wider bandwidth and larger number of antennas in 5G, the localization accuracy can attain meter-level \cite{isac_6g_ground}, and in 6G, the localization accuracy is expected to realize centimeter-level. This section presents the role key 6G enablers are expected to play in GAS localization systems, exploring their added value and highlighting research directions towards 6G \ac{GAS} localization.


\subsection{Reconfigurable Intelligent Surfaces}
\begin{figure*}[t]
	\centering
	\includegraphics[width=0.8\textwidth]{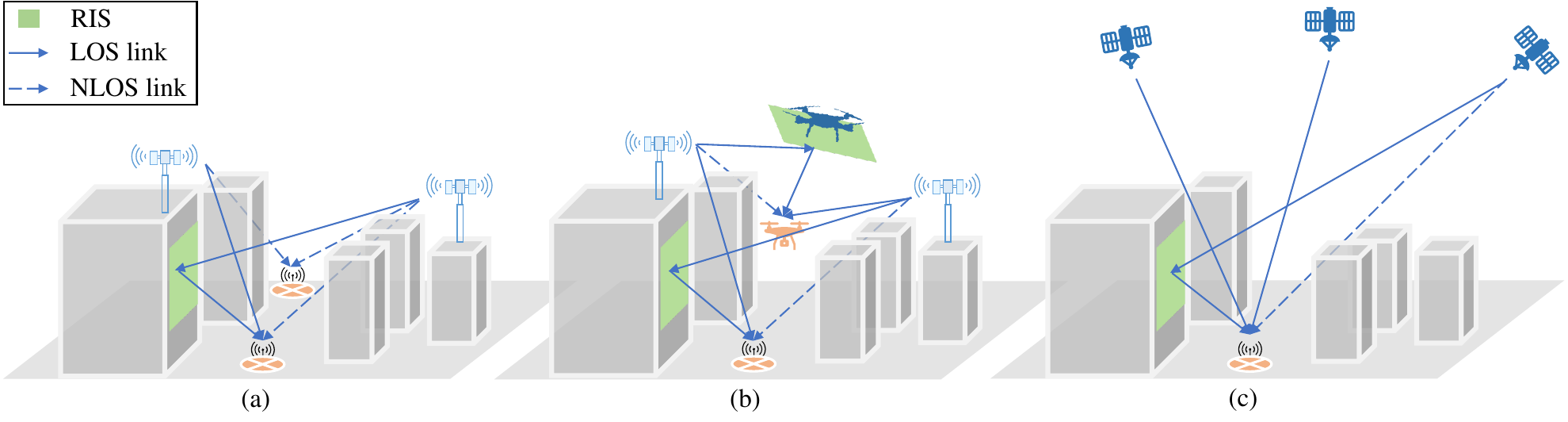}
	\caption{A representation of RIS in GAS.}
	\label{RIS_GAS}
\end{figure*}

The main motivation behind \ac{RIS}, for both positioning and communication purposes, aligns closely with that of integrated GAS networks, aiming to enhance the coverage, availability, and robustness of the solution. RIS offers a means to extend coverage by serving as supplementary anchors that can provide extra LOS positioning measurements, e.g., TOA, AOA, and AOD, for localization systems in 6G networks \cite{wymeersch2020radio,chen2022reconfigurable}. In cases where the LOS path between the UE and the \ac{TN}/\ac{NTN} is heavily degraded by the environment due to obstacles, the RIS alternative path can act as the main positioning path in the system. Furthermore, if the \ac{TN}/\ac{NTN} measurements were strong enough, then adding extra measurements to the fusion filter, in the form of RIS measurements, will enhance the localization accuracy further. It is worth noting that RIS measurements can be distinguished from other propagation paths by utilizing unique signatures, such as time encoding \cite{bjornson2022reconfigurable}. This is achieved by manipulating the phase or amplitude of the incident signals as they interact with the reconfigurable elements. These interactions are precisely controlled and can be finely tuned to create distinguishable temporal patterns in the signals reflected or refracted by the RIS.

RIS also introduces a distinctive advantage in high-precision positioning by providing an accurate estimation of horizontal and vertical AOA and AOD \cite{ma2023reconfigurable}. This is mainly achieved by performing a beam-sweep via the large number of reconfigurable elements spanning the RIS, as illustrated in Fig.~\ref{RIS_GAS}a. Such accurate angular measurements can facilitate fine estimation of the position of a UE in a \ac{SISO} system \cite{keykhosravi2022ris}. Hence, the requirement of having a 2D array at the UE, BS, or NTN for single anchor positioning can be relaxed in the presence of RIS. A noteworthy advantage of deploying the RIS technology, compared to deploying extra \ac{NTN}/\ac{TN} anchors, lies in its ability to improve positioning without introducing additional clock offsets or carrier offsets. Adding additional anchor nodes (terrestrial or non-terrestrial) to the system might complicate the processing chain due to synchronization challenges. On the other hand, RIS seamlessly integrates into existing systems without posing such hindrances. Additionally, the deployment and operation costs of RIS are considerably lower compared to TNs and NTNs \cite{liu2022path}. This characteristic streamlines the deployment of RIS-enabled positioning solutions, making them more practical and efficient to implement in 6G networks.

Within the realm of RIS technology, various configurations emerge, including passive RIS, active RIS, hybrid RIS, and simultaneous transmitting and reflecting (STAR)-RIS, also known as intelligent Omni-surface. In \cite{umer2023role}, the role of RIS in 6G localization has been comprehensively investigated in terms of different types of \acp{RIS}. Passive RIS is designed to reflect the incident signals without power amplification. Hence, they are more energy efficient than the other implementations but can only operate if the incident signal is strong enough. Active RIS implementations, on the other hand, hold the potential to amplify the power of \ac{NTN} signals, bolstering their efficacy for communications and localization purposes in degraded channel conditions \cite{schroeder2021passive}. As the name suggests, hybrid RIS comprises passive and active elements. Hybrid RIS can combine the simplicity and power efficiency of passive implementations while maintaining the favorable qualities of the active implementations \cite{nguyen_hybrid_2022}\footnote{The term \textit{hybrid} RIS is sometimes used to indicate the usage of some active RIS elements that can sample the incoming signal via an RF chain \cite{schroeder2021passive}. In our case, however, \textit{hybrid} RIS indicates the usage of some active elements that \textit{actively} amplifies the impinging signal before reflection.}. All of the aforementioned RIS implementations only consider reflecting the incident signal to any angle on the space in front of the RIS \cite{liu2021star}, which implies that both the user and the anchor node should be in front of the plane of the RIS to operate. STAR-RIS, on the other hand, proposed in \cite{liu2021star}, extends the operation of the RIS to the other side of the plane, enabling the signal to pass through the surface and be refracted to any angle on the other end. Hence, STAR-RIS presents an opportunity to extend the \ac{TN}'s and NTN's coverage to include indoor localization, effectively expanding the reach of the network's localization services. Yet, it is worth noting that increasing the reach of RIS may cause interference issues with other networks \cite{jiang_interference_2022}. Hence, managing such interference in RIS-equipped networks may pose challenges, necessitating careful coordination strategies to ensure that the benefits of RIS are realized without introducing undue disruptions.

As presented in the localization measurables in Section \ref{sec:locFun}, the uncontrollable multipath nature of the wireless channel is one of the major localization error sources. In principle, one of the key advantages of \ac{RIS}, from a localization perspective, is its ability to offer a controllable path in the channel model. The channel model between the transmitter and the receiver with a RIS found in between can be constructed by building upon the established model in (\ref{ch-model}):
\begin{equation}\label{ch-model-RIS}
    \boldsymbol{y}_{k,l}=\boldsymbol{W}_l^{\mathsf{H}}(\boldsymbol{H}_{k,l}+\boldsymbol{H}'_{k,l})\boldsymbol{f}_ls_{k,l}+\boldsymbol{n}_{k,l}\,,
\end{equation}
where $\boldsymbol{H}_{k,l}$ and $\boldsymbol{H}'_{k,l}$ comprise the uncontrollable and the semi-controllable RIS channels between the transmitter and the receiver, respectively. The semi-controllable channel constitutes the uncontrollable Tx-RIS $(\boldsymbol{H}_{\text{tr}_{k,l}})$ and RIS-Rx $(\boldsymbol{H}_{\text{ru}_{k,l}})$ channels and the controllable RIS response $\boldsymbol{\Omega}_l~=~\text{diag}(\boldsymbol{\vartheta}_l)~\in~\mathbb{C}^{N\times N}$, where $N$ is the number of RIS elements, and $\boldsymbol{\vartheta}_l \in \mathbb{C}^N$ is the RIS response vector. The elements of $\boldsymbol{\vartheta}_l$ are controllable and are chosen from $\boldsymbol{\Theta}$, the set of feasible RIS element configurations, and are formulated as follows:
\begin{equation}
\boldsymbol{\vartheta}_l=\boldsymbol{\gamma}_l\odot\boldsymbol{\omega}_l,
\end{equation}
where $\boldsymbol{\gamma}_l\in \mathbb{R}^N$ constitutes the RIS elements' gain/attenuation factors, and $\boldsymbol{\omega}_l~=~[e^{-\jmath 2 \pi f_c \tau_{1,l}},\dots, e^{-\jmath 2 \pi f_c \tau_{N,l}}]^{\top} \in \mathbb{C}^{N}$ are the corresponding phase shifts. Here, $\tau_{n,l}$ is the delay caused by the $n$-th RIS element during the $l$-th symbol.

\subsubsection{RIS-Aided Ground Anchors}
Utilization of significantly large RIS setups with ground anchors offers the potential for near-field localization by harnessing wavefront curvature \cite{ozturk2023ris}. This capacity leverages the RIS's ability to manipulate signal propagation at extremely close distances, enabling advanced positioning techniques that capitalize on the unique behavior of electromagnetic waves in near-field scenarios. Thanks to \ac{RIS}'s ability to partially control multipath propagation \cite{bjornson2022reconfigurable}, it is expected to have a positive impact on all localization \acp{KPI} when used with ground anchors. Key research challenges when using \ac{RIS} with ground anchors are related to the operation of the RIS itself, e.g., nonlinearity and mutual coupling \cite{bjornson2022reconfigurable}, and appropriate site selection, service access \cite{lu2021aerial}.

\subsubsection{RIS-Aided Aerial Anchors}
To overcome challenges related to RIS site selection, aerial deployment of RIS is explored in \cite{lu2021aerial}. Deploying \acp{RIS} on aerial anchors, as depicted in Fig.~\ref{RIS_GAS}b, is mainly driven by the extra \ac{DOF} obtained from the flexibility of anchors placement, which offers even more possibilities for \ac{LOS} links with aerial and ground users, not only outdoor but also indoor \cite{albanese2021first}. Moreover, given the limited payload of aerial anchors, RIS can act as an alternative to carrying heavy RF transceivers required for coverage extension, positively impacting the \ac{UAV} mission duration. Key challenges associated with RIS-aided aerial anchors mainly relate to the complexity of the localization system. In particular, this complexity arises from coordinating interference, incorporating \acp{RIS} with antennas needed for flying assets in a compact and energy-efficient way, and implementing effective controllers for the RIS configuration given that the channel might be changing aggressively \cite{alfattani2021aerial}. Moreover, considering the mobile nature of aerial anchors and their placement inaccuracy (cf. Section \ref{subsec:UAV_inAccu}), errors in the RIS placement and geometric layout are unavoidable in practice, necessitating the localization and calibration of the RIS itself \cite{zheng2023jrcup}.

\subsubsection{RIS-Aided Space Anchors}
\acp{RIS} can aid space anchors by deploying them on the ground as depicted in Fig.~\ref{RIS_GAS}c and as explored in \cite{wang2023beamforming}. Integrating \acp{RIS} on space anchors is investigated in \cite{tekbiyik2022reconfigurable}, where they are exploited to aid cooperation for inter-satellite THz links in the LEO constellation. While such implementation may indirectly aid several localization \acp{KPI}, its scalability and complexity require further investigation.

\subsection{Joint Communication, Localization, and Sensing}
To boost system integration and suitability in 6G, joint consideration of communication, localization, and sensing functionalities becomes prevailing. A comprehensive study on joint communication, localization, and sensing integrated in 6G radio technologies classifies them into levels. These levels are defined for communication as very short range (C1) and short-range (C2), for localization as high-accuracy (L1), low-latency (L2), and low-complexity (L3), and for sensing as monostatic (S1), and Bi-multi-static (S2). The requirements to integrate them are discussed in terms of signals, hardware architectures, and deployments. In addition to the integration, the assistance/enhancement among different functions plays a vital role in harnessing the added value of such integration, which spawns a variety of research topics, such as, among others, sensing-aided communications and localization-aided communications. In terms of multilateral assistance, the research challenges, as well as potential directions, can be summarized as follows. 
\begin{itemize}[leftmargin=*]
	\item \textit{Communication-aided sensing/localization:} The relevantly large communication bandwidth, e.g., FR1, FR2, and the expected GHz scale in future \ac{6G}, plays a key role in terms of delay/range resolvability and accuracy of both localization and sensing \cite{henkeucnc22}. Moreover, since communication networks are deployed aiming for wide coverage in \ac{JCAS}\footnote{We use the acronym JCAS to stay in line with the standard acronym used in the literature. Nonetheless, in this work, we stress that the sensing part of the acronym refers to the broad definition of sensing that includes passive sensing and active target localization.} systems, such coverage can also serve localization and sensing coverage. In addition, instead of depending solely on localization and sounding signals, e.g., \ac{PRS} and \ac{SRS}, anchors can jointly use communication signals for localization and sensing, which positively impacts localization latency, eliminating the localization latency bounds imposed by the period at which such signals are broadcast. As a targeted goal of \ac{JCAS}, sharing communication hardware in transceivers also for localization and sensing benefits the power efficiency of the system \cite{liu2022survey}.
	
	\item \textit{Localization-aided communication/sensing:} Communication systems are tied to location information in various ways, including distances, delays, velocities, angles, and predictable user behavior \cite{taranto14}. With accurate and ubiquitous location information, communication performance can be enhanced across all layers of the communication protocol stack, such as indicating \ac{SNR} and shadowing autocorrelation, predicting user mobility patterns, and securing driving operations. Moreover, wide-sense sensing naturally includes the positioning process. However, for multidimensional sensing, angular and Doppler information is needed, where localization can somehow increase sensing ability and reduce sensing complexity at the same time. In scenarios where narrow beams are used for communications, accurate knowledge of user location can aid in minimizing the number of adjacent beams employed for environment sensing and blockage prediction \cite{hersyandika2022guard}.

	\item \textit{Sensing-aided localization/communication:} Sensing, in the context of environment passive sensing, is generally considered together with the localization, e.g., \ac{SLAM} technique \cite{placed2023survey}. The positioning of the target can benefit from accurate environment mapping via static background removal. Sensing-aided communication has been under extensive study in recent years, where the realization of joint communication and sensing has been proposed and tested in spatial, temporal, and frequency domains \cite{liu2022survey}.
\end{itemize}

Towards practical implementation of \ac{JCAS} systems, the potential topics proposed in \cite{henk21pimrc} for JCAS systems include the following aspects: 1) Angle, range, and Doppler resolvability when employing high frequency such as mmWave and THz, 2) Directional beamforming and context awareness, 3) Joint waveforms and joint hardware, and 4) Algorithmic developments including model-based MIMO algorithms and data-driven AI. These general topics are also promising in the \ac{JCAS} system. However, in the following, we 
present tailored \ac{JCAS} topics for the various \ac{GAS} anchors.
\subsubsection{JCAS-Aided Ground Anchors}
Considering \ac{JCAS} systems in 6G, techniques such as \ac{SLAM} can be enabled by relying on communication signals. To this end, the waveform design of communication signals to jointly allow accurate SLAM is an important research direction. By employing JCAS in ground anchors, the localization performance can be boosted by identifying the scatterers in the environment, where a specific multipath can serve as a virtual anchor. The study in \cite{isac_6g_ground} shows through learning the locations of such virtual anchors, i.e., scatterers, the localization performance can achieve an accuracy of 10~cm, outperforming the 60~cm accuracy obtained when using localization systems solely. Thanks to the rich scattering in ground networks, multipath components serving as virtual anchors can help improve the localization accuracy that is constrained by the number of physical anchors.

\subsubsection{JCAS-Aided Aerial Anchors} Thanks to the umbrella view of aerial anchors on a given area, they offer the option to fuse \ac{RF}-based measurements, such as CSI, with vision-based sensing data captured from the equipped cameras. Under this context, a promising application resides in environmental reconstruction, which includes both static scatterers and mobile targets such as cars and humans. Thus, a data-fusion-based environment mapping is able to provide accurate location information of moving targets by using a swarm to form spatial diversity \cite{aerial_swarm} or a single moving aircraft to collect time-series information. Moreover, aerial anchors can be used in conjunction with ground anchors to form a bistatic \ac{JCAS} system, where the aerial anchor collects a ground \ac{BS}'s echoed signals for radar sensing \cite{hu2022trajectory}.

\subsubsection{JCAS-Aided Space Anchors} Jointly considering sensing/radar and communication in LEO satellites is a rising trend. Recent works such as \cite{leo_jcas} optimized the hybrid precoding to balance the radar beam pattern and energy efficiency of communication in a massive \ac{MIMO}-enabled \ac{LEO} satellite. It provides a solution to use sub-arrays to sense targets and serve users simultaneously. Besides, the LOS-dominant channel between space anchors facilitates the exchange of information free of multipath interference, e.g., via \ac{THz} narrow beams, which suggests space anchors can be easily synchronized, thus positively impacting localization accuracy and resource allocation in general.

\subsection{AI-Empowered Localization}
\Ac{AI} refers to a collection of algorithms and models employed in devices to enable them to cognitively plan and make optimum decisions. To train such algorithms and models, AI systems rely on \ac{ML}, which generates versatile models by exploiting relatively large amounts of data for model training \cite{jagannath2019machine,sun2019application}. This model training can be done in a supervised, unsupervised, semi-supervised, or reinforcement learning manner. AI-aided localization can be broadly categorized into two forms: A fully data-driven form and a complementary AI-enhanced model-driven form. 
\begin{itemize}[leftmargin=*]
	\item \textit{Data-driven}: In this form, AI-based algorithms are designed, taking raw measurements and \acp{RF} fingerprints, which may include power, time, and/or angle, to provide a position estimate \cite{sallouha2019localization,timoteo2020scalable}. Data-driven learning, in which data-trained models replace the full location estimation functionality, can potentially be beneficial in terms of computational complexity during the location estimation/inference phase \cite{sallouha2019localization}. However, the main drawback is the need for very large and representative datasets \cite{antwerpDset,li2019deep}. In order to reduce reliance on location-based labeled datasets, deep-reinforcement-learning-based unsupervised localization methods can be employed \cite{li2019deep}.
	When the statistics of the true and trained channels are not the same, blind data-driven methods can fail dramatically; in such cases, online learning is a promising solution at the cost of computational complexity and localization latency. Furthermore, the data-driven localization using measurements collected at the receiver can be extended to end-to-end learning for localization, jointly optimizing transmit beamformers and receiver-side algorithms, even in the presence of hardware impairments \cite{rivetti2023spatial}.
	
	\item \textit{AI-enhanced model-driven}: In this form, AI-based algorithms are designed to model the stochastic propagation environment, e.g., channel models \cite{aldossari2019machine}, or to calibrate and compensate for imperfections in localization measurables \cite{sallouha2021aerial}. One promising direction is the calibration in antenna arrays, e.g., ML-MUSIC, where \ac{ML} is used to calibrate the array imperfections \cite{de2022expert}. In addition, clock offset and hardware imperfections that make localization challenging are also relevant to AI-empowered localization \cite{sallouha2021aerial}. In addition, AI could be used to learn how to fuse multiple estimates in a weighted manner or make a final decision from noisy estimates obtained via different anchors in the GAS network.
\end{itemize}
Intuitively, AI-enhanced model-driven methods that tackle calibration and hardware imperfections are beneficial for all \ac{GAS} anchors, whereas the fully data-driven methods are more beneficial for ground and \ac{LAP} anchors compared to \ac{HAP} and space ones. In the following, we discuss the AI-empowered localization from each of the GAS anchors' perspectives.

\subsubsection{AI-Aided Ground Anchors}
Both data-driven \ac{AI} and AI-enhanced model-driven approaches can be exploited in scenarios with ground anchors. In fully data-driven methods, the rich multipath nature of the \ac{G2G} environments ensure sufficient variations in datasets features captured in the spatial domain, e.g., power, time, and angle features \cite{sallouha2019localization,timoteo2020scalable}. In \ac{G2A} scenarios, where the multipath effect is less pronounced, AI-empowered data-driven solutions can rely on feature variation in the spatial domain over relatively wide areas, e.g., tracking aircraft over ranges of several kilometers, at the cost of limited localization accuracy \cite{strohmeier2018k}. In AI-enhanced model-driven methods, \ac{ML} classifiers can be used to identify the LOS and NLOS signals received at the anchor, boosting the accuracy of time-based localization methods \cite{van2015machine}. \ac{ML}-based clock-drift modeling and offset compensation has been explored in \cite{sallouha2021aerial}, enabling \ac{TDOA}-based localization in widely distributed ground anchors. 

\subsubsection{AI-Aided Aerial Anchors}
\Ac{LAP} aerial anchors can exploit data-driven methods in \ac{A2G} localization, benefiting from the rich multipath ground environments and the mixed \ac{LOS}/\ac{NLOS} links compared to the predominantly \ac{LOS} links in the case of \acp{HAP}. In addition to the calibration and channel modeling advantages that can be obtained from AI-enhanced model-driven methods \cite{bithas2019survey}, 3D placement and trajectory design of aerial anchors can be optimized using reinforcement learning \cite{ebrahimi2020autonomous}. A critical issue that requires careful attention when employing \ac{ML} models in aerial anchors is the computational complexity and the corresponding power consumption of model training with large datasets. To address this issue, efficient methods for online learning are needed in order to work in resource-constrained aerial anchors \cite{bithas2019survey}. Moreover, federated learning is investigated in the literature to address the resource-constrained problem by distributing the computations load; however, it does not fully address the communication bottleneck, which is still an open research area, since AI models' weights still need to be transmitted \cite{zeng2020federated}. 

\subsubsection{AI-Aided Space Anchors}
The majority of works focusing on AI-aided space anchors address the resource allocation problem due to the need to optimize the scarce and expensive satellite resources \cite{azari2022evolution}. In data-driven-based localization, several works investigated using \ac{LEO} satellite images in 2D \cite{ahmadien2020predicting} or 3D \cite{sallouha2023rem}, aiding ground anchors in keeping up-to-date environment maps. Due to the relatively high speed of space anchors in the LEO orbit, space anchors can benefit from AI-enhanced model-driven methods, which enable orbit and clock/Doppler offset patterns prediction, improving target tracking accuracy \cite{mortlock2021assessing}.

\subsection{Cell-Free Paradigm}
\begin{figure}[t]
	\centering
	\includegraphics[width=0.49\textwidth]{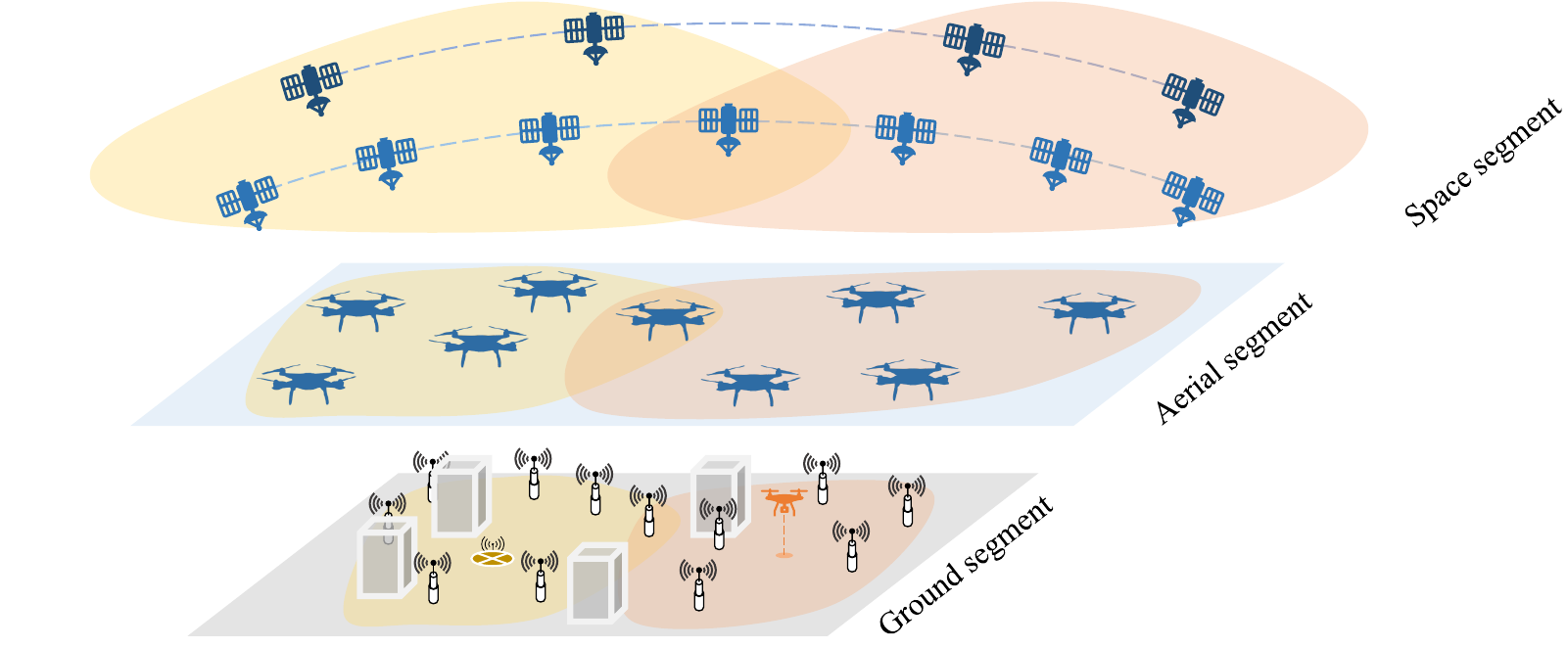}
	\caption{A representation of cell-free in the different segments of GAS networks.}
	\label{CF_GAS}
\end{figure}

Current 5G networks address the demand for higher spectral efficiency through densifying \ac{BS} (a.k.a.~\ac{AP}) deployment with small cells and increasing the number of antennas per base station, i.e., massive \ac{MIMO}. While this network-centric approach noticeably increases spectral efficiency, it makes the inter-cell interference problem more pronounced, which particularly impacts cell-edge users, e.g., they might be tens of dB weaker channel compared to cell-center users \cite{demir2021foundations,interdonato2019ubiquitous}. Intuitively, a given UE would only be affected by interference between its own cell and a set of neighboring cells; hence, it is only this corresponding cluster of \acp{AP} that needs to cooperate to alleviate inter-cell interference for this UE. This intuition is the foundation of the emerging \textit{user-centric \ac{CF}} paradigm, which adopts an alternative network architecture to address the inadequate cell-edge users' experience. In \ac{CF} networks, a user-centric cluster of \acp{AP} jointly operates to coherently serve \acp{UE} on the same time/frequency resource, using spatial multiplexing. In order to have sufficient spatial multiplexing capabilities, a massive number of distributed \acp{AP} are needed, introducing \ac{CF-mMIMO}, in which the word \textit{massive} implies more \acp{AP} than \acp{UE} \cite{demir2021foundations}. The architecture of \ac{CF-mMIMO} inherently implies (sub-)nanosecond synchronization among densely deployed distributed \acp{AP}, and potentially multiple antennas both in uplink and downlink, which benefit from reduced pathloss and macro-diversity mitigating shadowing/blocking \cite{chen2022survey}. Having a massive number of \acp{AP}, each acting as a localization anchor, as visualized in Fig.~\ref{CF_GAS}, offers unprecedented opportunities to realize a robust \ac{GDOP} across a wide area, positively impacting localization accuracy and coverage in all segments of \ac{GAS} networks. In addition, \ac{CF} as a network architecture could be a potential candidate to realize spatial multiplexing using the different segments of \ac{GAS} networks since \ac{LOS} \ac{MIMO} essentially requires a large antenna separation. The implementation of the cell-free architecture can take various forms in terms of using several \acp{AP} with a single or multiple antennas. The coherent transmission in cell-free networks can be done in time, phase, or both \cite{interdonato2019ubiquitous,ngo2017cell,zheng2023asynchronous}, whichever used directly influenced the localization system design. 
\begin{itemize}[leftmargin=*]
\item \textit{Time-coherent operation}: Operating in a time-coherent manner can be done using several single-antenna \acp{AP} that are synchronized by a GPS clock. Such tight synchronization provides accurate \ac{TOA} measurements with respect to each other, enabling target localization using \ac{TDOA} or \ac{RTT} \cite{sallouha2021aerial}.
\item \textit{Phase-coherent operation}: A more popular form of cell-free is the phase-coherent transmission \cite{demir2021foundations}, where phases among \acp{AP} are stable during each transmission period. In this form \acp{AP} are likely to have spatially distributed multiple antennas \cite{demir2021foundations}. From a localization perspective, this form of cell-free gives both time and phase information across the spatially distributed arrays.
\end{itemize}

Despite the enormous advantages of \ac{CF-mMIMO}, challenges related to synchronizing dynamic \acp{AP} clusters and resource allocation are open research questions \cite{demir2021foundations,interdonato2019ubiquitous}. In the following, we present \ac{CF} potential impact within \ac{GAS} anchors.

\subsubsection{Cell-Free-Aided Ground Anchors}
Beyond communication, \ac{CF} networks have also been shown advantageous for sensing \cite{sakhnini2022near} and localization \cite{de2022expert} using ground anchors. The essential benefit is the extra \ac{DOF} obtained from the relatively large number of cooperative distributed anchors, i.e., \acp{AP}, improving \ac{GDOP} of location estimates. The rather large number of \acp{AP} envisioned in \ac{CF} networks enables them to serve ground and aerial users jointly \cite{d2020analysis}. Cell-free networks could also be envisioned to provide localization of not only ground and aerial active targets but also passive ones; however, a detailed investigation is needed to understand whether the benefits outweigh the drawbacks from energy and computational perspectives.

\subsubsection{Cell-Free Aided Aerial Anchors}
Similar to ground anchors, employing a \ac{CF} network architecture using aerial anchors benefits localization performance, offering extra \acp{DOF}. In general, synchronization is a critical challenge in \ac{CF} networks \cite{interdonato2019ubiquitous}, limiting the accuracy of time-based localization methods. On the one hand, aerial anchors can exploit the high probability of LOS links to employ over-the-air synchronization. However, on the other hand, the mobile nature of aerial anchors leads to even more dynamic \ac{AP} clusters, necessitating frequent synchronization. This same mobile nature of aerial anchors can be exploited to optimize the aerial \acp{AP} placement in a UAV-based \ac{CF} to serve mobile \ac{UE} \cite{diaz2022cell,wang2022cell}.

\subsubsection{Cell-Free-Aided Space Anchors}
The concept of building distributed antenna arrays for the reception of signals from space is essentially used in radio astronomy as a powerful method to increase \ac{RSS} \cite{jin2008analysis}. While these deployments use distributed ground stations, the concept could be extended to the aerial and space segments if anchor density increases there. In such cases, the spatial multiplexing needs to be done in 3D, e.g., by having multi-segment anchor cooperation \cite{riera2022scalable}, which in turn also serves target localization in 3D. \ac{CF-mMIMO} using ultra-dense \ac{LEO} satellite constellations is introduced in \cite{abdelsadek2021future}. However, the authors addressed the network architecture only from pilot assignment, beamforming, and handover management. The localization problem using \ac{CF-mMIMO} using ultra-dense \ac{LEO} satellite constellations is an open research question.

\subsection{THz Band}
\Ac{THz} communications are considered as a 6G enabler, going beyond \ac{mmWave} systems, covering communications in the band of 0.1-10~THz \cite{moltchanov2022tutorial}. Moving from mmWave in 5G to THz 6G systems, higher frequencies, larger bandwidths, and larger array sizes with smaller footprints are expected. The high frequencies at the \ac{THz} band increase pathloss, reduce multipath components, and the corresponding high bandwidths provide relatively high path delay resolvability. These characteristics make THz localization most fit for ground anchors. In \cite{chen2022tutorial}, the authors presented a detailed tutorial on radio localization in the \ac{THz} band, where mainly ground anchors are explored. A comparison of localization merits and challenges between mmWave and THz is provided. This comparison includes several aspects, such as array size, array type, hardware imperfections, synchronization, and propagation effect. In the following, we discuss \ac{THz} from the \ac{GAS} anchors' perspective. 

\subsubsection{THz-Aided Ground Anchors}
Existing works on \ac{THz} localization address system structures, localization algorithms, and simulation platforms. In particular, \ac{AOA}-based localization is attracting considerable research focus due to the narrow beams used in \ac{THz} that result in a very high angle resolvability \cite{chen2022tutorial}. Moreover, exploiting \ac{RIS} and THz is a promising research direction, enabling near-field localization and channel estimation \cite{pan2023ris}. 

\subsubsection{THz-Aided Aerial Anchors}
Similar to ground anchors, THz can boost the \ac{AOA}-based localization using \ac{LAP} aerial anchors. However, the difficulties that aerial anchors face in maintaining their position, e.g., wobbling because of wind, can lead to frequent beam misalignment, negatively influencing the localization stability \cite{azari2022thz}. In addition, THz can aid UAV-to-UAV communications, ensuring low overhead resources and interference management. 

\subsubsection{THz-Aided Space Anchors}
In space anchors, THz is mainly explored for inter-LEO satellite communications \cite{tekbiyik2022reconfigurable}, which may indirectly benefit the localization performance of ground and aerial anchors, e.g., by improving resource allocation and interference management between space anchors.

\subsection{6G KVIs}
The 6G enablers discussed in this section introduce new \acp{DOF} and design aspects, promising applications with unprecedented demands in terms of localization \acp{KPI} such as extended reality and digital twin \cite{behravan2022positioning}. \Acp{KVI} are adopted in \ac{6G} networks in order to broaden the assessment of future wireless networks beyond the technical aspects-focused \acp{KPI}, to also include forward-looking societal aspects \cite{wymeersch20236g}. They aim to better capture the spirit of the sustainable development goals defined by the United Nations \cite{united2022sustainable}. In the following, we present vital \acp{KVI} related to localization in \ac{GAS} networks.

\subsubsection{Trustworthiness}
Localization security, integrity/robustness, and user privacy can be embedded in the trustworthiness \ac{KVI} \cite{wymeersch20236g}. In GAS localization systems, where the distance between the anchors and targets is relatively very large, e.g., as in the case of aerial and space anchors, localization signals are more prone to jamming and spoofing \cite{pirayesh2022jamming,yue2023low}.
While maintaining the localization system's integrity is important to ground targets, its degradation typically has a broader impact on the aerial targets (due to lower shadowing), and it could lead to serious or even life-threatening consequences.
Due to the many recently proposed and/or launched LEO constellations, a huge opportunity to strengthen our \ac{PNT}-dependent infrastructure arises. The various radio signals from GAS anchors can be used to increase the diversity gain, boosting localization system robustness by exploiting redundancy to detect and eliminate faults. Moreover, with the large number of \ac{GAS} anchors, jamming and spoofing attacks can be partially mitigated by detecting attacks and switching to anchors or systems that are not jammed or spoofed \cite{yue2023low}. 
Despite this opportunity, when using \acp{SOP} from \ac{NTN} anchors, there is no guarantee that the used signal is always correct and consistent. Any lack of consistency in the used \ac{SOP} might impede the navigation service, which is particularly critical for aerial users as guaranteeing a position fix at all times is essential. To combat this shortcoming, cooperation with the constellation operator can be considered. This concept, called fused LEO GNSS \cite{Iannucci2020}, requires only a small part of the downlink capacity of modern mega-constellations to provide reliable \ac{PNT} services with a higher accuracy than traditional \ac{GNSS}. In addition, using \ac{LEO}'s \acp{SOP} in conjunction with another navigation system, e.g., \ac{GNSS}, LEO-\ac{PNT}, aerial-based or ground-based systems, will strengthen both systems' robustness and security flaws detection.

\subsubsection{Sustainability}
Wireless networks' sustainability includes energy efficiency, reusability, and maintenance. There are two aspects of sustainability in 6G \ac{GAS} localization systems, namely, using localization to ensure sustainable communications and making the localization process sustainable in itself. In the former, accurate user location information facilitates a more efficient resource allocation, e.g., energy-efficient and swift beam alignment \cite{xiao2022overview}, which is critical for battery-limited aerial anchors. In the latter, accurate localization systems should be designed to meet \acp{KPI}, but with constraints on allocated resources and energy consumption. Relevant examples include energy-constrained trajectory design for aerial anchors and the number of LEO satellites allocated to localize a target. Regarding reusability, \ac{JCAS} is expected to play a major role by designing waveforms and hardware that support multi-functionalities.

\subsubsection{Inclusiveness}
While the localization coverage as a \ac{KPI} can be used to assess the geographical area in which accurate localization is available, inclusiveness goes beyond coverage to also include the accessibility and affordability of the localization system \cite{wymeersch20236g}. \ac{GAS}-based localization, mainly using aerial and space anchors, is expected to have a significant impact on the localization inclusiveness aspect by offering global connectivity and localization coverage using off-the-shelf equipment \cite{ardito2019performance}.  

\subsubsection{Harmonized Operation}
In Release-17 of \ac{3GPP}, two new FR1 bands for \ac{NTN} are included, namely, \ac{NTN} 1.6~GHz and n256 NTN 2~GHz. Several additional frequencies within those ranges are being discussed within \ac{3GPP} for additional \ac{NTN} spectrum \cite{lin2022overview}. Moreover, in the long term, FR3 band (7-24~GHz) is emerging \cite{cui20236g}, along with bands above 10~GHz are targeted as well (around 17~GHz for UL and 27~GHz for DL), offering more frequency options to use for improving localization performance. As the spectrum gets more crowded, it becomes more difficult to avoid interference. An interesting and important topic for research is how we could make GAS localization networks interference-resilient.
Given the high output power requirements in GAS networks, it becomes difficult to meet tight spectrum mask requirements. Current ongoing research on out-of-band radiation for Massive MIMO and \ac{CF-mMIMO} is also of interest here, and an important topic \cite{liu2022power}, as out-of-band radiation also gets beamformed, especially in a LOS scenario, which occurs in high probability in aerial and space segments.

\subsection{Key Takeaways}
6G enablers, which include \ac{RIS}, \ac{JCAS}, \ac{AI}, \ac{CF}, and \ac{THz} are expected to play important roles in pushing localization system KPIs to unprecedented levels. A great amount of research effort was put into studying RIS-aided localization systems, where the majority of works focus on using RIS with ground anchors. Investigating RIS integration with aerial or space anchors is an interesting open research question, where relatively long distances and high anchor mobility are open challenges. \ac{JCAS} is a 6G enabler that offers a unified cost-efficient communication, sensing, and localization framework. While ongoing research efforts on \ac{JCAS} focus on waveform design and hardware architecture, the network architecture, and deployment scenarios when considering single or multiple \ac{GAS} segments are open research problems. Regarding network architecture, \ac{CF}, as a potential architecture in 6G networks, can boost the localization performance by exploiting the massive deployment of \acp{AP} acting as anchors. However, time and phase synchronization, as well as resource allocation, are open research questions in \ac{CF} networks, even in communication use cases \cite{chen2022survey,larsson2024massive}. To tackle synchronization and calibration challenges in wireless networks, \ac{AI}-empowered solutions are gaining considerable attention. \ac{AI}-empowered model-driven localization solutions can help address calibration and imperfections at all 3 GAS anchors; however, more effort needs to be invested in obtaining labeled datasets, which is a challenging task, particularly for aerial and space anchors. Another 6G enabler is the \ac{THz} band. \ac{THz} localization solutions enable remarkably accurate time and angle measurements. However, due to the relatively high pathloss, the use of \ac{THz} in \acp{NTN} will potentially focus on inter-anchor links, e.g., for synchronization and measurements fusion. Finally, \acp{KVI}, such as trustworthiness and sustainability, are envisioned to complement existing \acp{KPI}, opening new performance evaluation aspects.
\section{Conclusion} \label{sec:conc}
In this tutorial paper, we addressed the localization problem in GAS networks. Localization fundamentals, considering the 3D space of GAS networks, have been detailed, including coordination systems, localization measurables, methods, and KPIs. Moreover, we presented the system model, design aspects, and special considerations for ground, aerial, and space anchors, covering both ground and aerial targets' localization. The promising potential and advantages of using each of the GAS anchors have been explored, and key results have been discussed, giving a quantified perspective. We explored the role of the main 6G enablers, namely, \ac{RIS}, \ac{JCAS}, \ac{CF-mMIMO}, \ac{AI}, and \ac{THz}, in the localization problem in GAS networks, highlighting the prospective impact these enablers have on the different localization aspects, KPIs, and KVIs.

%


\bibliographystyle{ieeetr} 
\bibliography{NTN_Loca}
\balance

\end{document}